\documentclass[usegraphicx,useAMS]{mn2e}
\newcommand{\be}{\begin{equation}}
\newcommand{\ee}{\end{equation}}
\newcommand{\ben}{\begin{eqnarray}}
\newcommand{\een}{\end{eqnarray}}
\newcommand{\bc}{\begin{center}}
\newcommand{\ec}{\end{center}}

\begin{document}

\title[Microlensing of $\gamma$-ray blazars]
{Gravitational microlensing of $\gamma$-ray blazars }
\author[Torres, Romero, Eiroa, Wambsganss, \& Pessah]
{Diego F. Torres$^{1}$, Gustavo E. Romero$^{2,\;3}$,  Ernesto F.
Eiroa$^{4}$, \and Joachim Wambsganss$^{5}$,  and Mart\'{\i}n E.
Pessah$^{6}$
\\  $^1$
Physics Department, Princeton University, NJ 08544, USA\\ $^2$
Instituto Argentino de Radioastronom\'{\i}a (IAR),  C.C. 5, 1894
Villa Elisa, Buenos Aires, Argentina\\ $^{3}$ Max-Planck-Institute
f\"ur Kernphysik, Postfach 103980, D-69029 Heidelberg, Germany\\
$^4$ Instituto de Astronom\'{\i}a y F\'{\i}sica del
Espacio, C.C. 67, Suc. 28, 1428, Buenos Aires, Argentina\\
$^5$ Universit\"at Potsdam, Institut f\"ur Physik, Am Neuen Palais
10,
14469 Potsdam, Germany\\
$^6$ Steward Observatory, University of Arizona, AZ 85721, USA}

\maketitle

\begin{abstract}
We present a detailed study of the effects of gravitational
microlensing on compact and distant $\gamma$-ray blazars. These
objects have $\gamma$-ray emitting regions which are small enough
as to be affected by microlensing effects produced by stars lying
in intermediate galaxies. We compute the gravitational
magnification taking into account effects of the lensing and show
that, whereas the innermost $\gamma$-ray spheres can be
significantly magnified, there is little magnification either for
very high $\gamma$-ray energies or for lower (radio) frequencies
(because these wavelengths are emitted from larger regions). We
analyze the temporal evolution of the gamma-ray magnification for
sources moving in a caustic pattern field, where the combined
effects of thousands of stars are taken into account using a
numerical technique. We propose that some of the unidentified
$\gamma$-ray sources (particularly some of those lying at high
galactic latitude whose gamma-ray statistical properties are very
similar to detected $\gamma$-ray blazars) are indeed the result of
gravitational lensing magnification of background undetected
Active Galactic Nuclei (AGNs). This is partly supported from a
statistical point of view: we show herein as well, using the
latest information from the Third EGRET Catalog, that
high-latitude $\gamma$-ray sources have similar averaged
properties to already detected $\gamma$-ray AGNs. Some differences
between both samples, regarding the mean flux level, could also be
understood within the lensing model. With an adequate selection of
lensing parameters, it is possible to explain a variety of
$\gamma$-ray light curves with different time scales, including
non-variable sources. The absence of strong radio counterparts
could be naturally explained by differential magnification in the
extended source formalism. The spectral evolution of the sources
during microlensing events is calculated, revealing specific
features that can be used to test the models with the next
generation of both, orbital and ground-based $\gamma$-ray
telescopes.\\\mbox{}
\end{abstract}

\section{Introduction}

The first extragalactic $\gamma$-ray source detected was the
quasar 3C273, which was observed by the COS-B satellite in a
particularly active state in the 1970s (Swanenburg et al. 1978).
Since then, many Active Galactic Nuclei (AGNs) have been detected
at high energies, most of them belonging to the blazar class (e.g.
Mukherjee 2001). The Third EGRET Catalog of point-like sources
lists currently 67 detections labeled as AGNs (Hartman et al.
1999). Notwithstanding, a large number of $\gamma$-ray sources,
scattered along
the entire sky, remain unidentified at present.\\

The unidentified $\gamma$-ray sources at low latitudes are
probably related to several distinct galactic populations (Romero
2001). Among them there might be pulsars (e.g. Kaspi et al. 2000,
Zhang et al. 2000, Torres et al. 2001d; Camilo et al. 2001;
D'Amico et al. 2001), SNRs in interaction with molecular clouds
(Combi \& Romero 1995, Sturner et al. 1996; Esposito et al. 1996;
Combi et al. 1998; Combi et al. 2001, Butt et al. 2001, Torres et
al. 2002b), stellar-size black holes (Punsly 1998a,b; Punsly et
al. 2000), X-ray transients (Romero et al. 2001), persistent
microquasars (Paredes et al. 2000, Kaufman Bernad\'o et al. 2002),
and massive stars with strong stellar winds (Benaglia et al.
2001). Some of these kinds of stellar objects present statistical
positional correlation with unidentified EGRET sources (far from
what is expected as a random result, e.g. Romero et al. 1999,
Torres et al. 2001b). Pulsars, however, remain as the only
confirmed low-latitude population, since pulsed $\gamma$-ray
radiation has been already detected for at least six different
sources (Thompson et al. 1999, Thompson 2001; Kaspi et al. 2000),
five of them included in the Third EGRET Catalog (Hartman et al. 1999). \\

Gehrels et al. (2000) have shown that the mid-latitude sources are
different from the bright population of unidentified sources along
the Galactic plane. Some of the detections ($5^0<|b|<30^0$) are
thought to be associated with the Gould Belt (Grenier 2000,
Gehrels et al. 2000), a starburst region lying at $\sim600$ pc
from Earth. Few other sources, at higher latitudes, could be the
result of electrons being accelerated at the shock waves of
forming clusters of galaxies (Totani \& Kitayama 2000). However,
for many of the high-latitude sources, no other explanation seems
to be available than they are AGNs yet undetected at lower
energies. This is particularly clear when one looks at the
variability levels of the associated light curves: models
requiring large acceleration region, like clusters of galaxies,
would produce non-variable sources, contrary to what is found for
most of the high latitude sources.
\\

All identified 67 EGRET AGNs are also strong radio sources with
flat spectra, as expected from synchrotron jet-like sources where
the $\gamma$-ray flux is the result of inverse Compton scattering
(Mattox et al. 1997). However, no strong radio source appears
within the contours of the unidentified high latitude EGRET
sources. In this paper, we shall develop a model, briefly outlined
by Torres et al. (2002a), which focuses precisely on that
difference and provides an explanation of why some of the high
latitude unidentified sources might not be detected at low
frequencies. {\it The main feature of such a model is that it will
account for the $\gamma$-ray properties of the high latitude
gamma-ray detections resorting to differential gravitational
lensing magnification of background, high-redshift, AGNs with
otherwise undetected $\gamma$-ray emission. Since these objects
have different sizes at different wavelengths, differential
microlensing effects will lead to a magnification of the innermost
$\gamma$-ray emitting regions, whereas the radio emission will be
largely unmagnified, therefore remaining under the detection threshold.}\\

The gravitational light deflection effect by compact objects on
background sources is commonly  called microlensing (e.g.
Paczy\'nski 1986). A source would be affected by different
magnifications, depending on its position. Typically, source,
lens, and observer move relative to each other, and therefore,
this translates into a variable flux measured for the background
source. Observationally, there are two interesting regimes of
microlensing. Local microlensing deals with the light deflection
effects by stars inside  the Milky Way disk on stars in the
Galactic bulge. Here the probability for a microlensing event is
of order $\times 10^{-6}$. This means that it is necessary to
monitor millions of stars in order to see a few occurrences. But
despite this small probability, various teams have been very
successful in detecting this kind of events in recent years (for a
review, see Paczy\'nski 1996). The other interesting regime of
microlensing is usually called quasar microlensing, but it can be
applied to any other compact source at moderate to high redshift.
In this case, an intervening galaxy provides the surface mass
density in stars (or other compact objects) which act as
microlenses on the background quasar (for a review, see Wambsganss
2001). Recently, this kind of microlensing has been suggested for
other astrophysical sources as well, e.g. gamma-ray bursts
(Williams \& Wijers 1997), gamma-ray burst afterglows (Garnavich,
Loeb \& Stanek 2000; Mao \& Loeb 2001; Koopmans \& Wambsganss
2001), and superluminal shocks in extragalactic radio sources
(Romero et al 1995, Koopmans \& de Bruyn 2000). The gamma-ray
sources discussed here are another type of astrophysical objects
for which
microlensing possibly plays an important role.\\

We recall that gravitational light deflection is basically an
achromatic phenomenon, being a geometric effect predicted by
General Relativity, i.e. the deflection angle does not depend on
the energy of the photon. However, it is nevertheless possible to
have chromaticity effects when the size of the source changes with
the observing wavelength. A large source is typically less
affected by a microlensing magnification than a small source
(Wambsganss \& Paczy\'nski 1991). This size-induced chromaticity
will be an essential ingredient of our model.\\

The structure of the paper is as follows. In Section 2 we present
a statistical study of the main characteristics of the sample of
unidentified $\gamma$-ray sources at high latitudes, including
variability. In Section 3 we review the formalism for
extragalactic gravitational microlensing in both the point-like
and extended source cases. Section 4 comments on the inner
structure of $\gamma$-ray emitting AGNs considered here as
background sources. Section 5 contains our results for the flux
and spectral evolution of single microlensing events in a variety
of situations and different types of host galaxies. Section 6
deals with the optical depth problem and the expected number of
microlensing events. Section 7 presents our results for full
numerical modeling of caustic patterns for different galactic
lensing parameters. Precise predictions for the light curves at
different frequencies are also shown there. We finally close with
some concluding remarks in Section 8.

\section{Unidentified $\gamma$-ray sources at high latitudes}

\subsection{Sample and photon spectral index}

Previous population studies using the Second EGRET Catalog have
already remarked that part of the sample of high latitude
unidentified sources is consistent with an isotropic population, a
fact that supports an extragalactic origin for these detections
(\"Ozel \& Thompson 1996). In what follows, we shall make a
comparison between the properties of identified $\gamma$-ray AGNs
in the Third EGRET Catalog and high-latitude unidentified sources.
We shall choose the lower cut-off in latitude as $|b|=30^0$, in
order to avoid possible contamination from Gould Belt sources.
There are 45 3EG unidentified sources within this latitude range;
we provide details on these sources in Table \ref{t21}.\\

\begin{table*}
\caption{The 45 unidentified sources considered in the analysis.
We list their 3EG Catalog name, their Galactic coordinates,
spectral index, variability index, and the values of
$\left<F\right>$ used to define $I$ in Eq. (\ref{rr}). We also
provide the 3EG  P1234 fluxes, $F$. $\left<F\right>$ and $F$ are
in units of $10^{-8}$ photons cm$^{-2}$ s$^{-1}$. The columns
labeled $\tau$, $\tau_{min}$, and $\tau_{max}$, give the central
value of Tompkins' (1999) index for variability and their 68\% CL
lower and upper limit deviations, respectively. The pulsar
population has $\left<\tau\right><0.1$, whereas typical AGNs have
$\left<\tau\right>\sim 0.7$. Extreme upper limits for $\tau$,
whose maximum is 10000, imply possible strong variability. }
\label{t21}
\begin{center}
\begin{tabular}{lrrcllcccc}
3EG       &  $l$ &  $b$ &  Spectral &  $I$ & $\tau$ &$\tau_{min}$
& $\tau_{max}$ &
$\left<F\right>$ &  $F$\\
JSource      &   &   &  index &     &
 & \\
\hline
0245$+$1758 &  157.62 &  $-$37.11 &  2.61 &  2.74 &    2.63    &  0.73    & 2287& 11.3 &  8.8\\
0404$+$0700 &  184.00 &  $-$32.15 &  2.65 &  1.50 &    0.34    &  0.00     & 1.65  & 13.5 &  11.1\\
0512$-$6150 &  271.25 &  $-$35.28 &  2.40 &  2.34 &       0.00   &   0.00   &  0.55 & 10.8 &  7.2\\
0530$-$3626 &  240.94 &  $-$31.29 &  2.63 &  1.62 &  17.8 &  15.8 & 0.61    &  0.15   &  2.28\\
0808$+$4844 &  170.46 &  32.48 &  2.15    &  0.90 &   0.00    &  0.00   &  0.39& 12.2 &  10.7\\
0808$+$5114 &  167.51 &  32.66 &  2.76    &  1.53 &   0.00    &  0.00   &  0.73&13.5 &  8.7\\
0910$+$6556 &  148.30 &  38.56 &  2.20    &  1.79 & 0.49     & 0.00   &  1.14& 9.2 &  5.9\\
1457$-$1903 &  339.88 &  34.60 &  2.67    &  2.72 &     0.42    &  0.00     &3.64&13.9 &  8.1\\
1504$-$1537 &  344.04 &  36.38 &  --      &  2.73 &    10.33   &   1.21   &  9999.& 12.9 &  8.8\\
1600$-$0351 &  6.30 &  34.81 &  2.65      &  7.26 &  68.17    &  0.00   &  9999.& 9.7 &  9.9\\
1621$+$8203 &  115.53 &  31.77 &  2.29    &  0.34 &   0.00     & 0.00    & 0.29&11.5 &  7.4\\
1733$+$6017 &  89.12 &  32.94 &  3.00     &  1.82 &    0.39     & 0.00    & 1.38& 16.9 &  8.7\\
1958$-$4443 &  354.85 &  $-$30.13 &  --   &  7.43 &   58.02     & 5.85 &    9999. & 11.3 &  6.4\\
2034$-$3110 &  12.25 &  $-$34.64 &  3.43  &  5.26 &  2.88      &0.89    & 155. & 6.8 &  5.2\\
2219$-$7941 &  310.64 &  $-$35.06 &  2.50 &  1.37 &   0.00    &  0.00   &  0.51& 19.8 &  13.5\\
2243$+$1509 &  82.69 &  $-$37.49 &  --    &  11.11&   3.42   &   0.88  &   3097 & 7.4 &  9.9\\
2248$+$1745 &  86.00 &  $-$36.17 &  2.11  &  2.20 &  1.07   &   0.43    & 3.98& 20.2 &  12.9\\
2255$+$1943 &  89.03 &  $-$35.43 &  2.36  &  5.54 &     2.31     & 0.80  &   48.6 &14.2 &  5.8\\
0038$-$0949 &  112.69 &  $-$72.44 &  2.70 &  2.71 &   0.00      &0.00   &  0.89 &15.3 &  12.0\\
0118$+$0248 &  136.23 &  $-$59.36 &  2.63 &  2.28 &      5.17  &    0.90 &    9999.& 11.5 &  5.1\\
0130$-$1758 &  169.71 &  $-$77.11 &  2.50 &  0.10 &   0.00    &  0.00    & 0.38& 13.5 &  11.6\\
0159$-$3603 &  248.89 &  $-$73.04 &  2.89 &  1.26 &   0.00   &   0.00    & 1.16 &11.9 &  9.8\\
0215$+$1123 &  153.75 &  $-$46.37 &  2.03 &  3.67 &     10.06   &   1.19  &   9999.& 8.0 &  9.3\\
0253$-$0345 &  179.70 &  $-$52.56 &  --   &  7.99 &   16.44    &  1.38 &    9999.&  5.0 &  6.2\\
0348$-$5708 &  269.35 &  $-$46.79 &  --   &  4.76 &    6.60   &   1.29 &    9999.& 5.8 &  3.8\\
0917$+$4427 &  176.11 &  44.19 &  2.19    &  1.69 &   0.00     & 0.00 &  0.34& 18.6 &  13.8\\
1009$+$4855 &  166.87 &  51.99 &  1.90    &  1.14 &    0.00     & 0.00 &    0.60 &7.7 &  4.8\\
1052$+$5718 &  149.47 &  53.27 &  2.51    &  2.02 &    0.21      &0.00  &   0.74& 6.8 &  5.0\\
1133$+$0033 &  264.52 &  57.48 &  2.73    &  4.44 &   0.71      &0.16    & 2.00& 9.1 &  3.7\\
1134$-$1530 &  277.04 &  43.48 &  2.70    &  3.35 &    2.85    &  1.11 &    51.5& 17.2 &  9.9\\
1212$+$2304 &  235.57 &  80.32 &  2.76    &  5.65 &   78.82   &   0.00  &   9999.& 6.3 &  3.3\\
1219$-$1520 &  291.56 &  46.82 &  2.52    &  3.16 &   1.78    &  0.74    & 13.7 & 8.9 &  4.1\\
1222$+$2315 &  241.87 &  82.39 &  --      &  1.95 &  61.09    &  2.44  &   9999. &6.9 &  5.7\\
1227$+$4302 &  138.63 &  73.33 &  --      &  3.20 &   61.09    &  2.44  &   9999. &6.7 &  4.6\\
1234$-$1318 &  296.43 &  49.34 &  2.09    &  2.03 &      0.42  &    0.12 &    0.81 &11.0 &  7.3\\
1235$+$0233 &  293.28 &  65.13 &  2.39    &  1.04 &     0.23  &    0.00   &  0.65& 10.5 &  6.8\\
1236$+$0457 &  292.59 &  67.52 &  2.48    &  1.70  & 0.00    &  0.00  &   1.45 & 7.3 &  6.5\\
1310$-$0517 &  311.69 &  57.25 &  2.34    &  1.28 &      2.94   &   1.69 &    7.92 &11.4 &  7.9\\
1323$+$2200 &  359.33 &  81.15 &  1.86    &  5.17 &   2.69     & 0.93   &  46.8 &10.1 &  5.2\\
1337$+$5029 &  105.40 &  65.04 &  1.83    &  2.85 &     0.54   &   0.00  &   1.35& 10.0&  9.2\\
1347$+$2932 &  47.31 &  77.50 &  2.51     &  1.10 &    0.48    &  0.00  &   1.45 &15.3 &  9.6\\
1424$+$3734 &  66.82 &  67.76 &  3.25     &  1.90 &  0.01     & 0.00  &   9999.& 18.0 &  --\\
2241$-$6736 &  319.81 &  $-$45.02 &  2.39 &  1.25 &   0.00    &  0.00   &  1.09 &16.6 &  --\\
2251$-$1341 &  52.48 &  $-$58.91 &  2.43  &  5.17 &   9.49 &     1.58 &    9999. &9.7 &  6.5\\
2255$-$5012 &  338.75 &  $-$58.12 &  2.79 &  1.59 &  0.41 &0.00 &
1.46& 12.7& 9.2\\ \hline\hline
\end{tabular}
\end{center}

\end{table*}

\begin{figure*}
\includegraphics[width=8cm,height=11cm]{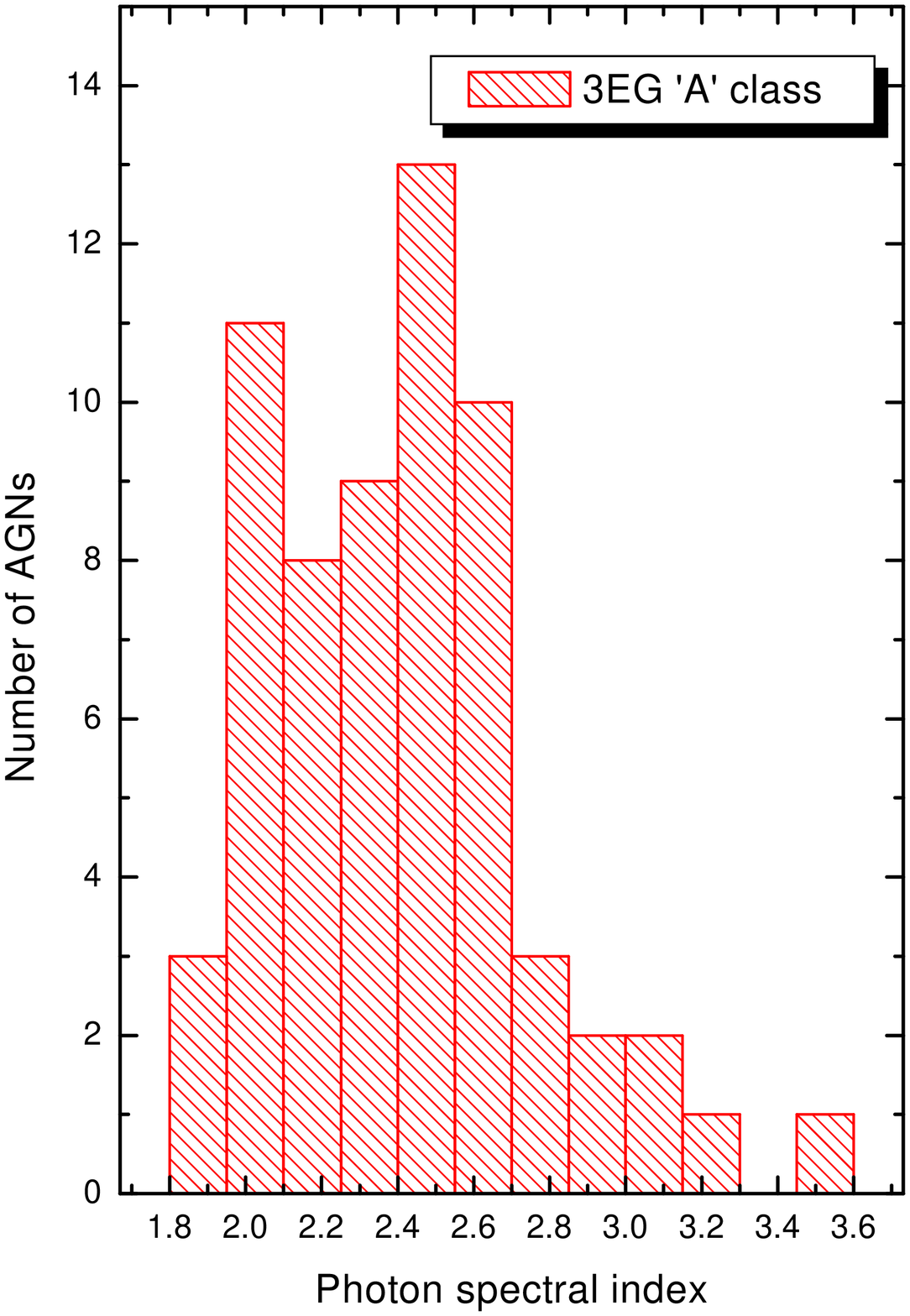}
\includegraphics[width=8cm,height=11cm]{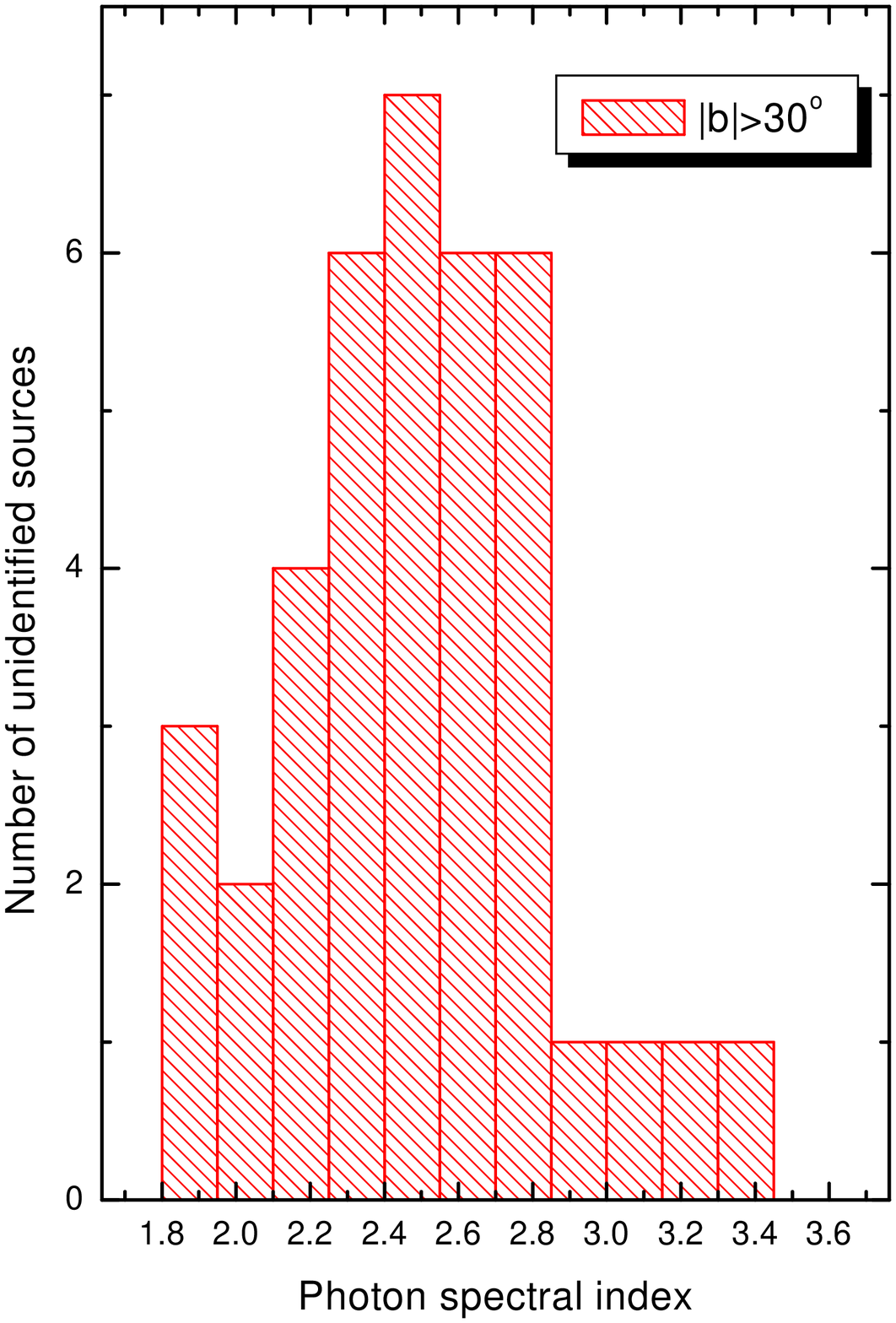}
\caption{Photon spectral index comparison. The left panel shows
the distribution for the 67 detected AGNs, dubbed A, in the Third
EGRET Catalog. The right panel shows the corresponding
distribution for the $|b|>30^{\rm o}$ unidentified sources. }
\label{spec}
\end{figure*}

Figure \ref{spec} shows the distribution of the $\gamma$-ray
photon spectral index for both sets of sources, 45 unidentified
EGRET sources and 67 detected EGRET AGNs. The mean value of the
photon index is 2.36$\pm$0.36 for AGNs, and 2.49$\pm$0.34 for the
unidentified detections. They are compatible within the
uncertainties and, on average, steeper than what is observed for
low latitude sources, which are thought to belong to our own
Galaxy.

\subsection{Variability}

The study of the time variability of $\gamma$-ray sources is of
fundamental importance. Several models for $\gamma$-ray sources in
our Galaxy predict non-variable emission during the time scale of
EGRET observations. AGNs, on the contrary, are expected to present
a variable flux emission. Variability is, in case we were able to
quantify it with some degree of precision, a powerful tool to
probe the nature of the sources. Visual inspection of the flux
evolution through the different viewing periods is obviously a
first indication of the variability status of any given source.
However, fluxes are usually the result of only a handful of
incoming photons, experimental errors are sometimes huge and their
origin uncertain, and consequently more reliable ways of
quantizing the flux evolution should be devised: these are known
as variability indices. Two such indices have been recently
introduced in the literature and applied to 3EG sources so far
(Tompkins 1999 (index $\tau$), Torres et al. 2001a (index $I$)).
In general, statistical results from these two indices are well
correlated (see Torres et al. 2001c for a discussion). Here we
shall adopt the index $I$, previously used in blazar variability
analysis (Romero et al. 1995) and applied to some of the 3EG
sources by Zhang et al. (2000) and Torres et al. (2001a,c) as our
main quantitative evaluation of variability, although the results
for $\tau$ are also given in Table 1. The basic idea behind the
index $I$ is to directly compare the flux variation of any given
source with that shown by known $\gamma$-ray pulsars, which are
assumed to be an intrinsically non-variable population. This
index, contrary to Tompkins' index $\tau$ (Tompkins 1999), uses
only the publicly available data of the 3EG Catalog.\\

\begin{figure*}
\includegraphics[width=8cm,height=11cm]{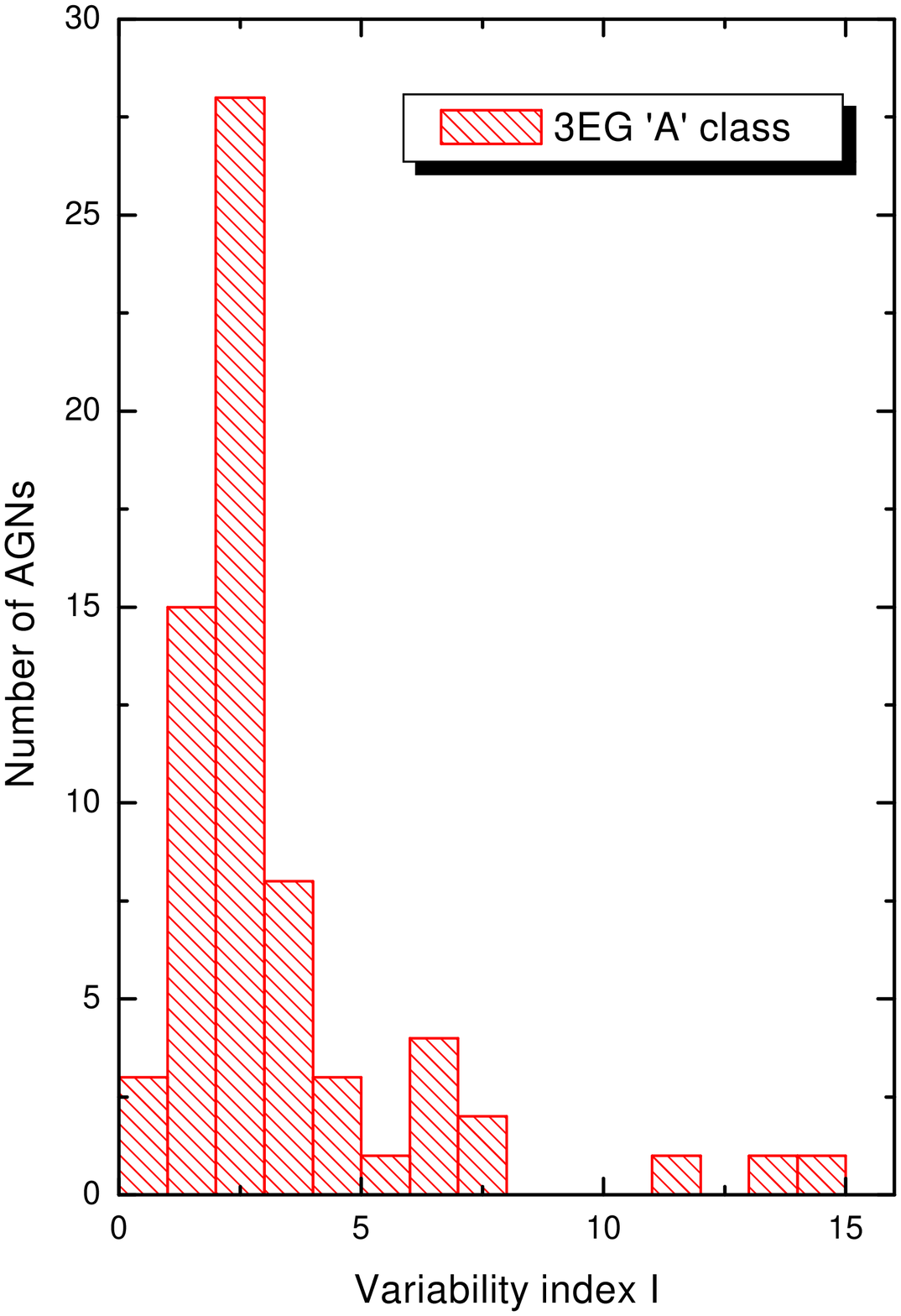}
\includegraphics[width=8cm,height=11cm]{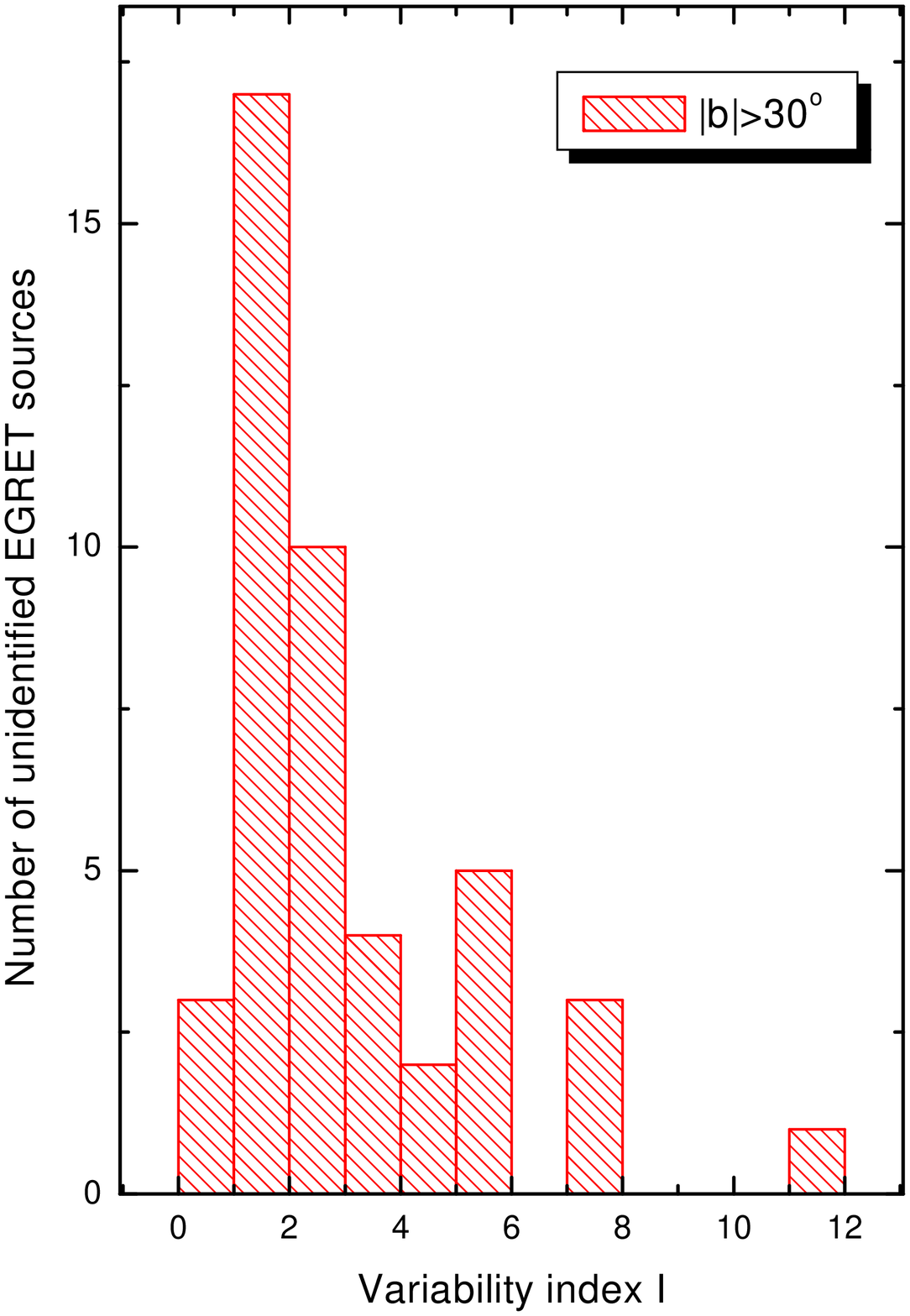}
\caption{Variability index comparison. The left panel shows the
distribution for the 67 detected AGNs, dubbed A, in the Third
EGRET Catalog. The right panel shows the corresponding
distribution for the $|b|>30^{\rm o}$ unidentified sources. }
\label{Idis}
\end{figure*}

Let us recall the basic elements that are used to define the
$I$-index. Firstly, a mean weighted value for the EGRET flux is
computed: \be \left< F \right> = \left[ \sum_{i=1}^{N_{{\rm vp}}}
\frac{F(i)}{\epsilon(i)^2} \right]\times \left[
\sum_{i=1}^{N_{{\rm vp}}} \frac 1{\epsilon(i)^{2}}
\right]^{-1},\ee where $N_{{\rm vp}}$ is the number of single
viewing periods, $F(i)$ the observed flux in the $i^{{\rm
th}}$-period, and $\epsilon(i)$ the corresponding error. For those
observations in which the significance ($\sqrt{TS}$ in the EGRET
catalog) is greater than 2$\sigma$, the error is $\epsilon(i) =
F(i)/\sqrt{TS}$. For those observations which are in fact upper
bounds on the flux, it is assumed that both $F(i)$ and
$\epsilon(i)$ are half the value of the upper bound. The
fluctuation index $\mu$ is defined as: \be \mu =100\times
\sigma_{{\rm sd}}\times \left< F \right>^{-1} . \label{rr}\ee In
this expression, $\sigma_{{\rm sd}}$ is the standard deviation of
the flux measurements. This fluctuation index is also computed for
the confirmed $\gamma$-ray pulsars in the 3EG catalog, assuming
the physical criterion that pulsars are non-variable $\gamma$-ray
sources. The averaged statistical index of variability, $I$, is
then given by the ratio \be I=\frac{\mu_{{\rm
source}}}{<\mu>_{{\rm pulsars}}}.
\ee Once the index is defined, we need to clarify the thresholds
for variability. Following Torres et al. (2001c), clearly variable
sources will be those with $I>5$, possibly variable sources will
have $2.5<I<5$, non-variable sources will have $I<1.7$, and the
remaining sources will be considered as dubious cases. These are
very conservative cut-offs: $I>5$ means that we are asking for the
value of $I$ to be 8$\sigma$ above that of pulsars in order to
classify a source as variable.
\\

In Figure \ref{Idis} we compare the $I$-index distribution for the
samples under analysis. The mean value for AGNs (left panel) is
3.3$\pm$2.6. A possible peak in the plot is seen at $I=2.5$, which
represents a value $4\sigma$ above that presented by pulsars. An
apparently extra peak appears at $I\sim 6$. Clearly, most of the
AGNs are likely variable sources. The mean for the unidentified
sources (Figure \ref{Idis}, right panel) is also high:
3.0$\pm$2.3. There are, again, apparently two distinct
peaks in the distribution, located at $I\sim 2$ and $I\sim 6$.\\

\begin{figure*}
\includegraphics[width=8cm,height=13cm]{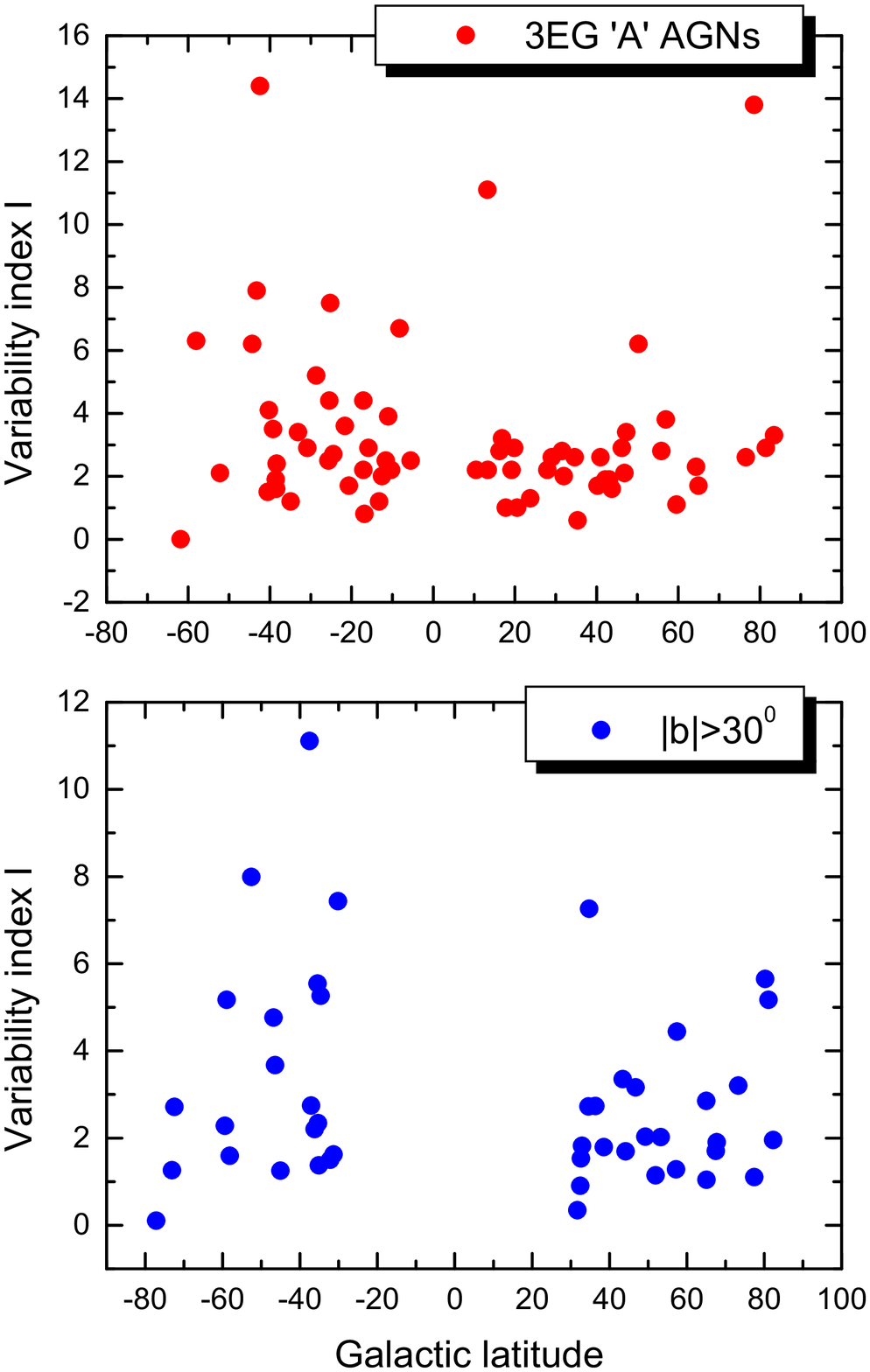}
\includegraphics[width=8cm,height=13cm]{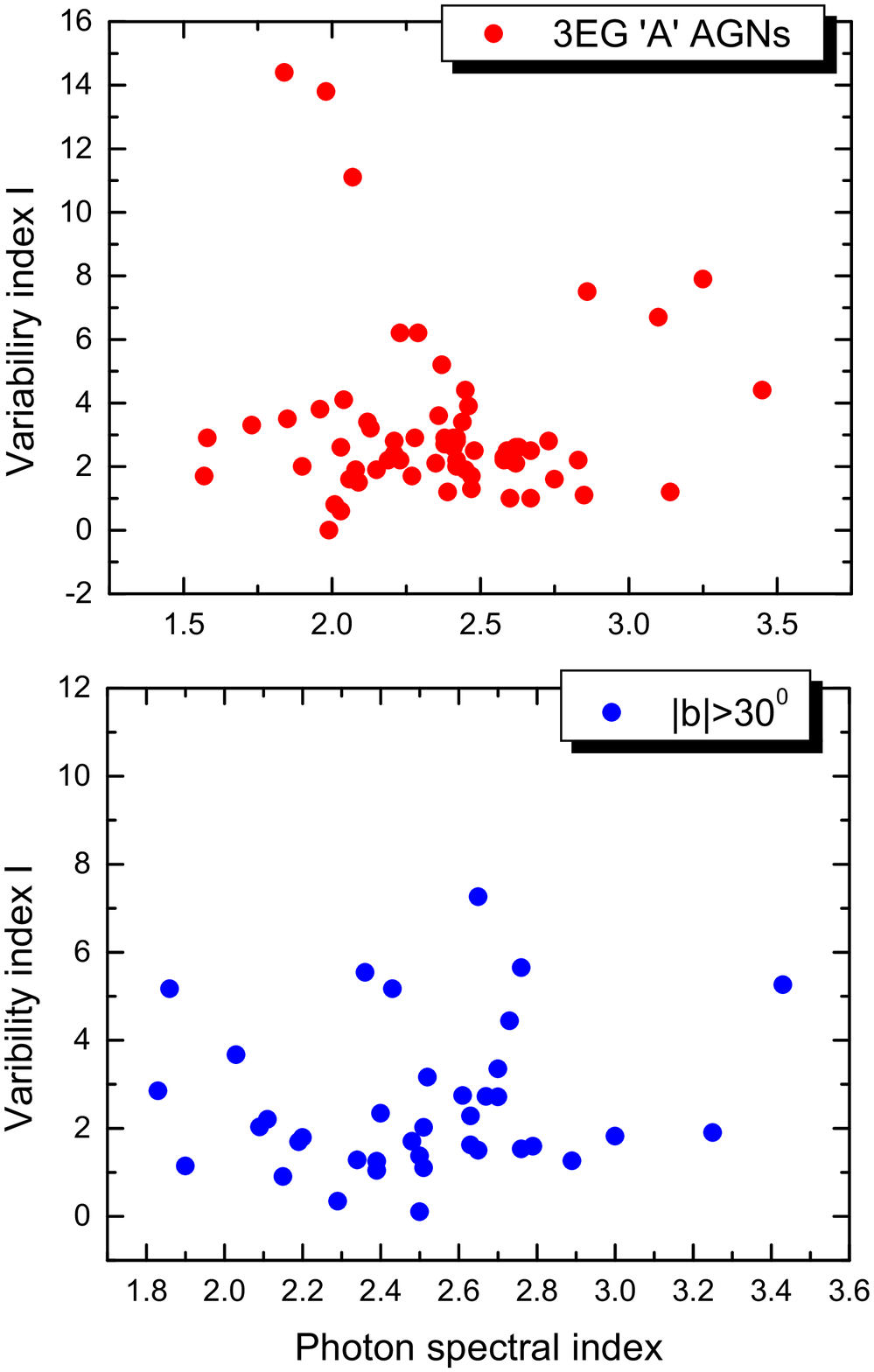}
\caption{Variability index $I$ (Torres et al. 2001) versus
Galactic latitude (left) and versus photon spectral index
(right).} \label{Ib}
\end{figure*}

In Figure \ref{Ib} (left panel) we show the variability index $I$
versus the Galactic latitude. The constraint we are imposing on
our sample of unidentified sources (to have $|b|>30^0$) can be
clearly noticed in the bottom plot. There is not a clear
dependence of the variability index with latitude, neither for
AGNs nor for unidentified sources. The same happens in the plots
of Figure \ref{Ib} (right panel), where we show the variability
index versus the photon spectral index. An apparent trend of
increasing the variability status for the steepest sources,
already noticed by Torres et al. (2001a) and Reimer (2001), is
shown in this figure. However, this is not conclusive since
results for a Spearman Rank test are in the range of a few percent
for this to be a random phenomenon. An overall characteristic of
Figures \ref{Idis} and \ref{Ib} is that both samples look quite
similar, with no apparent strong deviation from each other shown
in terms of variability or photon spectral index distributions.

\begin{figure*}
\includegraphics[width=8cm,height=13cm]{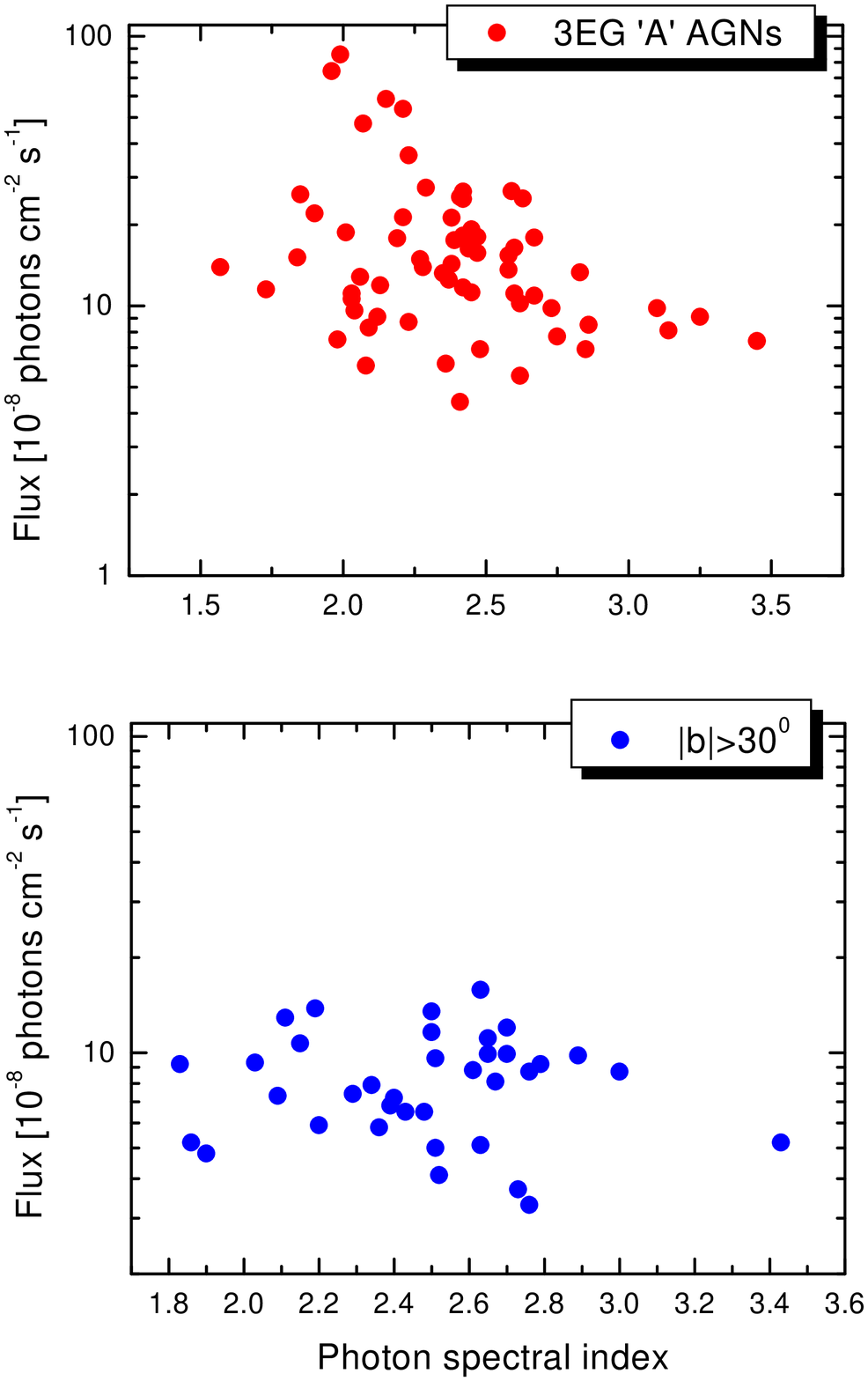}
\includegraphics[width=8cm,height=13cm]{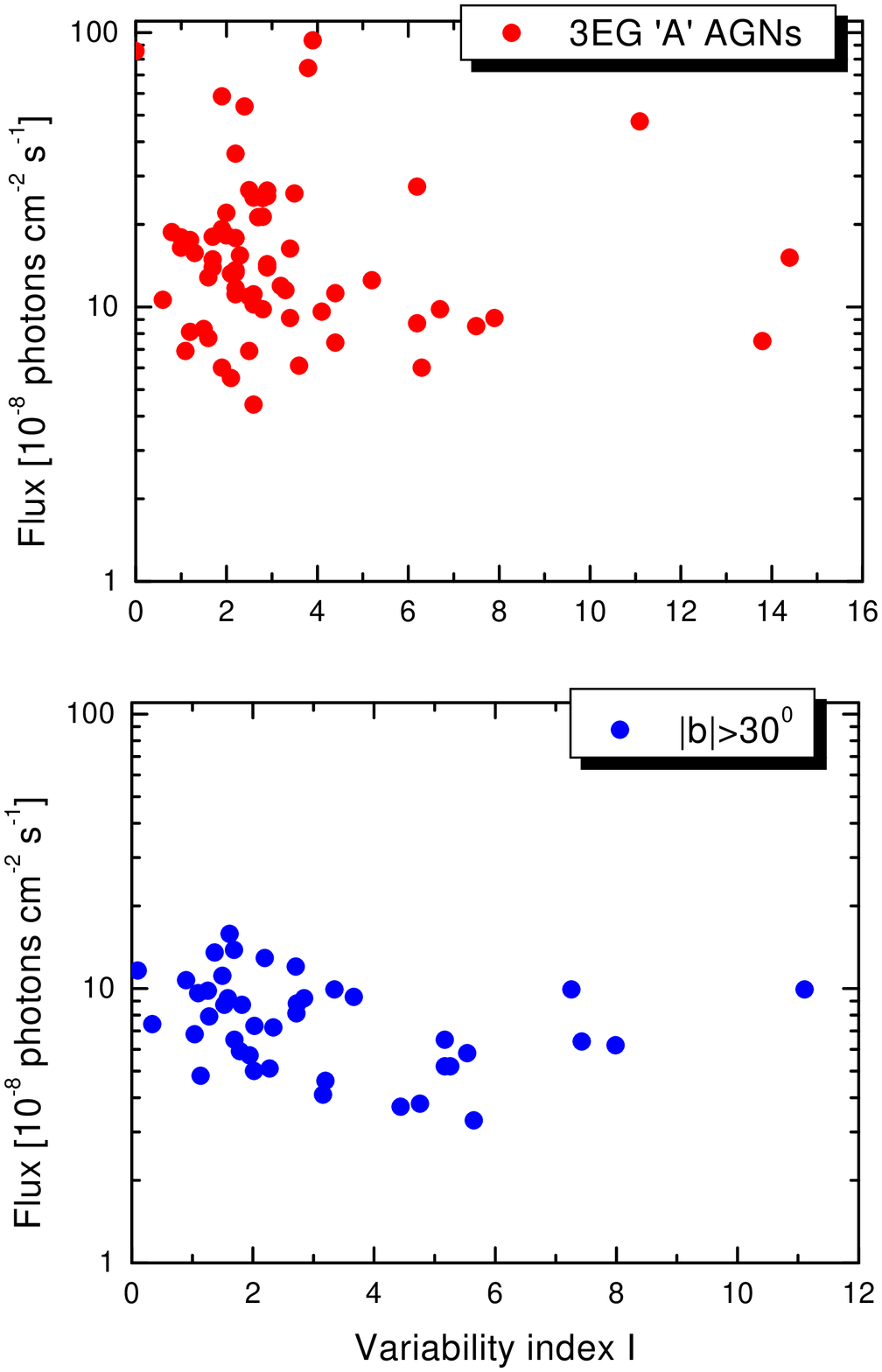}
\caption{Source fluxes (P1234, Hartman et al. 1999) versus photon
spectral index (left) and  variability index (right).}
\label{fgama}
\end{figure*}

\subsection{Fluxes and possible radio counterparts}

In Figure \ref{fgama} we show the EGRET averaged flux as a
function of the photon spectral index and the variability index,
respectively. We notice that, although there is no apparent
difference in the form of the distribution for both samples under
consideration, there is a clear contrast on the flux values:
Whereas most identified $\gamma$-ray AGNs have fluxes above
$10^{-7}$ photons cm$^{-2}$ s$^{-1}$, most of the unidentified
sources present lower values. This is consistent with what was
presented by Gehrels et al. (2000) for sources at latitudes
$|b|>5^0$.\\

\begin{figure*}
\includegraphics[width=8cm,height=11cm]{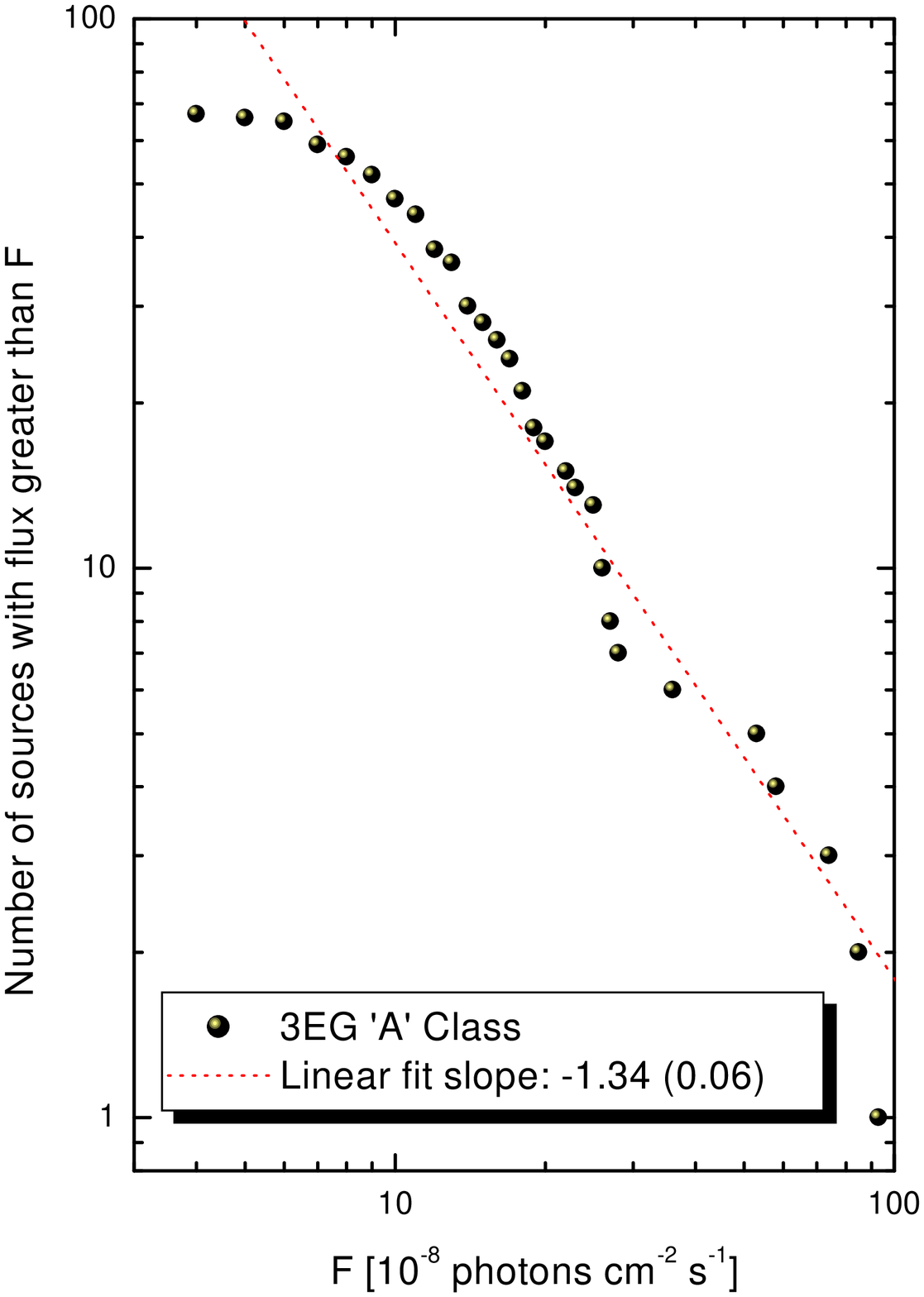}
\includegraphics[width=8cm,height=11cm]{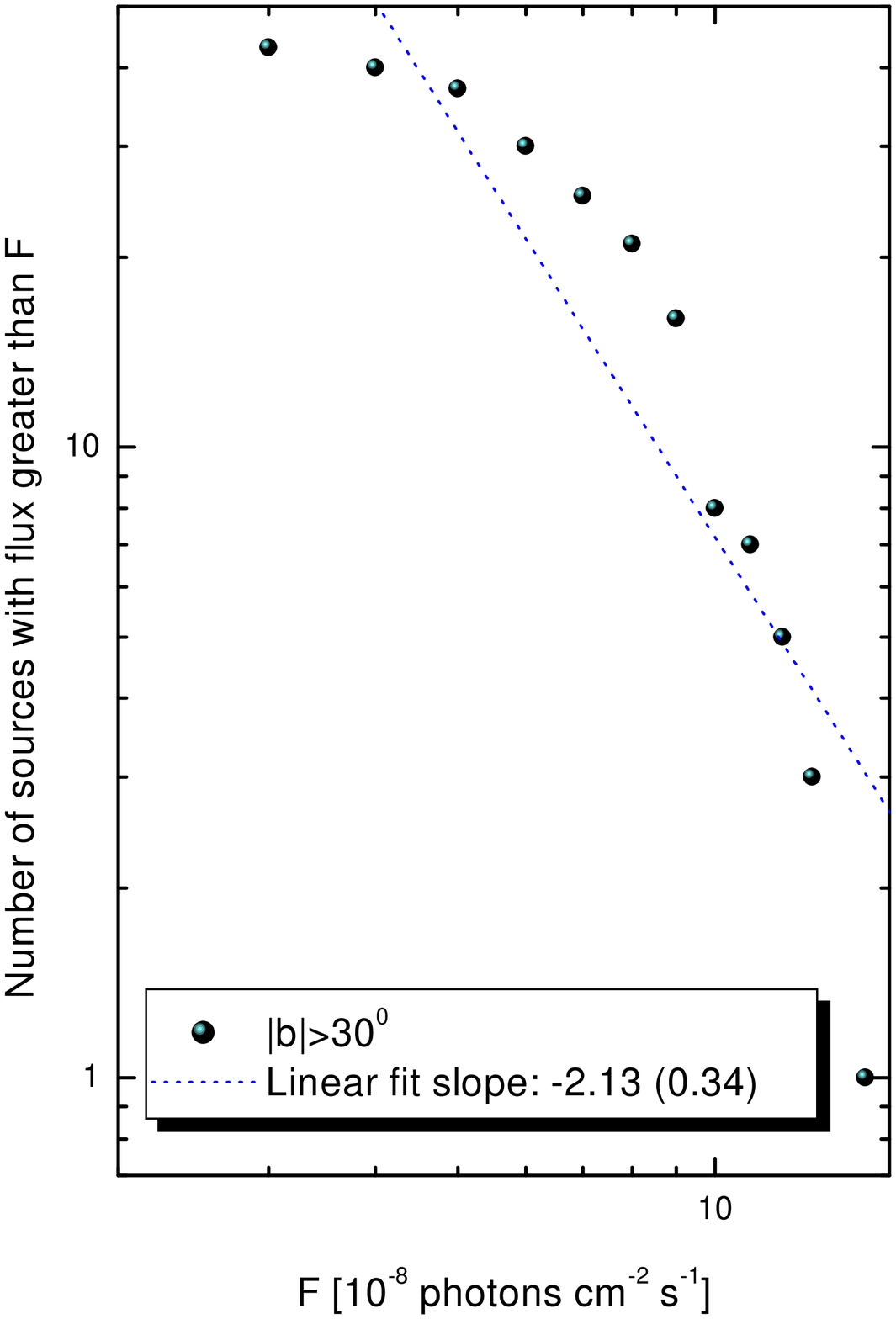}
\caption{Log N--Log $F$ comparison. The value of the linear fit
slope (and error within parentheses) is shown for each case. $F$
is the 3EG P1234 flux.} \label{log}
\end{figure*}

This difference in the flux values is also translated into the Log
N -- Log $F$ plots we present in Figure \ref{log}. It can be seen
there that the linear fits differ significantly. AGNs present a
fit close to what is expected for an isotropic and uniform
population ($F^{-3/2}$), and also similar to what was found using
the 2EG Catalog (\"Ozel \& Thompson 1996). The unidentified
sources, however, present a steeper dependence. The difference in
the flux levels is also shown in the $x$-axis. However, the
analysis of the result for the sample of unidentified sources
should be done with extra care, since the errors are far larger,
as well as  the number of sources considered is smaller.
Additionally, AGNs present an apparent lack of sources at $F\sim
30\; \times 10^{-8}$ photons cm$^{-2}$ s$^{-1}$, which should be
confirmed or falsified by future observations.\\

Reimer \& Thompson (2001) studied in detail the log N -- log $F$
plots obtained from 3EG sources, but including also those sources
with lower confidence level (which did not appear in the published
version of the 3EG catalog).  They found that there is a very
pronounced  contrast between average and peak flux representation
in a log N -- log $F$ diagram for the sources above $|b|>30^0$.
This is due to the fact that sources at high latitudes are mostly
detected only in some (or in many cases, only in one) viewing
periods (see below), when they show their peak flux, leaving  the
average over the four phases of the experiment in a much lower
value. The differences in fluxes between the peak detections of
both distributions, although still present, are not so strong as
the ones presented in the P1234 averaged values. As we shall see
below, sources showing large fluxes only in one viewing period
could be particularly suitable to be explained by microlensing of
gamma-ray blazars.\\

The differences in the Log N -- Log $F$ plots can be pointing
towards one of the two following possibilities:
\begin{enumerate}
  \item We are looking at (at least) two different populations; for
instance, AGNs and a new halo class of high-energy objects.
  \item These samples are formed mostly by the
  same kind of objects (AGNs) but they present different $\gamma$-ray flux
levels. This difference could be produced as an extrinsic effect
when the sources are farther away from us than those which produce
the most energetic detections.
\end{enumerate}

\begin{figure}
\begin{flushleft}
\includegraphics[width=8cm,height=11cm]{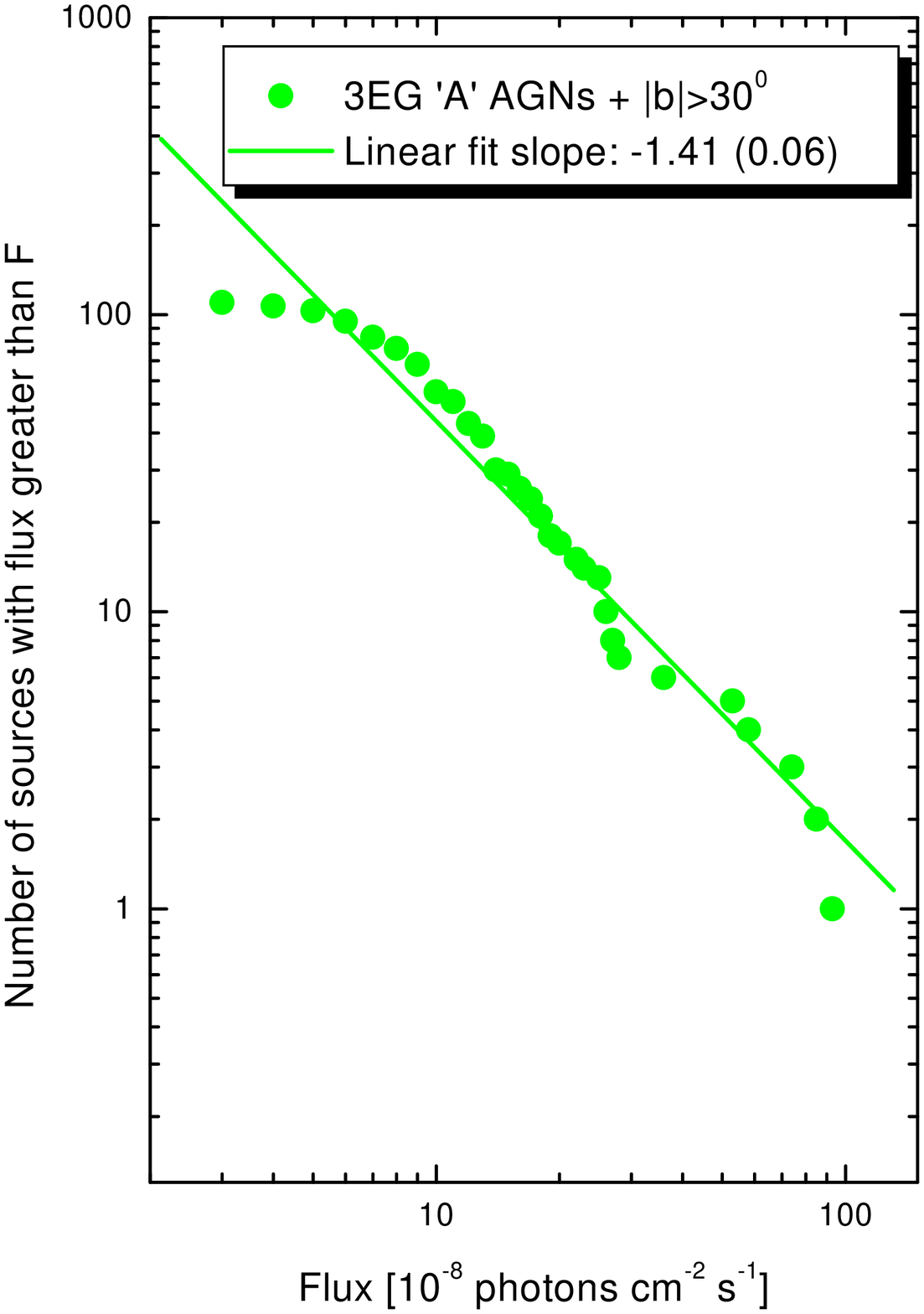}
\end{flushleft}
\caption{Log N--Log $F$ for the combined samples. The value of the
linear fit slope (and error within parentheses) is shown.}
\label{logall}
\end{figure}

In this paper we shall explore a plausible situation in which the
second possibility prevails. If behind both samples there is
actually a single class of extragalactic objects, the combined Log
N -- Log $F$ plot should approximately follow a $N\propto
F^{-3/2}$ law. The result, given in Figure \ref{logall}, confirms
this and it is close to the isotropic and uniform expectation (see
also the comments by Reimer \& Thompson 2001). This seems to be
suggesting that the sample of unidentified sources under
consideration is formed by many weak AGNs, which at the same time
presents a low (i.e. under the thresholds of the corresponding
surveys) radio emission. We note that the apparent bump at $F\sim
30\; \times 10^{-8}$ photons cm$^{-2}$ s$^{-1}$ continues to
appear in the combined plot, since no unidentified sources present
such
relatively high $\gamma$-ray fluxes.\\

\begin{figure}
\begin{flushleft}
\includegraphics[width=8cm,height=11cm]{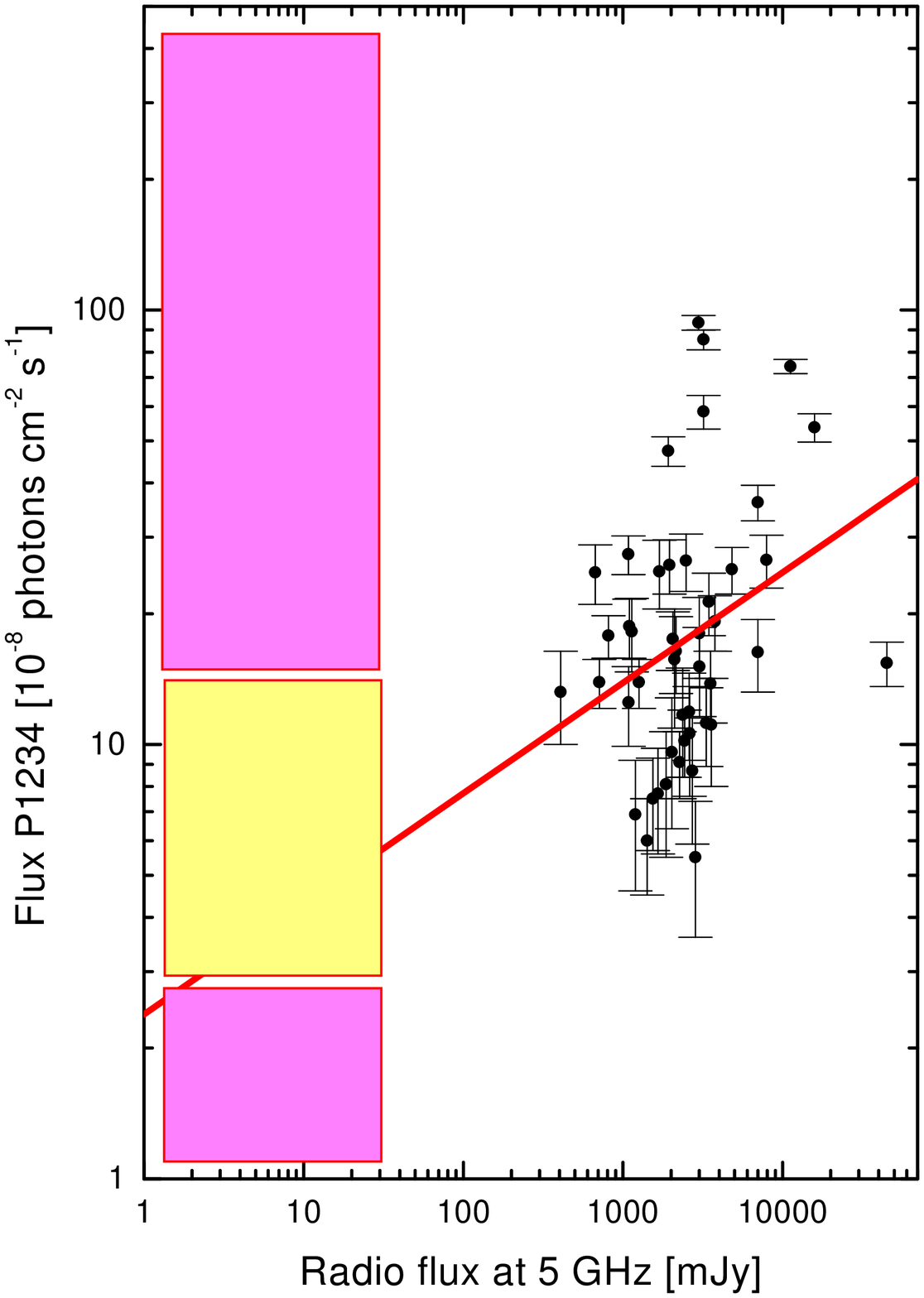}
\end{flushleft}
\caption{Radio and $\gamma$-ray fluxes of the 46 most likely AGN
detections. The big vertical box signals the threshold sensibility
in the radio surveys used to search for counterparts ($\sim 30$
mJy). The middle box signals the range of $\gamma$-ray fluxes for
the unidentified sources considered in these paper. }
\label{radio}
\end{figure}

Recently, Mattox et al. (2001) have presented a quantitative
analysis of potential radio identifications for all 3EG sources.
They used radio surveys at 5 GHz, as it was done previously for
the 2EG Catalog (Mattox et al. 1997), and evaluated an a priori
probability for these associations to be physical, based on the
positional offsets and radio fluxes of the proposed counterparts.
They found that 45 out of the 67 3EG sources classified with `A'
by Hartman et al. (1999) were among the EGRET identifications with
the highest probability of being correct. Only one extra possible
association in the list of these most likely identifications was
not dubbed `A' in the 3EG Catalog. For each of these 46
associations, they have compiled radio fluxes at 5 GHz, and when
available, also those obtained with VLBI. In Figure \ref{radio} we
show the $\gamma$-ray flux of each of these 3EG sources (note that
we plot the P1234 EGRET flux of the 3EG source from Hartman et al.
1999, not the flux of the AGN quoted, for instance, in Mattox et
al. 1997) as a function of the radio flux at 5 GHz of the likely
counterparts. Most of the detections present a radio flux above 1
Jy, and there is an apparent trend to become more radio loud when
the observed $\gamma$-ray flux is higher. This is the expected
behaviour when the emitted $\gamma$-rays have their origin in
inverse Compton interactions of the same particle population
(leptons) that generates the radio emission, targeting a soft
photon field. We have superposed a linear function to the
radio-gamma data that is also shown in Figure \ref{radio}. There,
we indicate with a dashed line, the extension of this linear fit
to the region where there are no 3EG `A' AGN sources. The big
vertical box signals the threshold for detectability in the radio
surveys used to search for counterparts ($\sim 30$ mJy). The
middle, lighter colored box, signals the range of $\gamma$-ray
fluxes for the unidentified sources considered in this paper. It
is apparent, then, that if weak AGNs were to approximately follow
the linear fit, we could find several $\gamma$-ray sources without
significant radio flux. Many of them could be those unidentified
sources we are studying here. In addition, if $\gamma$-ray sources
are affected by a differential gravitational lensing effect, this
process, as we shall show below, would enhance only the
$\gamma$-ray emission, keeping the radio fluxes at low levels.
This mechanism, then, would be in agreement with what is shown in
Figure \ref{radio}, provided the associated sources are within the
middle box on the left. Of course, this cannot apply to all
unidentified sources because, otherwise, it would result in a hole
in the source distribution between the already detected AGNs and
the candidates, at radio flux levels of $\sim 100$ mJy. It should
be remembered, anyhow, that sources at high latitude are
preferentially identified by their peak flux, which can be much
higher than the average. Although population studies are a
powerful tool to study the nature of the unidentified detections,
they should be supplemented by a source-by-source analysis.

\subsection{Light curves}

\begin{figure*}
\begin{center}
\includegraphics[width=8cm,height=11cm]{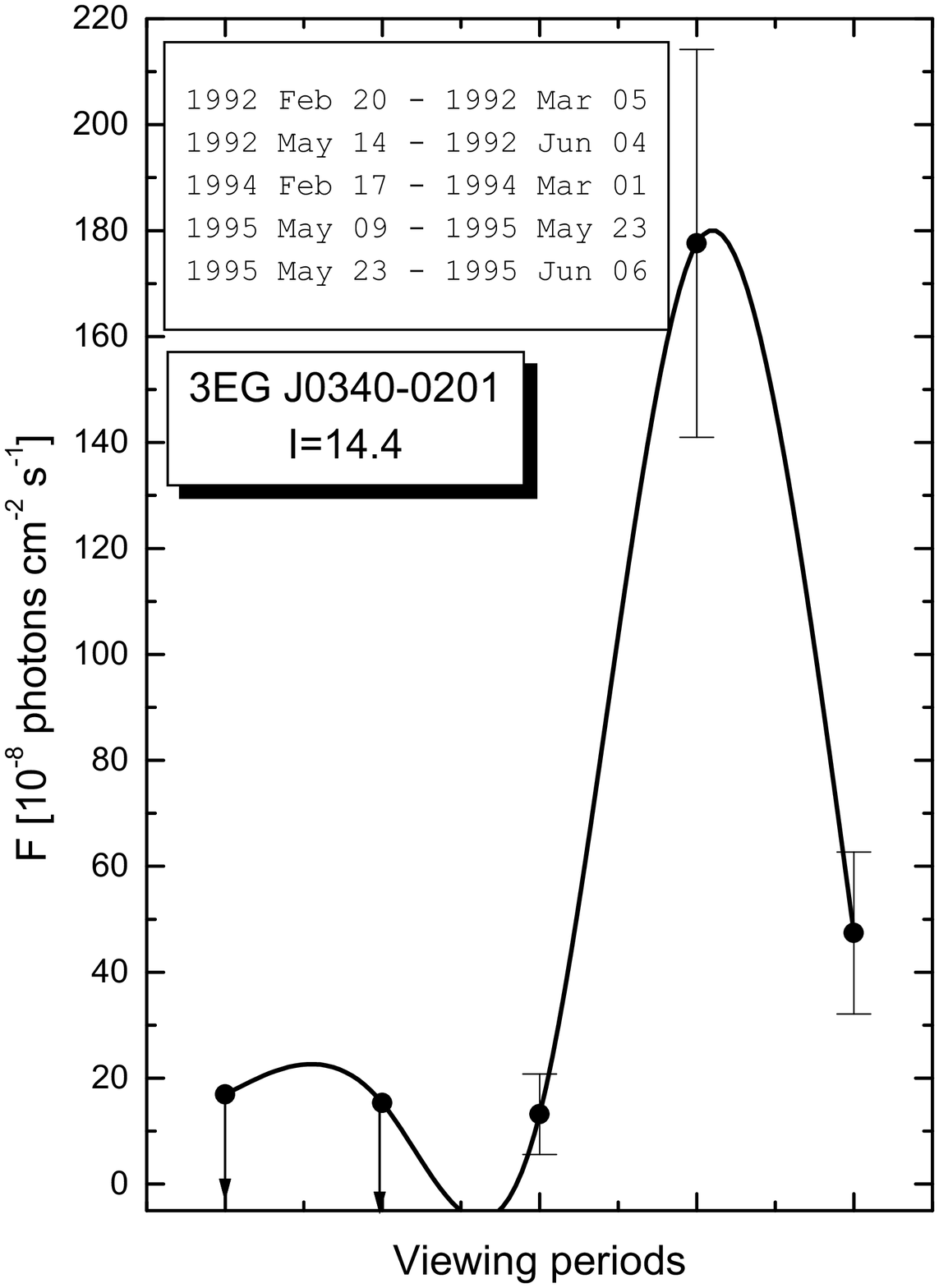}
\includegraphics[width=8cm,height=11cm]{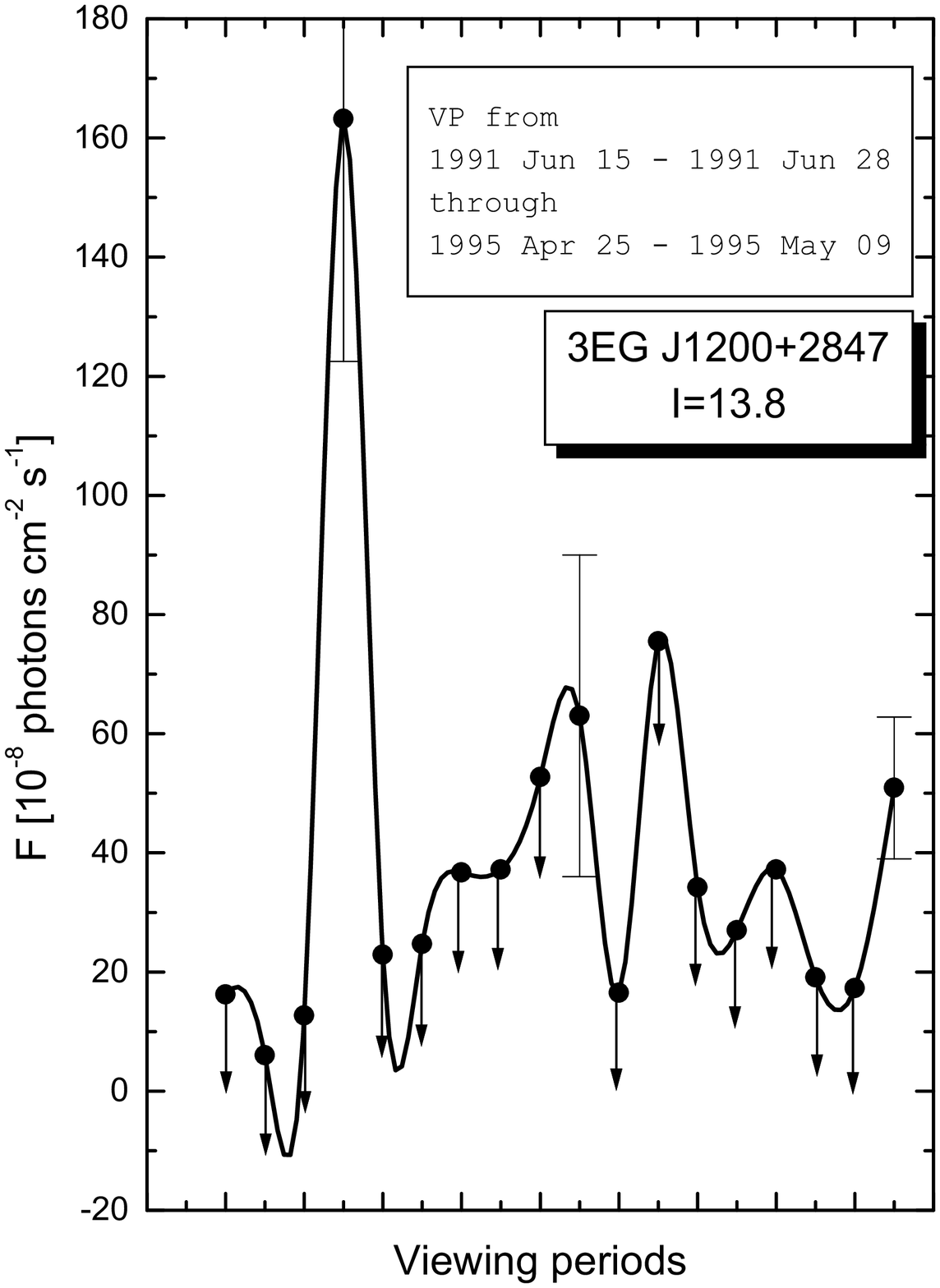}
\includegraphics[width=8cm,height=11cm]{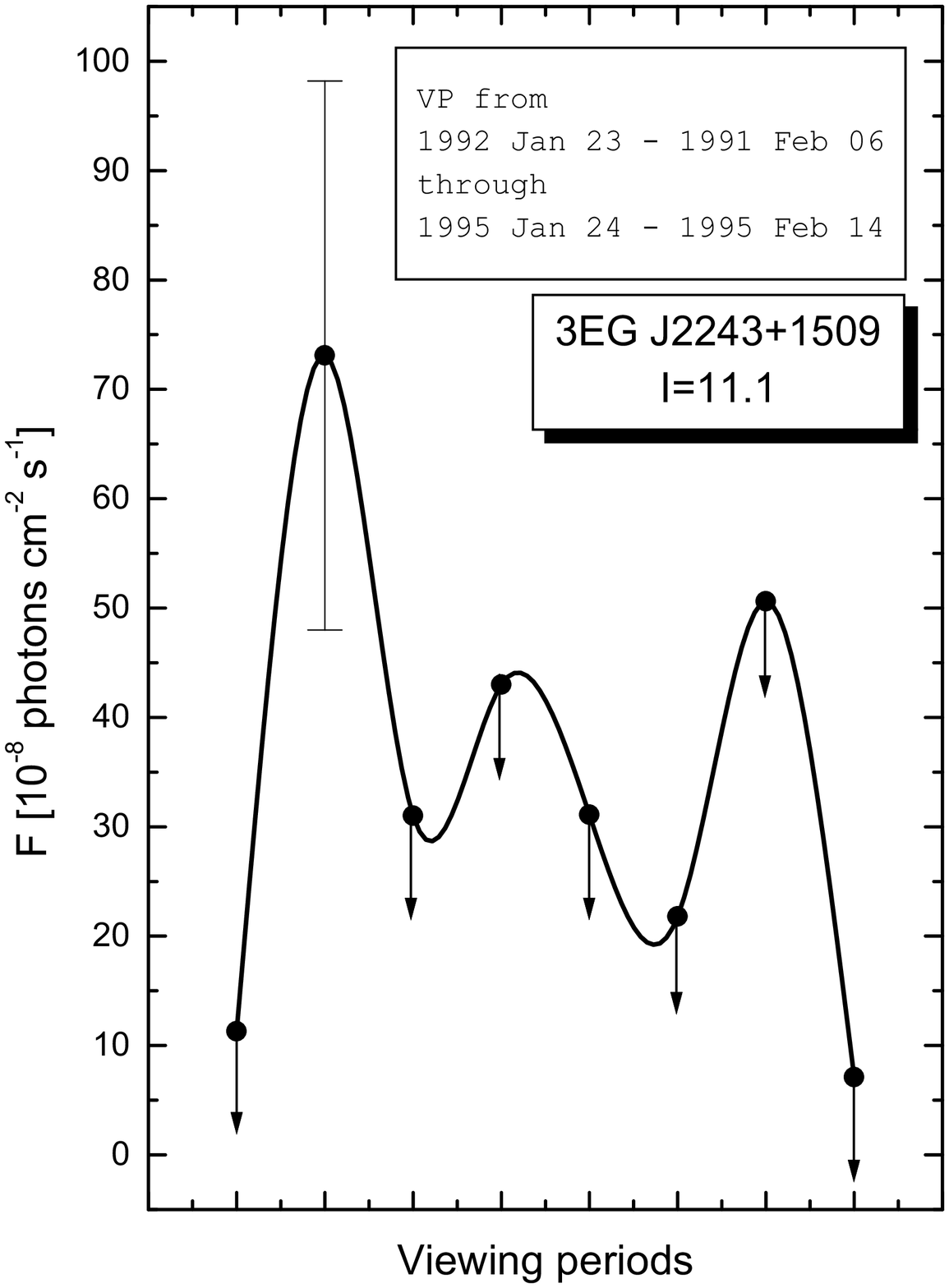}
\includegraphics[width=8cm,height=11cm]{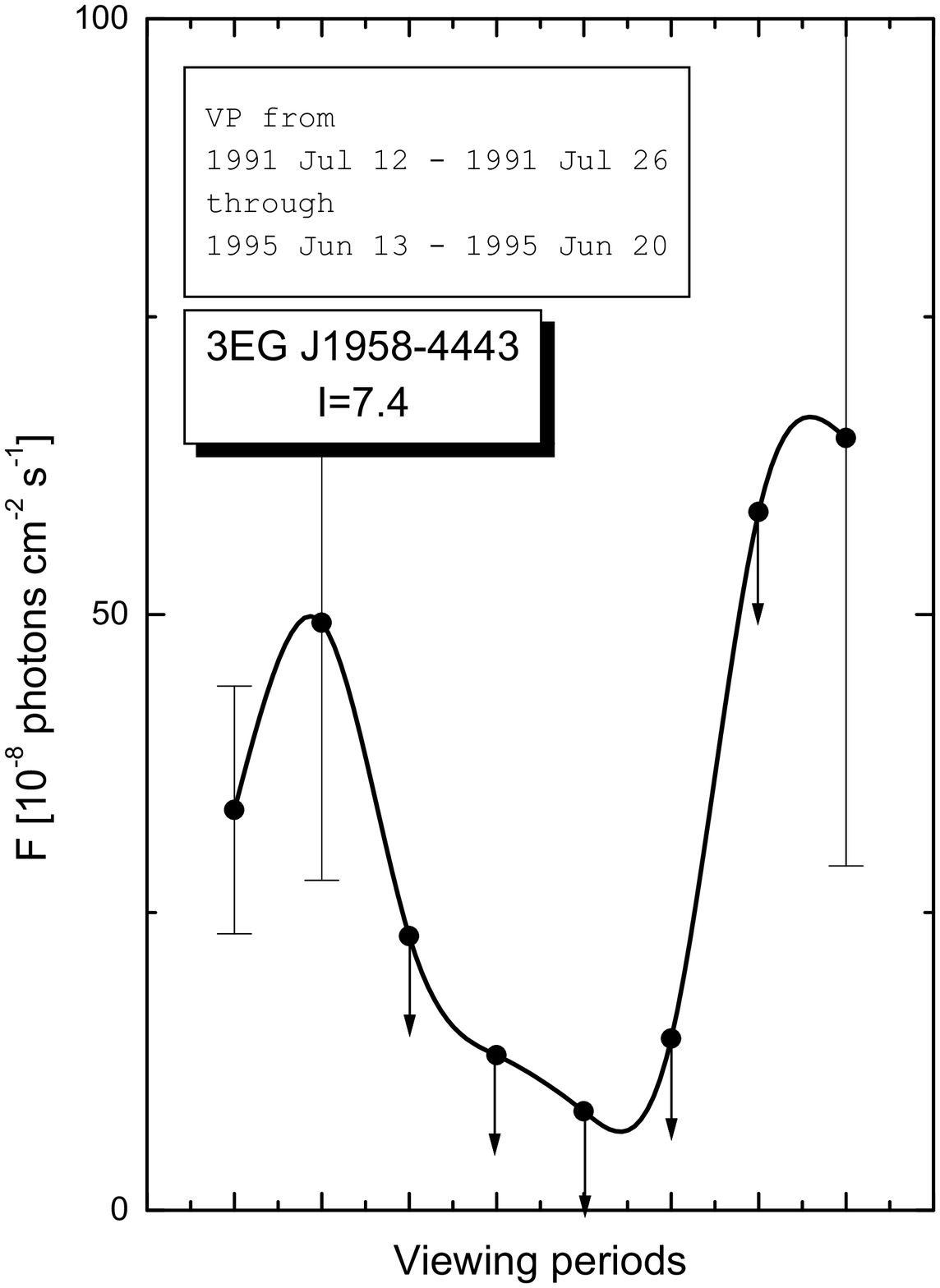}
\end{center}
\caption{Examples of $\gamma$-ray light curves of 3EG `A' AGNs
(upper panel) and unidentified sources with $|b|>30^0$ (lower
panel). Cases chosen among the most variable ones are shown here.
Points with arrows represent upper limits on the flux. Spline
curves do not represent actual fits. } \label{curvas}
\end{figure*}

The most interesting candidates in order to pursue further study
seem to be those sources presenting the highest levels of
variability. Most AGNs are, as we have already said, variable
sources, even with $\gamma$-ray emission fluctuating from
detection to non-detection in successive viewing periods. We show
some examples of typical light curves in Figure \ref{curvas}. This
kind of light curves are also shown by many unidentified sources,
what is consistent with the high levels of variability they
present.

\subsection{Assessment}

To finish this section we remark our main conclusion up to here,
namely, that {\it it is likely that some of the high-latitude
unidentified sources be not more than otherwise undetected AGNs,
presenting a low or nil (below any current detection threshold)
radio flux.} General gamma-ray characteristics of both
$\gamma$-ray blazars and high-latitude unidentified sources are
very similar. There remains, however, the question of why, whereas
most $\gamma$-loud blazars present radio flux at the Jy level, the
unidentified sources have no strong radio counterpart at all. If
the same mechanism for $\gamma$-ray production operates in both
groups of objects, why are they so different at lower energies? We
shall argue in the next sections that extrinsic effects can result
in such a behaviour.

\section{Microlensing}

\subsection{Point-like source and point-like lens}

Let us consider a background and weak $\gamma$-ray emitting
blazar, whose GeV flux is well below the EGRET sensitivity
threshold and whose radio flux is at the mJy level. We shall first
use the Chang \& Refsdal (1984) model. Assuming a galaxy is
interposed in the line of sight, the lens equation in the lens
plane for a point source is (e.g. Schneider et al. 1992):
\begin{equation}
{\bf r}-{\bf r}_{0}-R_{E}^{2}\frac{{\bf r}}{r^{2}}-\left(
\begin{array}{cc}
\kappa +\gamma & 0 \\
0 & \kappa -\gamma
\end{array}
\right) {\bf r}-{\bf d}_{0}=0  ,\label{p1}
\end{equation}
where the coordinate system is centered on the microlens (a star
within the galaxy), with the orientation of the orthonormal basis
$\{{\bf e}_{1},{\bf e}_{2}\}$ chosen to diagonalize the quadrupole
matrix; the source is at ${\bf r}_{0}$ and the image position is
${\bf r}$. The third term in equation (\ref{p1}) arises from the
deflection in the lens plane due to the microlens, here considered
as a point mass $M$. $R_{E}$ is the usual Einstein radius
\begin{equation}
R_{E}=\sqrt{\frac{4GM}{c^{2}}\frac{D_{ol}D_{ls}}{D_{os}}} ,
\label{p2}
\end{equation}
with $D_{os}$ the observer-source distance, $D_{ol}$ the
observer-lens distance and $D_{ls}$ the lens-source distance. The
fourth and fifth terms in Eq. (\ref{p1}) arise from the deflection
imprinted by the galaxy as a whole. In the fourth term, $\kappa $
and $\gamma $ are the focusing and the shear of the galaxy, at the
lens position, respectively. ${\bf d}_{0}$ depends on the
deflection imprinted by the galaxy as a whole at the location of
the microlens; its only effect is to change the unperturbed source
position ${\bf r}_{0}$ by a constant. We shall ignore ${\bf
d}_{0}$, assuming a displaced source position ${\bf s}={\bf
r}_{0}+{\bf d}_{0}$ in the lens plane.\\

Defining new coordinates ${\bf X}$ and ${\bf Y}$, in the lens and
the source plane respectively, as
\begin{equation}
{\bf X}=\frac{\sqrt{\left| 1-\kappa +\gamma \right| }}{R_{E}}{\bf
r} \label{p3},
\end{equation}
and
\begin{equation}
{\bf Y}=\frac{1}{R_{E}\sqrt{\left| 1-\kappa +\gamma \right| }}{\bf
s}, \label{p4}
\end{equation}
the lens equation becomes
\begin{equation}
{\bf Y}=\varepsilon \left(
\begin{array}{cc}
\Lambda & 0 \\
0 & 1
\end{array}
\right) {\bf X}-\frac{{\bf X}}{\left| {\bf X}\right| ^{2}} ,
\label{p5}
\end{equation}
where
\begin{equation}
\varepsilon =sign\left( 1-\kappa +\gamma \right),  \label{p6}
\end{equation}
and
\begin{equation}
\Lambda =\frac{1-\kappa -\gamma }{1-\kappa +\gamma } . \label{p7}
\end{equation}
The solution of equation (\ref{p5}) can be found reducing the
problem to a fourth order equation for $X^{2}$ (Schneider et al.
1992). In order to do it we introduce new coordinates \be X_1=X
\cos\phi , \label{00}\ee \be X_2= X\sin \phi,\label{000}\ee so we
obtain \be Y_1=\epsilon \Lambda X_1 - \frac{X_1}{X^2}=
\left(\epsilon \Lambda -\frac 1{X^2}\right) X \cos \phi , \ee and
\be Y_2=\epsilon X_2 - \frac{X_2}{X^2}= \left(\epsilon -\frac
1{X^2}\right) X \sin \phi . \ee Solving for $\cos \phi$ and for
$\sin \phi$, we get \be \label{cos} \cos \phi = \frac
{Y_1}{X(\epsilon \Lambda - 1/ X^2)}, \ee and \be \label{sin} \sin
\phi = \frac {Y_2}{X(\epsilon - 1/ X^2)}.\ee This readily implies
\be \frac{Y_1^2}{X^2 (\epsilon \Lambda - 1/ X^2)^2} + \frac
{Y_2^2}{X^2(\epsilon  - 1/ X^2)^2}=1,\ee that translates into
\begin{eqnarray}
\Lambda ^{2}X^{8}-\left[ 2\varepsilon \Lambda \left( \Lambda
+1\right) +Y_{1}^{2}+\Lambda ^{2}Y_{2}^{2}\right] X^{6} + \nonumber \\
\left[ \Lambda ^{2}+4\Lambda +1+2\varepsilon \left(
Y_{1}^{2}+\Lambda Y_{2}^{2}\right) \right] X^{4}- \nonumber
\\ \left[ 2\varepsilon \left( \Lambda +1\right)
+Y_{1}^{2}+Y_{2}^{2}\right] X^{2}+1=0 . \label{p8}
\end{eqnarray}
Here, $X=\left| {\bf X}\right| $ and the sub-indices stand for the
different components of the respective vectors. Equation
(\ref{p8}) can be solved by any standard method. Since it is a
fourth order equation, we can have zero, two or four real
solutions to the lens equation. The number of images is then given
by the number of real and positive solutions of this equation. For
any real and positive solution, $X^{2}$, we can take the positive
value of $X$ and obtain the position of the image using the
definitions given in Eqs. (\ref{cos}--\ref{sin}) and
(\ref{00}--\ref{000}) above.\\

The magnification $A=I_{obs}/I_{0}$ for any image is in turn given
by the determinant of the Jacobian matrix. This can be computed
using the chain rule for derivatives \be \frac {\partial {\bf
s}}{\partial {\bf Y}} \frac{\partial {\bf Y}} {\partial {\bf r}} =
\frac {\partial {\bf s}}{\partial {\bf Y}} \frac{\partial {\bf
Y}}{\partial {\bf X}} \frac{\partial {\bf X}} {\partial {\bf
r}},\ee that finally yields
\begin{equation}
A=\left| \det \left( \frac{\partial {\bf s}}{\partial {\bf
r}}\right) \right| ^{-1}=\frac{1}{\left| 1-\kappa +\gamma \right|
}\left| \det \left( \frac{\partial{\bf Y} }{\partial {\bf
X}}\right) \right| ^{-1} .\label{p10}
\end{equation}
Since
\begin{equation}
\frac{\partial {\bf Y}}{\partial {\bf X}}=\left(
\begin{array}{cc}
\varepsilon \Lambda +\frac{X_{1}^{2}-X_{2}^{2}}{X^{4}} & \frac{2X_{1}X_{2}}{%
X^{4}} \\
\frac{2X_{1}X_{2}}{X^{4}} & \varepsilon -\frac{X_{1}^{2}-X_{2}^{2}}{X^{4}}
\end{array}
\right) , \label{p11}
\end{equation}
then \ben A=\frac{1}{\left| 1-\kappa +\gamma \right| }\times
\hspace{5cm}\nonumber \\ \frac{\left( X_{1}^{2}+X_{2}^{2}\right)
^{2}} {\left| \Lambda \left( X_{1}^{2}+X_{2}^{2}\right)
^{2}+\varepsilon \left( 1-\Lambda \right) \left(
X_{1}^{2}-X_{2}^{2}\right) -1\right| }. \label{p12} \een The total
magnification for a point source is
\begin{equation}
A_{0}=\sum_{i=1}^{n}A_{i},  \label{p13}
\end{equation}
where $n$ is the number of images and $A_{i}$ are obtained
replacing the solutions of Eq. (\ref{p5}) in Eq. (\ref{p12}).\\

The curves for which the determinant of the Jacobian matrix is
zero are called critical curves. For the Chang-Resfdal lens, these
critical curves are Cassini ovals given by the equation
\begin{equation}
\Lambda \left( X_{1}^{2}+X_{2}^{2}\right) ^{2}+\varepsilon \left(
1-\Lambda \right) \left( X_{1}^{2}-X_{2}^{2}\right) -1=0  .
\label{p13b}
\end{equation}
The caustics are the mappings of the critical curves, through the
lens equation, onto the source plane. When the source crosses a
caustic, the number of images changes by two, and, if a point
source is on a caustic, the magnification diverges. In realistic
situations, sources are extended instead of point-like, and when
the source is on (or near) a caustic the magnification is finite
(although large).\\

When $\gamma =0$ (no shear case), $\Lambda =1$, and the lens
geometry becomes axially symmetric (notice also that the caustic
curve collapses to a point). Then, we can use the one-dimensional
lens equation:
\begin{equation}
Y=\varepsilon X-\frac{1}{X} , \label{p14}
\end{equation}
which has as solutions
\begin{equation}
X_{\pm }=\frac{\varepsilon Y}{2}\pm
\sqrt{\frac{Y^{2}}{4}+\varepsilon } \label{p15}
\end{equation}
for $Y^{2}>-4\varepsilon $. Hence, for $\varepsilon =1$ we always
have two solutions, and when $\varepsilon =-1$ there are two
solutions for $\left| Y\right| >2$ and none for $\left| Y\right|
<2$. The magnification of any image is
\begin{equation}
A_{\pm }=\frac{1}{\left| 1-\kappa \right| }\frac{X_{\pm
}^{4}}{\left| X_{\pm }^{4}-1\right| } , \label{p16}
\end{equation}
and the total magnification is given by
\begin{equation}
A_{0}=A_{+}+A_{-} . \label{p17}
\end{equation}
Replacing Eq. (\ref{p16}) in Eq. (\ref{p17}) and using Eq.
(\ref{p15}), we arrive at
\begin{equation}
A_{0}=\frac{1}{\left| 1-\kappa \right| }\frac{Y^{2}+2\varepsilon
}{\left| Y\right| \sqrt{Y^{2}+4\varepsilon }} . \label{p18}
\end{equation}
Some tricky --although straightforward-- algebra used to arrive at
the previous equation is described in the Appendix. In this case,
we have the critical curve at $X=1$, which gives a degenerate
caustic at $Y=0$ for $\varepsilon=1$, and a circular caustic of
radius $Y=2$ for $\varepsilon=-1$.

\subsection{Extended source}

For an extended circular source, the light curve is given by (see
e.g. Han et al. 2000):
\begin{equation}
A=\frac{\int_{0}^{2\pi }\int_{0}^{r_{s}}{\mathcal I}(r,\varphi
)A_{0}(r,\varphi )rdrd\varphi }{\int_{0}^{2\pi
}\int_{0}^{r_{s}}{\mathcal I} (r,\varphi )rdrd\varphi } ,
\label{e1}
\end{equation}
where $(r,\varphi )$ are polar coordinates in a reference frame
placed in the center of the source, $r_{s}$ is the radius of the
source and ${\mathcal I }(r,\varphi )$ is the surface intensity
distribution of the source. For a uniform distribution this
produces a simplified expression,
\begin{equation}
A=\frac{\int_{0}^{2\pi }\int_{0}^{r_{s}}A_{0}(r,\varphi
)rdrd\varphi }{\pi r_{s}^{2}}  . \label{e2}
\end{equation}
Introducing the dimensionless radial coordinate $R=r/R_{E}$, we
obtain
\begin{equation}
A=\frac{\int_{0}^{2\pi }\int_{0}^{R_{s}}A_{0}(R,\varphi
)RdRd\varphi }{\pi R_{s}^{2}}  \label{e3} ,
\end{equation}
where $R_{s}=r_{s}/R_{E}$ is the corresponding dimensionless
radius of the source.\\

\begin{figure}
\begin{center}
\vspace{1.2cm}
\includegraphics[width=8cm,height=8cm]{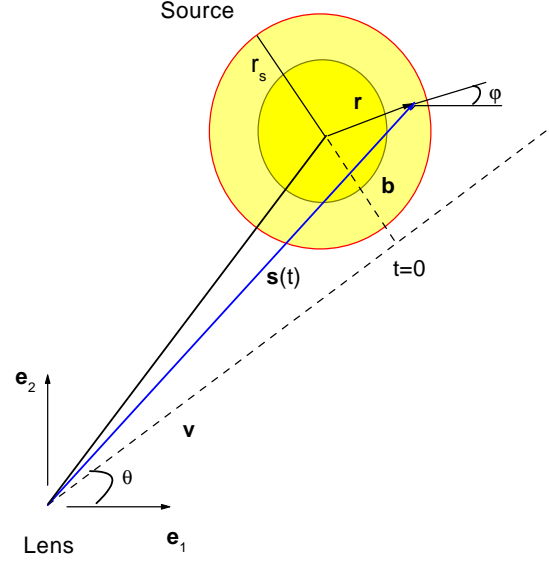}
\end{center}
\caption{Geometrical sketch of an extended source and physical
variables. All points of the source must be taken into account
when computing the magnification in a microlensing event.}
\label{f1}
\end{figure}

We assume that the microlens is moving with constant velocity
${\bf v}$ and we choose the origin of time $(t=0)$ as the instant
of closest approach between the lens and the source. Then, if the
center of the source is placed at $t=0$ in ${\bf
b}=(b_{1},b_{2})$, the position of any point of the source with
polar coordinates $(r,\varphi )$ is: \be s_{1}(t)=b_{1}-vt\cos
\theta +r\cos \varphi ,\label{ee4}\ee
\begin{equation}
s_{2}(t)=b_{2}-vt\sin \theta +r\sin \varphi  , \label{e4}
\end{equation}
where $v=\left| \bf{v}\right| $ , $\theta $ is the angle between $
{\bf v}$ and ${\bf e}_{1}$, $0< r < r_{s}$ and $0< \varphi < 2\pi
$. Figure \ref{f1} shows these geometrical
considerations.\\

Plugging Eqs. (\ref{ee4}-\ref{e4}) into Eq. (\ref{p4}), we obtain
\be Y_{1}=\frac{b_{1}-vt\cos \theta +r\cos \varphi
}{R_{E}\sqrt{\left| 1-\kappa +\gamma \right| }} ,  \label{ee5}\ee
\begin{equation}
Y_{2}=\frac{b_{2}-vt\sin \theta +r\sin \varphi }{R_{E}\sqrt{\left|
1-\kappa +\gamma \right| }}.  \label{e5}
\end{equation}
In units of the Einstein radius, these equations transform into
\be Y_{1}=\frac{B_{1}-T\cos \theta +R\cos \varphi }{\sqrt{\left|
1-\kappa +\gamma \right| }} , \label{ee6} \ee
\begin{equation}
Y_{2}=\frac{B_{2}-T\sin \theta +R\sin \varphi }{\sqrt{\left|
1-\kappa +\gamma \right| }},  \label{e6}
\end{equation}
where $T=vt/R_{E}$ , $B_{1}=b_{1}/R_{E}$ and $B_{2}=b_{2}/R_{E}$ .
When $\gamma =0$ (no shear), we can take $\theta =0$ and $
{\bf{B}}=(0,B_{0})$ in the above expressions without loosing
generality.

\section{$\gamma$-ray blazars as sources}

The fact that some $\gamma$-ray blazars have been observed to
flare dramatically on time scales of days imposes severe
constraints on the size of the emitting region. The optical depth
for intrinsic $\gamma+\gamma\rightarrow e^+ + e^-$ attenuation is
(e.g. Schlickeiser 1996):
\begin{equation}
\tau\simeq\sigma_T n_{\gamma} R= \frac{\sigma_T}{4\pi c
\left<\epsilon\right>} l, \label{tau}
\end{equation}
where $\sigma_T$ is the Thompson cross section, $n_{\gamma}$ the
$\gamma$-ray photon density, $R<c t_{\rm v}$ the source size
inferred from the intrinsic variability time scale, $<\epsilon>$
the mean photon energy, and $l$ the compactness parameter defined
as the ratio of the intrinsic source luminosity to its radius. The
optical depth can be written as:
\begin{equation}
\tau > 200 \frac{L_{48}}{t_{\rm v}/{1\;\rm day}} \label{tau2},
\end{equation}
with $L_{48}$ the luminosity in units of $10^{48}$ erg s$^{-1}$.
For typical values $t_{\rm v}\sim 1$ day and $L_{48}\sim1$, the
source is opaque, contrary to the observed fact that $\gamma$-ray
blazars present a power-law $\gamma$-ray spectrum over several
decades of energy. This rules out isotropic emission in the rest
frame. The emission, consequently, should be beamed. It is usually
thought to be produced in a relativistic jet through inverse
Compton scattering of lower energy photons (e.g. Blandford \&
Levinson 1995). The soft, seed photons for the Inverse Compton
process could originate as synchrotron emission from within the
jet, or they could come from the accretion disk surrounding the
central supermassive compact object, or they could be disk
radiation reprocessed in the broad line region. In the last two
cases the
seed photons are external to the jet itself.\\

In addition to these leptonic models, some hadronic alternatives
have been proposed in the literature. The $\gamma$-ray emission
would be produced in this case by relativistic protons interacting
with ambient matter, radiation fields, or the magnetic field of
the jet. For reviews and references the reader can see von
Montigny et al. (1995) and Mukherjee (2001).\\

Independently of how the $\gamma$-rays are produced, they must
traverse the strong X-ray field produced in the innermost region
of the accretion disk. The observed $\gamma$-ray photons cannot
originate from a too small radius, otherwise they will be absorbed
through pair creation in the disk photosphere (e.g. Becker \&
Kafatos 1995, Blandford \& Levinson 1995). This naturally leads to
the concept of $\gamma$-spheres in AGNs: for each $\gamma$-ray
photon energy there is a radius $r_{\gamma}$ beyond which the pair
production opacity to infinity equals unity (Blandford \&
Lenvinson 1995). The size of the $\gamma$-sphere will depend on
both the energy of the $\gamma$-ray photons and the soft photon
flux.\\

For an isotropic, power-law, central source of soft photons
scattered by free electrons in a warped disk, Blandford \&
Levinson (1995) obtain:
\begin{equation}
r_{\gamma }(E)\propto E^p,\label{g1}
\end{equation}
with $p$ depending on the details of the central source. A similar
result is obtained for pure disk emission (Becker \& Kafatos 1995,
Romero et al. 2000). Typically, $p\in[1,2]$. The larger
$\gamma$-spheres, then, are those for the higher photon energies.
This energy-dependency of the source size will naturally lead to
chromaticity effects during microlensing events.

\section{Light curves and spectra of single point lenses embedded in smoothly
distributed matter}

Now we focus our study on the lensing of an extragalactic
$\gamma$-ray source. For immediate use, we shall define a
reference source having a radius $r_{{\rm ref}}$ and a
$\gamma$-ray energy $ E_{{\rm ref}}$ such that
\begin{equation}
R_{\gamma }(E)=R_{{\rm ref}}\left( \frac{E}{E_{{\rm ref}}}\right)
^p \label{g3} ,
\end{equation}
where $R_{\gamma }(E)=r_{\gamma }(E)/R_{E}$ and
$R_{{\rm ref}}=r_{{\rm ref}}/R_{E}$.\\

We shall assume that the intensity of the source (without yet
being lensed) is uniform, and that its spectrum approximately
follows a power law
\begin{equation}
I_{0}(E)\propto E^{-\xi },  \label{g4}
\end{equation}
with $\xi \in (1.7,2.7)$ (Krolik, 1999). Then
\begin{equation}
I_{0}(E)=I_{{\rm ref}}\left( \frac{E}{E_{{\rm ref}}}\right) ^{-\xi
} , \label{g5}
\end{equation}
where $I_{{\rm ref}}$ is the intensity of the reference source.
The surface intensity distribution of the source finally ends
being
\begin{equation}
{{\mathcal I}}_{0}(E)=\frac{I_{0}(E)}{\pi \left( R_{\gamma
}(E)\right) ^{2}}. \label{g6}
\end{equation}
Since ${\mathcal I}_{0}(E)$ does not depend on $(R,\varphi )$ we
can write
\begin{equation}
A=\frac{\int_{0}^{2\pi }\int_{0}^{R_{\gamma }(E)}A_{0}(R,\varphi
)RdRd\varphi }{\pi \left( R_{\gamma }(E)\right) ^{2}} . \label{g7}
\end{equation}
Using that $A=I/I_{0}$ and Eq. (\ref{g5}), we have
\begin{equation}
I=AI_{0}=AI_{{\rm ref}}\left( \frac{E}{E_{{\rm ref}}}\right)
^{-\xi }. \label{g8}
\end{equation}
We define, for plotting purposes, the ratio
\begin{equation}
J\equiv \frac{I}{I_{{\rm ref}}}=A\left( \frac{E}{E_{{\rm
ref}}}\right) ^{-\xi }, \label{g9}
\end{equation}
as the intensity in units of $I_{{\rm ref}}$. We shall as well
define the dimensionless impact parameter $u=B/R_{\gamma }$, with
$B= \sqrt{B_{1}^{2}+B_{2}^{2}}$. Throughout the rest of this paper
we shall adopt $p =1.1$ , $\xi =2$, and a reference source with
dimensionless radius $R_{{\rm ref}}=2\times 10^{-3}$ at a
$\gamma$-ray energy of $E_{{\rm ref}}=0.1$GeV. Results are,
however, of a generic nature, and remain self-similar under
reasonable variations of these parameters.


\subsection{Galactic model}

We shall now assume a parameterization for the mass distribution
in the interposed galaxy. This will allow us to compute the values
of $\kappa_s $ and $\gamma $. In this Section, we assume for
illustrative purposes that all matter in the lensing galaxy is
smoothly distributed except for a single point lens. The full
non-linear treatment for large number of lenses will be done in
Section 7. If the center of the galaxy is located at ${\bf
x}_c=(b,0)$ in the lens plane, and if its total mass is described
by a surface density $\Sigma$ given by \be \Sigma =\frac {\Sigma_c
a_g} {\sqrt{a_g^2 + (x-d)^2 + y^2}},\label{sigma1}\ee where
$\Sigma _{c}$ is the density at the center of the galaxy, $a_{g}$
is the core radius of the galaxy, and $d$ is the distance of the
microlens to the center of the galaxy, then the focusing and shear
can be written as (Schneider et al. 1992, Romero et al. 1995):
\begin{equation}
\kappa_s =\frac{\Sigma _{c}}{\Sigma _{{\rm crit}}}\frac{a_{g}}{\sqrt{a_{g}^{2}+d^{2}}%
} , \label{g10}
\end{equation}
\begin{equation}
\gamma =\frac{\Sigma _{c}}{\Sigma _{{\rm crit}}}\left( \frac{a_{g}}{\sqrt{%
a_{g}^{2}+d^{2}}}-\frac{2a_{g}\sqrt{a_{g}^{2}+d^{2}}}{d^{2}}+\frac{2a_{g}^{2}%
}{d^{2}}\right) , \label{g11}
\end{equation}
where $\Sigma _{{\rm crit}}$ is the critical density, defined by
\be \Sigma _{{\rm crit}}\equiv \frac{c^{2}D_{s}}{4\pi
GD_{l}D_{ls}}.\ee Defining $m=\Sigma _{c}/\Sigma _{{\rm crit}}$
and $\chi =d/a_{g}$, we have
\begin{equation}
\kappa_s =m\frac{1}{\sqrt{1+\chi ^{2}}}  ,\label{g12}
\end{equation}
\begin{equation}
\gamma =m\left( \frac{1}{\sqrt{1+\chi ^{2}}}-\frac{2\sqrt{1+\chi
^{2}}}{\chi ^{2}}+\frac{2}{\chi ^{2}}\right) . \label{g13}
\end{equation}
When $\chi \ll 1$, $\gamma $ adopts a simpler expression,
\begin{equation}
\gamma =m\frac{\chi ^{2}}{4}\left( \frac{3}{2}\chi ^{2}-1\right).
\label{g14}
\end{equation}
If the microlens is placed at a distance from the center of the
galaxy far smaller than the core radius, $d\ll a_g$, then $\chi
\approx 0$ and the lensing parameters result in $\gamma =0$,
$\kappa_s=m$. We shall then adopt $\gamma =0$, $\kappa_s=0.8$, in
the zero shear case, as an example. To study other situations,
where $\gamma$ can not be neglected, we shall take, $m=2$, $\chi
=10$ $\Rightarrow$ $\kappa_s =0.20$, $\gamma=-0.16.$.

\subsection{Model results and analysis for a single point lens}

\begin{figure*}
\includegraphics[width=14cm,height=16cm]{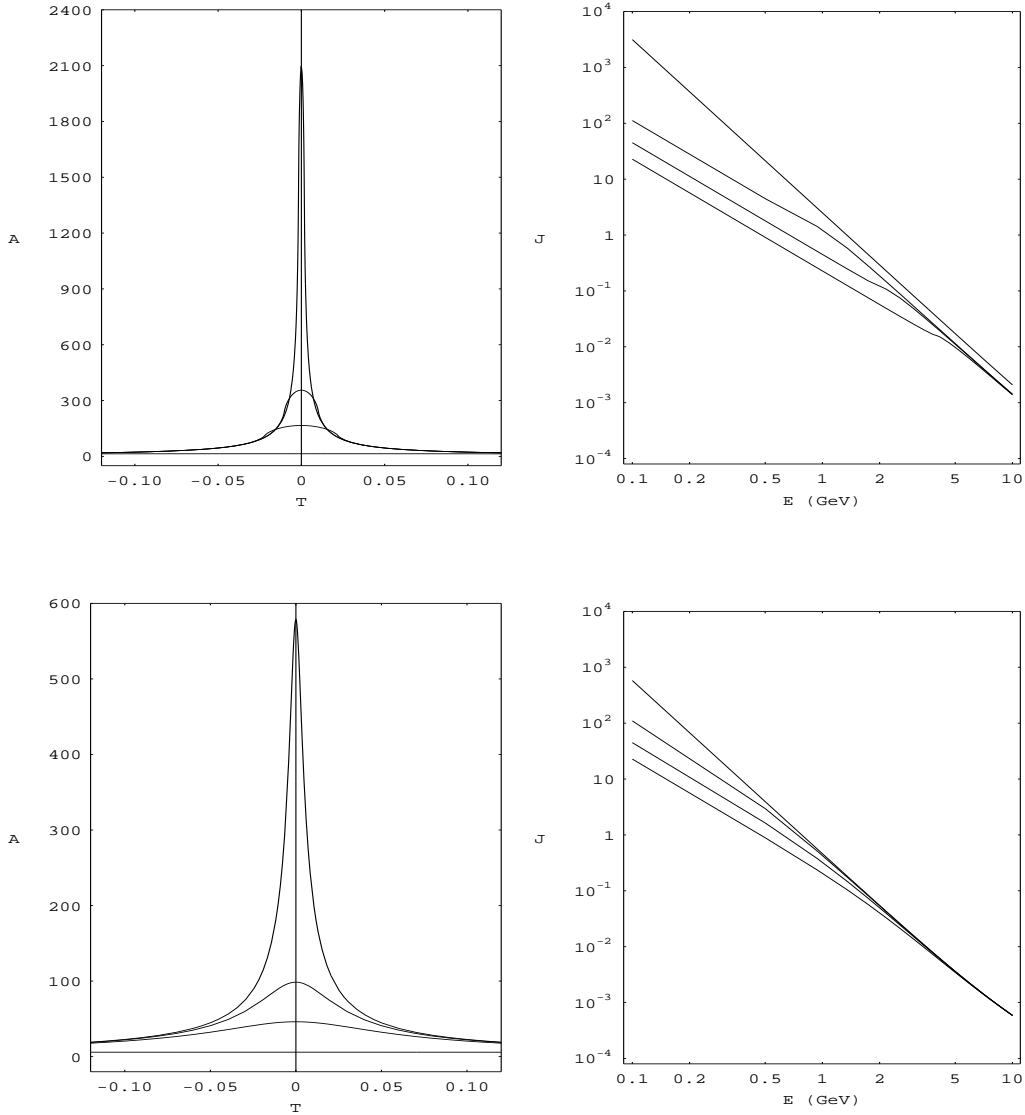}
 \caption{Lensing results for a single point lens with an impact parameter
$u=b/r_s=B/R_s=0.5$ (upper panel) and 2 (lower panel). Left: light
curves, from top to bottom $E=$0.1 GeV, 0.5 GeV, 1 GeV, 10 GeV.
Right: spectra, from top to bottom $T=vt/R_E$=0, 0.02, 0.05 and
0.10. Lensing parameters were chosen as  $\kappa_s=0.8$ and
$\gamma=0$.} \label{ff3}
\end{figure*}

Results for the first group of parameters (shear equal to zero)
are plotted in Figure \ref{ff3}, for different impact distances.
It can be seen there that {\it background AGN $\gamma$-ray
emission above 100 MeV can be magnified by a very large factor in
single events.} This magnification can make an otherwise unnoticed
background source to exceed the detection threshold, and to be
detected only when lensed. Note also the chromaticity effect: {\it
the $\gamma$-sphere corresponding to 10 GeV, whose size is similar
to the radio emitting regions of AGNs, is negligibly magnified,
while the lower energy curves -but still with energies in the
EGRET range, 100-500 MeV- all show significant magnifications}
This phenomenon also has a spectral signature, produced by the
differential magnification of the different $\gamma$-spheres (see
the discussion below). The magnification grows with the focusing.
The same happens, as expected, when the impact parameter is smaller.\\

A critical requirement for such a microlensing event to occur is
that the size of the background source projected onto the lens
plane is not larger than the Einstein ring of the lensing mass.
Only background sources whose size is a fraction of the Einstein
radius will then be significantly magnified. Since AGNs have
emission regions of different scales for different radiation
wavelengths, we expect a differential magnification, as observed
in Figure \ref{ff3}. The most internal regions of AGNs,
responsible for the $\gamma$-ray emission, have sizes about
10$^{14}$--10$^{15}$ cm (Blandford \& Levinson 1995). Then,  they
can be significantly magnified. At the same time, since radio
emission is originated far down the jet, the sizes of its emitting
region ($>10^{17}$ cm) exceeds the Einstein radius of the lenses
and lead
to the absence of counterparts at lower frequencies.\\



\begin{figure*}
\includegraphics[width=14cm,height=16cm]{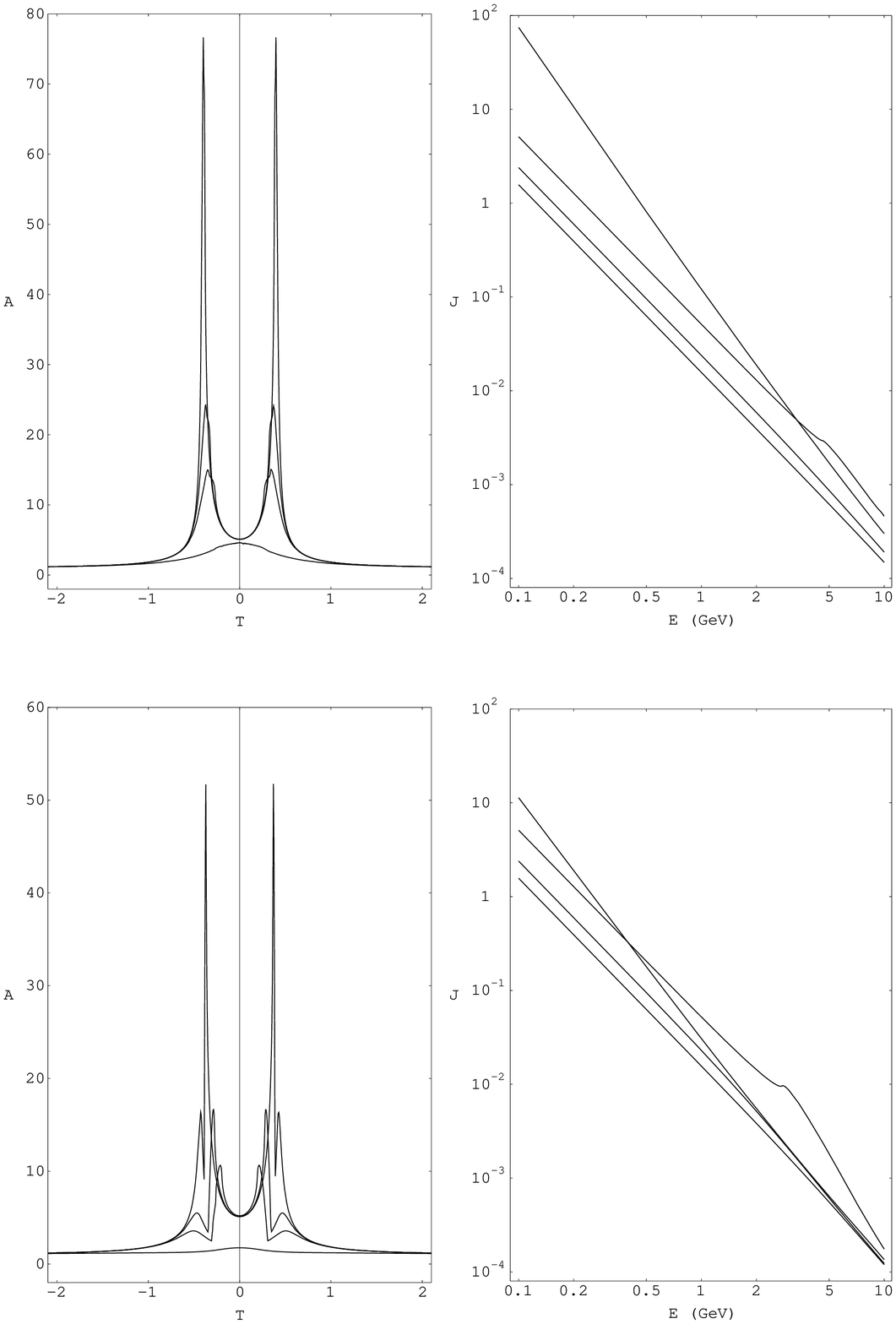}
\caption{Lensing results  for a single point lens with  an impact
parameter $u=$0.5 (upper panel) and 2 (lower panel). Left: light
curves, from top to bottom $E=$0.1 GeV, 0.5 GeV, 1 GeV, 10 GeV.
Right: spectra, from top to bottom $T=vt/R_E=$0.4, 0, 0.8 and 1.2.
Lensing parameters were chosen as $\kappa_s=0.20$ and
$\gamma=-0.16$ and $\theta=0^0$. } \label{f5}
\end{figure*}

In Figure \ref{f5}, we present a completely different behavior
produced when the lensing parameter $\gamma$, the shear, is
different from zero. We show the results for $\kappa_s=0.20$,
$\gamma=-0.16$, and $\theta=0^0$.
$\theta$ (see Figure \ref{f1}), is just the angle between the
direction of the lens movement and the ${\bf e_1}$ unit vector, in the source plane. \\

In the cases of Figure \ref{f5}, for the less energetic
$\gamma$-spheres with energies in the range 0.1--1 GeV, there
appears two (instead of one) enhancements of intensity, and with a
very peculiar time profile. In the case of impact parameter
$u=0.5$, the intensity first rise relatively slowly compared with
the immediately subsequent behavior, and suddenly decreases.
Afterwards, this effect repeats itself in a specular way. This
phenomenon is reminiscent of that found for putative matter
violating some of the energy conditions (e.g. Safonova et al.
2001, Eiroa et al. 2001), although in that case, a non-zero shear
is not needed in order to produce it and
the flux rising is sharper.\\

\begin{figure*}
\includegraphics[width=8cm,height=8cm]{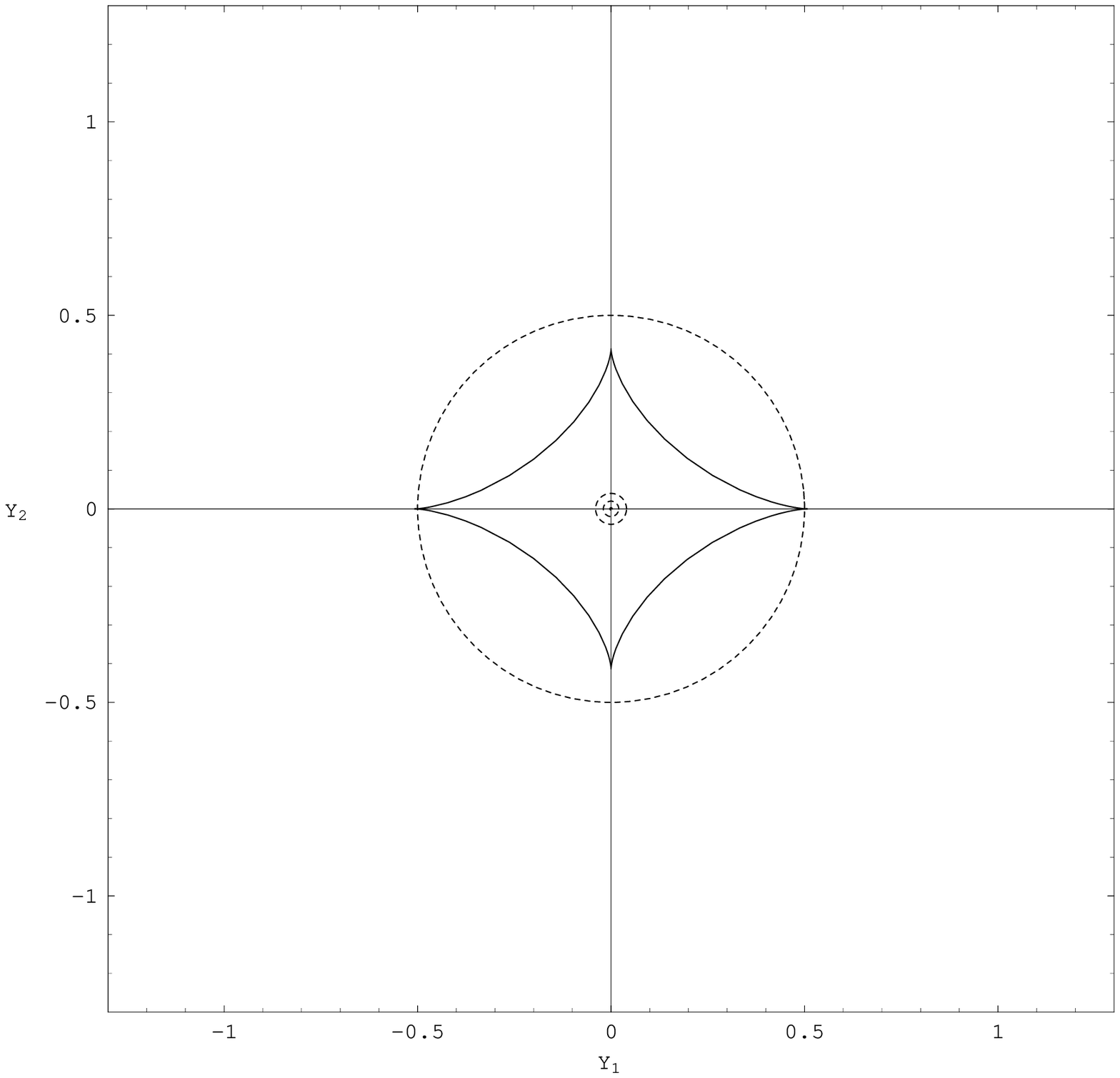}
\includegraphics[width=8cm,height=8cm]{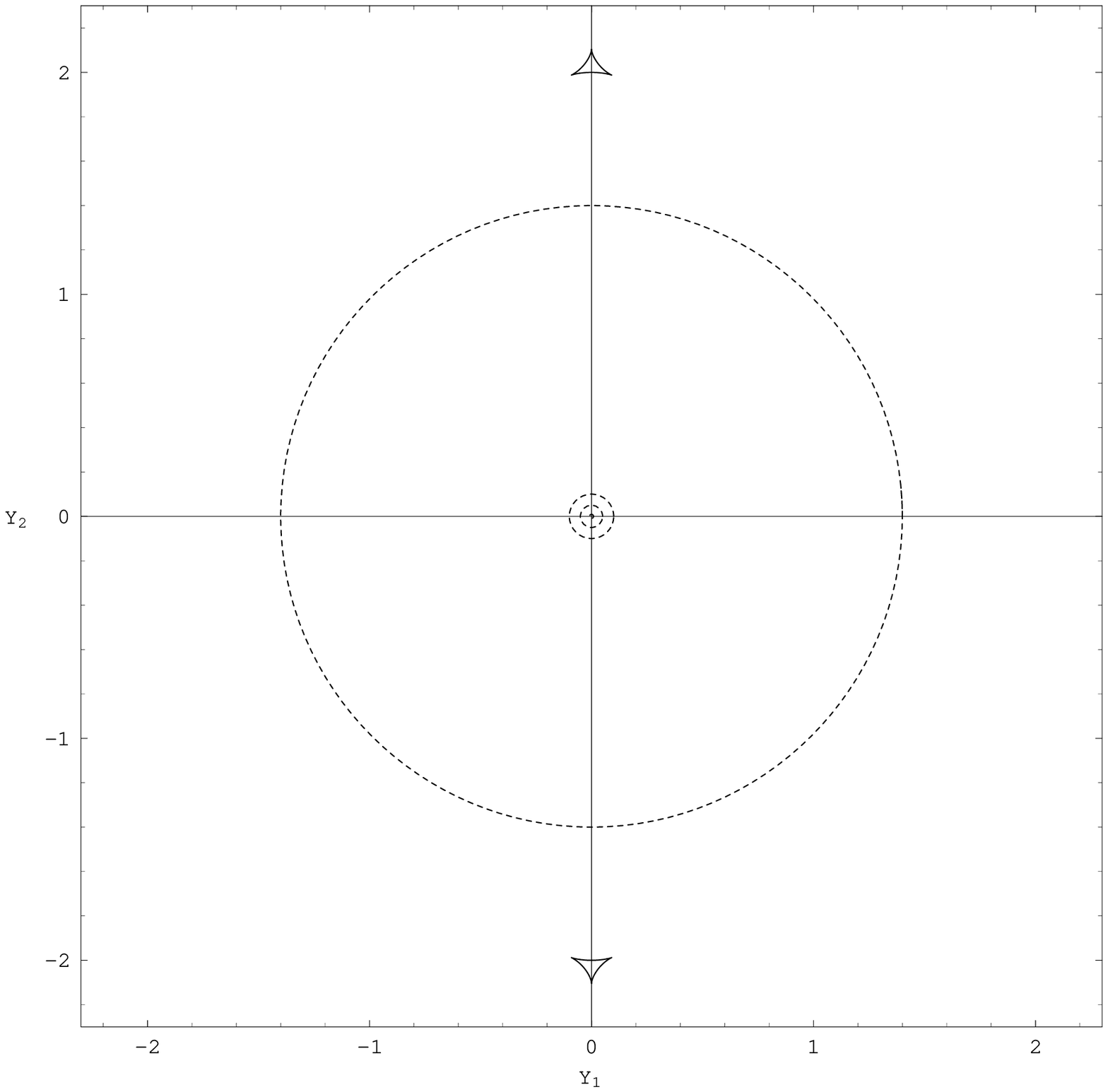}
\caption{Left: Caustics for $\kappa_s=0.20$ and $\gamma=-0.16$
(diamond shaped curve). Right: Caustics for $\kappa_s=0.89$ and
$\gamma=-0.34$ (triangle shaped curves). Sources inside the
caustic have four images and those outside have two. The dashed
circles are the gamma spheres, the bigger one corresponding to
$E=10$ GeV, then, 1, 0.5, and 0.1 GeV. } \label{f4-f7}
\end{figure*}

In Figure \ref{f4-f7}, we show the size and shape of the caustics
(in the source plane) for two specific lens-source configurations
and lensing parameters. We also present, for comparison, the sizes
of the $\gamma$-spheres for different energies. It is possible to
see, then, that the biggest $\gamma$-sphere here considered, the
one corresponding to $E=$10 GeV, is larger even than the caustic
size for the lensing parameters used in the left panel of the
Figure. It is for this $\gamma$-sphere that we see only one
intensity enhancement, translated into a peak of the
magnification. The reason is that light from different parts of
the source smooth down the enhancement produced by the only part
crossing the caustic. Therefore, for the biggest $\gamma$-ray
sphere, the caustic curve is seen as a point. However, for the
smaller $\gamma$-spheres,  the relative motion of the lens and
source makes the latter to encounter two caustics, leading to
different peaks in the time profile.  Contrary to the caustic
presented in the left panel of Figure \ref{f4-f7}, the lensing
parameters used for the right panel produce a caustic curve
actually formed by two separated pieces. When $\theta=0$ there
would not be a direct crossing of the caustic, even for the
largest $\gamma$-spheres. Two peaks profiles would be explained by
double caustic crossing. In this case, we would see two peaks even
for the largest $\gamma$-spheres.



\subsection{Spectral evolution}

\begin{figure}[t]
\begin{flushleft}
\includegraphics[width=8cm,height=9cm]{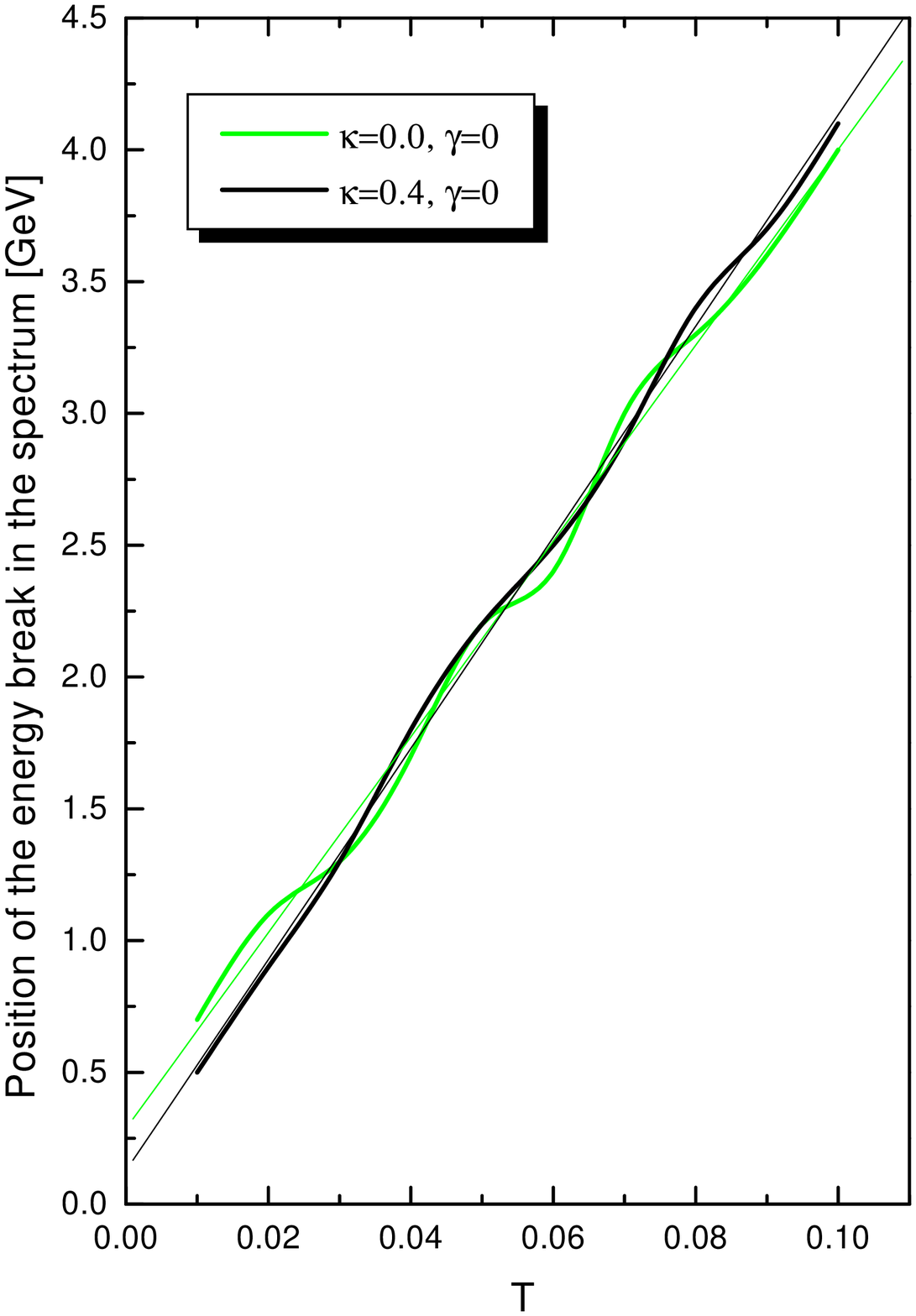}
\end{flushleft}
\caption{Temporal evolution of the spectral break in models with
zero shear. The thinner solid lines represent linear fits, showing
the general trend to increase the energy at which the spectral
break is produced as a function of time. The impact parameter in
this example is $u=0.5$. The x-axis is $T=vt/R_E$. } \label{break}
\end{figure}

The differential magnification makes its way to a spectral slope
change at medium energies (see Figures \ref{ff3} and \ref{f5},
right panels). This break, that we predict as a distinctive
feature for microlensing events of $\gamma$-ray blazars, together
with its peculiar time evolution --it is shifted towards high
energies as times goes by since transit-- can be used to
differentiate this from other phenomena. The temporal evolution of
the break for the
cases of zero shear is shown in Figure~\ref{break}. \\

The spectral evolution shown by lensing with non-zero shear is
notoriously modified. In Figure \ref{f5}, for instance, the
spectral evolution starts at $T=0$, then the intensity grows
towards higher energies because of the encounter of the right peak
in the left hand side of the left panel plots, and then continues
downward as the lens gets away from the caustic.  As it can be
seen in Figure \ref{f5} and successive ones, it is not possible to
plot in this case the evolution of the spectral break with time as
a continuous function. The break in the spectrum appears only once
in this energy range, for the earliest time, and then presumably
moves upwards, to energies corresponding to non-magnified
$\gamma$-spheres. The evolution of the spectrum itself (and not of
the position of the break) is what can help to decide in these
cases if the putative observations are associated to a
microlensing event.

\section{Optical Depth}

\subsection{Time scale}

The characteristic time of a microlensing event is the time that
the source takes  to cross the Einstein radius of the lens. It is
given by
\begin{equation}
t_{0}=\frac{R_{E}}{v} =
\left({\sqrt{\frac{4GM}{c^{2}}\frac{D_{ol}D_{ls}}{D_{os}}}}\right){v}^{-1},
\end{equation}
where $v$ is the lens velocity. For a cosmological source, using
the Friedmann-Lema\^{\i}tre-Robertson-Walker model of the
universe, with $z$ the redshift of the object, and the Hubble
parameter $H_0$ and the current ratio between the actual and the
critical density of the universe $\Omega_0$ being well constrained
in the range
\begin{equation} 50\frac{{\rm km}}{{\rm s\; Mpc}}< H_{0}< 100
\frac{{\rm km}}{{\rm s\; Mpc}},
\end{equation}
\begin{equation}
0.1< \Omega _{0}< 1,
\end{equation}
we have \be \label{pp} D(z_i,z_j)=\frac{2c}{H_0}
\frac{(1-\Omega_0-G_iG_j) (G_i-G_j)}{\Omega_0^2 (1+z_i)
(1+z_j)^2}, \ee with \be G_{i,j}= (1+\Omega_0 z_{i,j})^2.\ee
Taking as an example, $H_{0}=75$ km s$^{-1}$ Mpc$^{-1}$, $\Omega
_{0}=0.2$,  a lens mass of $M=0.1 M_{\odot }$, a velocity $v=5000$
km s$^{-1}$, $z=0.1$ for the lens, and $z=1$ for the source, we
obtain
\begin{equation}
t_{0}=1.5 \times 10^{7}\; {\rm s}\;\simeq 170\; {\rm days}
\end{equation}
The width of the peaks in the light curves (see Figure \ref{ff3})
is about
\begin{equation}
\Delta t\simeq 0.05 t_{0}\;\simeq 7.5 \times 10^{5}\; {\rm s}
\simeq 9\; {\rm days}.
\end{equation}
These numbers would change to 1202 days and 60 days, respectively,
if the lensing were produced by a 5$M_{\odot}$ star. It is
interesting to compare such numbers with typical timescales of
high-latitude sources determined through the EGRET experiment. The
Third EGRET catalog was constructed during a period of six years,
dividing the total time span in viewing periods with a duration of
about  15 days. Then, when the mass of the lensing object is
sub-solar, each peak in a microlensing light curve can be
completely within a single EGRET viewing period: the phenomenon
can lead, in principle, to a very variable source, with
$\gamma$-ray fluxes varying from detection to upper limits in
consecutive viewing periods. On the contrary, for a 5$M_{\odot}$
star, a single peak would last several viewing periods and the
$\gamma$-ray source could appear as a steady, non-variable
detection.

\subsection{Linear size of the source}

We have mentioned before that in order for a microlensing event to
occur, the linear size of the source, $x$, should be less than the
Einstein radius, \be x < 2 R_E \frac{D_{os}}{D_{ol}}.\ee For the
same typical redshifts considered in the previous section \be R_E
= 2.23  \times 10^{16} \left(\frac{M}{M_\odot}\right)^{1/2} {\rm
cm},\ee  and we can then obtain the following relationship between
the source size and the mass of the lens: \be x<6.37  \times
10^{17} \left(\frac{M}{M_\odot}\right)^{1/2} {\rm cm}.\ee For a
source with $x=10^{14} {\rm cm}$ strong magnification occurs for
stars with masses $M/M_\odot \geq 2 \times 10^{-3}$. This makes
most of the stars MACHO-like objects in a galaxy able to produce
strong gravitational lensing effects upon the innermost regions of
background active galactic nuclei. The smaller masses can give
rise to very rapid events (Romero et al. 1995).

\subsection{Expected number of microlensing cases in $\gamma$-ray catalogs}

The concept of optical depth, $\Gamma$, was originally introduced
in gravitational microlensing studies by Ostriker and Vietri
(1983), and it was later applied by Paczy\'nski (1986).  If
$\Gamma$ is smaller than unity, it provides a measure of the
probability of microlensing. Alternatively, $\Gamma$ can be
defined as the ratio of the surface mass density of microlensing
matter to the critical mass density of the galaxy (Paczy\'nski
1986). The value of $\Gamma$ depends on the model adopted for the
matter
distribution along the line of sight.\\

It is usually assumed that the a priori probability of finding a
small group of distant, gravitationally magnified objects is below
1\%. Recent results (Wyithe \& Turner 2002), taking into account
the clustering of stars in interposed galaxies, give for the a
priori probability of finding magnified sources in random
directions of the sky values between 10$^{-2}$ -- 10$^{-3}$. In
those directions where there is gravitational lensing, the
probability of having large local values of optical depth is high.\\

The high surface mass density associated with the core of normal
galaxies along with the usual assumption that most of this mass is
under the form of compact objects naturally leads to high optical
depths for microlensing. For instance, in the case of the lensed
quasar Q2237+0305, where four images are well-resolved, lensing
models indicate values of $\tau\sim0.5$ (Schneider et al 1988;
Wambsganss \& Paczy{\'n}ski 1994), which are corroborated by the
detections of microlensing-based optical variability with
relatively high duty cycles (e.g. Corrigan et al. 1991, Wozniak et
al. 2000, Witt \& Mao 1994). Other lensed sources display even
higher
duty cycles (e.g. Koopmans \& de Bruyn 2000).\\

In our case, the number of potential compact $\gamma$-ray emitting
background sources is large. The last version of the V\'eron-Cetty
\& V\'eron's (2001) Catalog --which is still very incomplete at
high redshifts-- contains more than $10^3$ already identified
blazars, in addition to more than $10^4$ quasars and other less
energetic AGNs. The GLAST mission itself is expected to pinpoint
about 10$^4$ $\gamma$-ray emitting blazars with unparalleled
resolution (Gehrels \& Michelson 1999);  also the number of
unidentified sources at high latitudes is expected to be large. If
the actual total number of $\gamma$-ray emitting blazars is in
excess of, say, 10$^7$, (1 blazar out of 10 000 normal galaxies)
they could produce many of the expected detections by GLAST. Even
when considering reduced probabilities for microlensing, scaling
as $\tau/A^2$ with $\tau$ being the local optical depth and $A$
the magnification, an interesting number of detections could be
potentially ascribed to microlensing.\\

 A crude estimation of the
number of possible $\gamma$-ray sources produced by microlensing
can be obtained as the product of three factors: {\it random
lensing probability} $\times$ {\it local lensing probability}
$\times$ {\it number of background sources}, i.e. approximately $5
\times 10^{-3} \times 1/A^2 \times 10^7$. The uncertainty in the
previous expression, however, is large. We can only roughly
estimate the total number of background sources, but the value of
$A$ they need to become visible i.e. with fluxes above the
sensibility of EGRET and/or GLAST, will depend on the luminosity
function of $\gamma$-ray emitting AGNs, which is unknown.
According to the different sensitivities of both instruments, we
could expect, perhaps, a number of detections of $\sim 10$ and
$\sim 100$, respectively.

\subsection{Number of events per light curves}

We shall now estimate the number of microlensing events expected
for a $\gamma$-ray blazar with a galaxy interposed in the line of
sight. For the galaxy mass distribution we shall adopt the model
used by Griest (1991) --see also Eq. (\ref{sigma1}) above--, where
the density profile is given by \be \rho(r) = \rho_0
\frac{a^2}{a^2+(x^2+y^2+z^2)},\ee and where $\rho_0$ is the mass
density in the center of the galaxy and $a$ is the core radius of
the halo. We shall consider that the entire microlens population
is at the same distance from the observer. The projected surface
mass density of the microlensing at a radius
$r^2=x^2+y^2=constant$ from the center of the galaxy will be \be
\Sigma(r) = \int_{-\infty}^{\infty} dz \; \rho_0\;
\frac{a^2}{a^2+(x^2+y^2+z^2)} =\frac{\pi \rho_0
a^2}{\sqrt{a^2+r^2}}.\ee Assuming that $\Sigma(r) $ is constant
for the smallest impact parameters and equal to
$\Sigma_c=\Sigma(r=0)$, we obtain \be \Gamma = \frac{\Sigma_c \pi
u^2 R_E^2 }{M} = B^2 \kappa_s ,\ee with $\kappa_s=\Sigma
_{c}/\Sigma _{{\rm crit}}$. Usually $B$ is taken as unity. This
would happen, for instance, when the distant quasar and the galaxy
in which the lenses reside are perfectly aligned (as in the case
of PKS 0537-441, see Romero et al. 1995 for discussion). In this
case, all lenses are close to the center of the galaxy, and the
previous approximation is valid. After some algebra, and taking
derivatives, we may write this latter expression as \be d\Gamma =
\frac{d\Sigma_c}{\Sigma_{\rm crit}} B^2 =
\frac{\Sigma_c}{\Sigma_c} \frac{d\Sigma_c}{\Sigma_{\rm crit}} B^2
= \Gamma \frac{d\Sigma_c}{\Sigma_c}. \label{ga}\ee In order to
make further estimates, we need to assume some parameterization
for the mass function of the lenses. The standard choice is a
power law of the form $N(M) \propto M^{-\alpha}$ (e.g. D'Antona \&
Mazzitelli 1986). It is likely, however, that a single power law
will not suffice for all different types of stars. The value of
$\alpha$, also known as the Salpeter index, is usually taken as
2.35 (Salpeter 1955), although more massive stars may require a
steeper index, whereas it could probably be flatter for the less
massive stars. As was done in Surpi et al. (1996), we normalize
the mass function to yield a total surface density $\Sigma_c
=\kappa_s \Sigma_{\rm crit}$ at the center of the galaxy. The
differential surface density in objects with masses between $M$
and $M+dM$ at the center would then be \be \frac{d\Sigma_c}
{\kappa_s \Sigma_{{\rm crit}}}= \frac{M N(M) dM} {\int_{M_{\rm
min}}^{M_{\rm max}} M N(M) dM }  . \ee Using this last expression
in Eq. (\ref{ga}) we find \be d\Gamma =
\frac{\Gamma}{C(\alpha,M_{\rm min},M_{\rm max})} \left(
\frac{M}{M_\odot}\right)^{1-\alpha} d\left(
\frac{M}{M_\odot}\right), \label{ge}\ee with
\begin{eqnarray}
C(\alpha,M_{\min},M_{\max})&=& \frac{\left[ \left(
\frac{M_{\max}}{M_{\odot}}\right) ^{2-\alpha}-\left(
\frac{M_{\min}}{M_{\odot}}\right) ^{2-\alpha}\right] }{2-\alpha},\nonumber \\
&&\;\;\;\;\;\;
{\rm for}\;\; \alpha\neq2, \nonumber \\
&=& \ln\left(  \frac{M_{\max}}{M_{\min}}\right)  , \;\;{\rm
for}\;\; \alpha=2.
\end{eqnarray}
Following Subramanian \& Gopal-Krishna (1991), if the lenses have
the same mass and the observing period is $\Delta t$, the expected
number of events in the case of a background source moving with
velocity $v$ will be \be N=\Gamma \left( 1+ \frac{2 \Delta t}{\pi
B t_0} \right),
\end{equation}
where $\Gamma $ is the optical depth and $B=b/R_E$. Then, using
Eq. (\ref{ge}), we obtain \ben dN = \left( \frac{2 v \Delta t}
{\pi B R_E} + 1 \right) \times \hspace{3.5cm} \nonumber\\ \mbox{}
\;\;\;\;\;\;\;\;\;\;\;\;\;\frac{\Gamma}{C(\alpha,M_{\rm
min},M_{\rm max})} \left( \frac{M}{M_\odot}\right)^{1-\alpha}
d\left( \frac{M}{M_\odot}\right). \label{gi}\een It is convenient
to write the previous expression in more useful units. We define
the dimensionless parameter \be {\cal D}= \frac {1}{\pi} \frac {30
{\rm days}}{\sqrt{GM_\odot}} c^2
\sqrt{\frac{D_{os}}{D_{ol}D_{ls}}},\ee to finally obtain \ben dN =
{\cal D} \frac{\sqrt{\kappa_s \Gamma}}{C} \frac{\Delta t}{30 {\rm
days} } \frac vc \left( \frac{M}{M_\odot}\right)^{1/2-\alpha}\!\!
d\left( \frac{M}{M_\odot}\right) \nonumber \\ \mbox{} \;\;\;\;\;+
\frac{\Gamma}{C} \left( \frac{M}{M_\odot}\right)^{1/2-\alpha}\!\!
d\left( \frac{M}{M_\odot}\right) . \label{go}\een For typical
values, say $z_s \sim 0.9$ and $z_l \sim 0.1$, ${\cal D} \sim
2.1$. Eq. (\ref{go}) can now be used to obtain the number of
events per lens mass interval as wellas the total number of
microlensing events.\\

For the assumed power law mass distribution of lenses between
$M_{\min}< M< M_{\max}$, the number of expected microlensing
events by stars with masses in the range $(M_{1}, M_{2})$,
included in the total mass range $(M_{\min}, M_{\max})$, during
$\Delta t$ days of observations is \ben N_{M_{1}-M_{2}}^{\Delta
t}=\frac{D(\alpha,M_{1},M_{2})\Gamma}{C} + \hspace{3cm} \nonumber \\
2.16\sqrt{\kappa_s \Gamma}\frac{v}{c}
\frac{D(\alpha,M_{1},M_{2})}{C(\alpha,M_{\min},M_{\max})}\frac{\Delta
t}{30{\rm days}}, \een where $D$ is given by
\begin{eqnarray}
D(\alpha,M_{1},M_{2}) &=& \frac{2\left[  \left(
\frac{M_{2}}{M_{\odot}}\right) ^{\frac32-\alpha}-\left(
\frac{M_{1}}{M_{\odot}}\right) ^{\frac32
-\alpha}\right]}{3-2\alpha},\nonumber \\
&&\;\;\;\;\;\;\;\; {\rm for}\;\;\alpha\neq\frac{3}{2}, \nonumber \\
&=& \ln\left(  \frac{M_{2}}{M_{1}}\right), \;\;{\rm for}\;\;
\alpha=\frac{3}{2},
\end{eqnarray}
and we have assumed the previously mentioned redshifts ($z_s\sim
0.9$ and $z_l\sim0.1$) to fix the numerical coefficient. Clearly,
the total number of events will strongly depend (apart from the
expected influence of $\Gamma$ and $\kappa_s$) on the relative
source-lens velocity and the Salpeter index $\alpha$.
\\

\begin{figure}
\includegraphics[width=8cm,height=11cm]{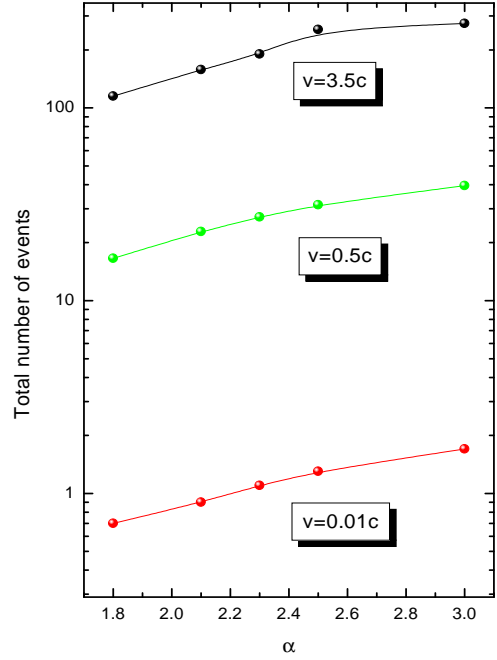} \caption{Number of events as a function of
the Salpeter index $\alpha$ for the case of apparently super and
sub-luminal movements of the source in the lens plane. Lensing
parameters are $\gamma=0$, $\kappa_s=0.4$, and $\Gamma=0.2$. For
$v\ll c$, the time scale of one event is comparable with EGRET
observing time. For $v\sim c$ the time scale of a single event is
comparable with one EGRET viewing period.} \label{f90}
\end{figure}

Blazars where the bulk of the high-energy emission is produced in
a superluminal component with apparent velocity $v>c$ in the lens
plane will produce $\gamma$-ray sources with the highest levels of
variability. In Figure \ref{f90} we present three curves showing
the total number of microlensing events in a 5 years period for
the case of the source being a blazar superluminal component, a
subluminal ($v=0.5 c$) component, and for a fixed core source with
a foreground lens moving with $v\sim 1/100 c$. We find that during
the EGRET observing time, and in the case of the lensing of a
superluminal source, hundreds of events can be
expected.\footnote{``Events'' refers to single spikes within the
light curves whereas ``sources'' to the number of EGRET detections
in different parts of the sky that could be ascribed to this
mechanism.} These $\gamma$-ray sources can strongly fluctuate from
one viewing period to another, leading to high levels of
variability. Instead, AGNs whose velocities in the lens plane are
much smaller than $c$ will be only mildly affected by lensing,
yielding few events for usual values of $\Gamma$. These cases
would produce $\gamma$-ray light curves shifting from detection to
upper limits in separated viewing periods,  even producing only
one $\gamma$-ray detection in a period of five years. Depending
ultimately on the mass of the lens, this unique event could
produce a steady non-variable source (seen along consecutive
viewing periods), or one which is seen only during a single
viewing period. The peculiarities of each event will ultimately
determine the time scales involved. Table \ref{tab:eg} shows the
results for the total number of events for different lensing
parameters, whereas Table \ref{tab:eg2} shows the distribution of
the events by lensing mass. It is interesting to note that most of
the events are produced by MACHO-like, sub-stellar mass objects.
These objects present small Eintein radius, indicating that only
the innermost $\gamma$-spheres will be magnified. In addition,
this would entail smaller time scales, and within the galactic
model here considered, more variable $\gamma$-ray sources.

\begin{table}
\caption{Total number of events for different lensing parameters
and Salpeter indices produced in a period of five years. Two cases
(bottom panels) show the results for very low optical depths: even
in that cases, significant number of events are expected if the
sources have apparently superluminal velocities. } \label{tab:eg}
\begin{center}
\begin{tabular}{ccccccc}
$\kappa_s$ & $\Gamma$ & $v/c$  & $M_{\rm min}/M_\odot$ & $M_{\rm max}/M_\odot$ & $\alpha$ & $N$ \\
\hline
0.4 & 0.2 & 3.5 & 0.1 & 20 & 1.8 & 115.1 \\
&--&--&--&--&2.1& 158.8 \\
&&--&--&--&2.3& 189.9 \\
&&&--&--&2.5& 255.1 \\
&&&&--&3.0& 273.9 \\\hline
0.4 & 0.2 & 0.016 & 0.1 & 20 & 1.8 & 0.7 \\
&--&--&--&--&2.1& 0.9 \\
&&--&--&--&2.3& 1.1 \\
&&&--&--&2.5& 1.3 \\
&&&&--&3.0& 1.7 \\\hline
0.4 & 0.2 & 0.5 & 0.1 & 20 & 1.8 & 16.6 \\
&--&--&--&--&2.1& 22.8 \\
&&--&--&--&2.3& 27.2 \\
&&&--&--&2.5& 31.4 \\
&&&&--&3.0& 39.5 \\\hline
0.6 & 0.2 & 0.5 & 0.1 & 20 & 1.8 & 20.3 \\
&--&--&--&--&2.1& 27.9 \\
&&--&--&--&2.3& 33.5 \\
&&&--&--&2.5& 38.6 \\
&&&&--&3.0& 48.2 \\\hline
0.2 & 0.4 & 0.9 & 0.1 & 20 & 1.8 & 29.9 \\
&--&--&--&--&2.1& 41.1 \\
&&--&--&--&2.3& 49.3 \\
&&&--&--&2.5& 56.9 \\
&&&&--&3.0& 71.7 \\\hline
0.4 & 0.4 & 0.016 & 0.1 & 20 & 1.8 & 1.1 \\
&--&--&--&--&2.1& 1.5 \\
&&--&--&--&2.3& 1.8 \\
&&&--&--&2.5& 2.1 \\
&&&&--&3.0& 2.6 \\\hline
0.8 & 0.6 & 0.5 & 0.1 & 20 & 1.8 & 40.8 \\
&--&--&--&--&2.1& 56.2 \\
&&--&--&--&2.3& 67.2 \\
&&&--&--&2.5& 77.5 \\
&&&&--&3.0& 96.9 \\\hline

0.4 & 0.01 & 0.01 & 0.1 &20 & 2.5 & 0.1\\
--  & --   & 0.1 & -- & -- & -- & 1.4\\
--  & --   & 0.5 & -- & -- & -- & 7.0\\
--  & --   & 3.5 & -- & -- & -- & 48.9\\ \hline

0.4 & 0.001 & 0.01 & 0.1 &20 & 2.35 & 0.0\\
--  & --   & 0.1 & -- & -- & -- & 0.4\\
--  & --   & 0.5 & -- & -- & -- & 2.0\\
--  & --   & 3.5 & -- & -- & -- & 14.0\\ \hline

\hline
\end{tabular}
\end{center}

\end{table}

\begin{table}
\caption{Distribution of the number of lensing events by mass for
a galaxy with $\kappa_s=0.4$, $\gamma=0$, and $\Gamma=2$. The
Salpeter index is $\alpha=2.1$ for the first panel, $\alpha=2.3$
for the second, and $\alpha=2.5$ for the third one. The relative
source velocity in the lens plane is chosen as $v=3.5c, 0.1c$ and
0.01$c$, from top to bottom.} \label{tab:eg2}
\begin{center}
\begin{tabular}{cccc}
Mass range $(M_{\odot})$& $N_{10^{-2}-10^{-1}}^{5{\rm years}}$ &
$N_{10^{-1}-1}^{5{\rm years}}$ & $N_{1-20}^{5{\rm
years}}$ \\
 \hline
\hline

$1-20$ & -- & -- & 69.4 \\ $10^{-1}-20$ & -- & 124.1 & 347
\\ $10^{-2}-20$ & 303.2 & 76.1 & 21.3 \\\hline

$1-20$ & -- & -- & $2.2$ \\ $10^{-1}-20$ & -- & 4.8 & 0.8 \\
$10^{-2}-20$ & 13.7 & $2.1$ & 0.4 \\\hline

$1-20$ & -- & -- & $0.3$ \\ $10^{-1}-20$ & -- & $0.8$ & $0.0$\\
$10^{-2}-20$ & $2.6$ & $0.2$ & 0.0 \\\hline \hline

\end{tabular}
\end{center}
\end{table}

\subsection{Critical assessment}

Although the Chang-Refsdal scheme used in the previous section is
a very important tool in the understanding of the model herein
proposed, it cannot be used to claim precise predictions for the
magnifications. For high values of $\kappa_s$ and $\gamma$ a more
complex caustic pattern should be considered in order to get
reliable light curves and the correct magnifications. The
quantitative predictions for the amount of magnification and the
form of the light curves could actually change in more detailed
simulations of a galaxy core (where the effects of thousands of
stars are considered simultaneously). The rest of this paper is
devoted to develop full numerical models of the situation
presented from an analytical point of view in the previous
sections .

\section{Magnification maps}

\begin{figure}[t]
\vspace{1.5cm}
\begin{center}
\includegraphics[width=6cm,height=9cm]{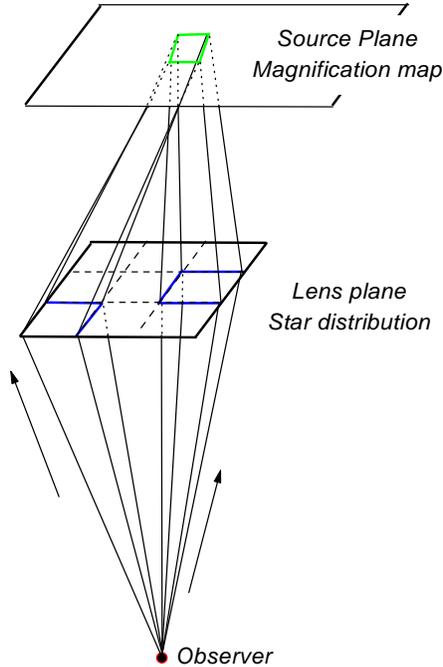}
\end{center}
\caption {Light rays are traced back towards the source. Each of
them is deflected in the lens plane. Different parts of the lens
plane can generate deflections converging onto the same pixel in
the source plane. There, the magnification map --the number of
rays per pixel-- is made. Once this is done, a source with a given
size moves in this magnification plane and generates a light
curve. } \label{graf}
\end{figure}

As we have seen, the parameters that describe a microlensing
scenario are the dimensionless surface mass density $\kappa$
--expressed in units of the critical surface mass density-- and
the external shear $\gamma$ (cf. Kayser, Refsdal \& Stabell 1986;
Schneider \& Weiss 1987). The former --often also called
convergence or optical depth-- describes the amount of matter in
front of the source. The latter is a tidal force caused by matter
outside the light bundle. In order to simulate the effect of a
particular combination of $\kappa$ and $\gamma$, point lenses are
distributed randomly according to the given surface mass density.
If we would replace each point lens by a disk with radius equal to
its Einstein radius, the total fraction of the sky that is covered
by the sum of all these disks  is equal to the optical depth.\\

For the ray-shooting simulations, a large number of light rays (of
order $10^9$) are followed {\it backwards} from the observer
through the field of point lenses. The number of rays determine
the resolution of the numerical simulation. These rays start out
in a regular (angular) grid. In the lens plane, the deflections
according to the individual lenses are then superposed for each
individual ray $i$, \be \bar \alpha_i=\sum_{j=1}^{n} \bar
\alpha_{ij}=\frac {4G}{c^2}\sum_{j=1}^{n} M_j \frac{\bar
r_{ij}}{r_{ij}^2}.\ee Here
 $M_j$ is the mass of the point lens $j$, $\bar r_{ij}$ is the
 projected vector distance between the position of the light ray
 $i$ and the lens $j$, and $r_{ij}$ is its absolute value. It is the
 computations of these individual deflections what requires
 most of the time of the simulation.
The effect of the external shear is included as well. The
deflected rays are then followed further to the source plane.
There, they are collected in small pixels. The number of rays per
pixel (on average $\sim 100$ for a region typically of $2500
\times 2500$ pixels) is proportional to the magnification at this
position. A two-dimensional map of the ray density --a
magnification pattern, also referred to as caustic pattern-- can
then be produced. The magnification as a function of (source)
position is indicated by colors. Sharp lines correspond to
locations of very high magnification, i.e. the caustics. In Figure
\ref{graf} we schematically show the geometry of the shooting
technique.\\

The minimal number of lenses that have to be considered depends on
the focusing and shear values, as well as on the ratio between the
diffuse and the total flux. The diffuse flux ($\epsilon$) is that
coming from rays that are deflected into the receiving area from
stars far outside the region where microlenses are considered, and
should be consistently low. An approximated expression for the
number of lenses to be included in each magnification map is
(Wambsganss 1999) \be N_* \sim
\frac{3\kappa^2}{(1-\kappa)^2-\gamma^2} \frac 1\epsilon \ee which
entails values from several hundreds (for $\kappa < 0.4$) up to
several hundred thousands (for $\kappa \sim 1$) stars, in the case
of zero shear and $\epsilon = 0.01$. Since for computing the map
with high resolution, this would imply literally several million
billions of operations, an intelligent numerical routine should
come to the rescue to make this problem feasible. The technique is
called hierarchical tree code. Basically, it groups the stars as a
function of their distance to the light ray. Since their influence
decreases as gravity does, with $r^{-2}$, one does not need to
compute the deflection from all the stars involved with the same
level of precision to obtain an overall sensitive result. A
detailed account of the numerical technique can be found in the
paper by Wambsganss (1999).
\\

In general, a smooth out distributed surface mass density
$\kappa_c$ also contributes to the deflection as well, and the
general microlensing equation to be solved is
\begin{equation}
y=\left(
\begin{array}{cc}
1-\kappa -\gamma & 0 \\
0 & 1-\kappa +\gamma
\end{array}
\right) \bar x-\kappa_{c} \bar x .\label{ppp1}
\end{equation}
In observational situations $\kappa_c$ and $\gamma$ are obtained
from macrolensing simulations of the resolved source. Rays
representing a square ($x_1=x_2$) are then mapped onto a rectangle
with a side ratio
$T=(1-\kappa-\kappa_c-\gamma)/(1-\kappa-\kappa_c+\gamma)$. But, as
we would like the receiving --and not the shooting-- area to be a
square (the pixel) the shooting field (i.e. the area in the lens
plane in which rays are mapped) is chosen to be a rectangle
of size $T^{-1}$.\\

The fact that source, lens and, observer are moving relative to
each other, means that the source is affected by a variable
magnification as a function of time: it moves through the
magnification pattern and so we measure a variable flux. When a
source crosses a caustic, formally two very bright new
(micro-)images appear or disappear. However, their angular
separation is much smaller than the resolution of any telescope.
Therefore, we can only measure the combined total brightness,
which can produce dramatic `jumps' in the observed flux. Assuming
geometrical optics and a point source, the amount of the
magnification is infinite. However, since any real source is
finite, the magnification stays finite as well. The exact amount
of magnification depends on the source size: the smaller it is,
the higher the magnification. In order to determine the light
curve for a finite source, its brightness profile has to be
convolved with this two-dimensional magnification pattern.\\

\begin{figure*}
\begin{center}
\end{center}
\caption{Magnification map for lensing with parameters
$\kappa=0.5$ and $\gamma=0.0$. For details, see text. The
corresponding file is $magpat_k50_g00.gif$.} \label{mapk50g0}
\end{figure*}

\subsection{Results of the numerical simulations}

In Figure \ref{mapk50g0} we show the magnification map for the
case in which $\kappa=0.5$ and $\gamma=0.0$; the brighter the
region, the stronger the magnification. The characteristic
critical lines of the Chang-Refsdal model appear in this map, as
well as the diamond shaped structures due to the close-by star. In
the bottom left of the panel, we show the size of the source for
three different energies. The innermost pixel (the center of the
circles) represent the size of the lowest energy $\gamma$-sphere,
that corresponds to $E=100$ MeV. The second circle is the size of
the 1 GeV $\gamma$-sphere, whereas the largest one is the
corresponding size of the $E=10$ GeV emitting region. It is clear
the origin of the differential magnification: the $\gamma$-spheres
will be affected according to their size while moving in the
caustic pattern. \\

\begin{figure*}
\begin{center}
\includegraphics[width=16cm,height=15cm]{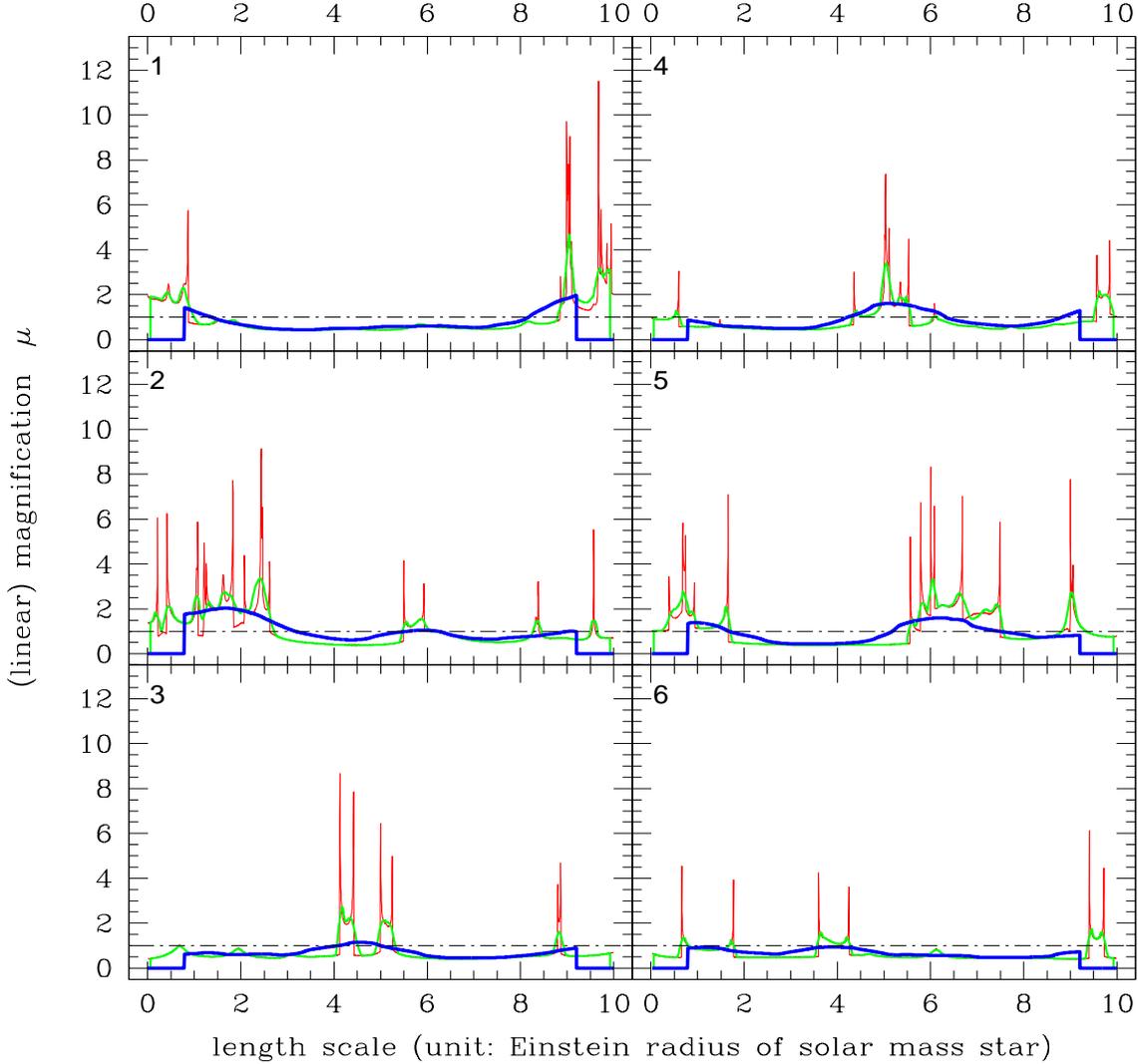}
\end{center}
\caption{Light curves for different source trajectories. Numbers
corresponds to those given in Figure  \ref{mapk50g0}. Darker lines
corresponds, respectively, to regions emitting photons of 100 MeV,
1, and 10 GeV, and whose emitting sizes are depicted in the bottom
left corner of Figure \ref{mapk50g0}, being the innermost point
the less energetic $\gamma$-ray sphere. } \label{curvask50g0}
\end{figure*}

The numbered lines in Figure \ref{mapk50g0} represent different
source trajectories. Numbers 1 to 6 are trajectories common to all
maps we shall present, and are defined a priori in our code. The
light curves for each of these six trajectories are given in
Figure \ref{curvask50g0}. We can see that the overall effect
described with the Chang-Refsdal model appears here as well. In
lighter color we show, for each trajectory, the corresponding
light curve for the $E=100 $ MeV emitting region. Darker lines
correspond to regions emitting photons of 1 and 10 GeV. The
latter, in all cases, are smoother versions of the former and,
always, the magnification is weaker for these regions. There is a
typical factor of 10 more magnification for the innermost regions
than for the larger $\gamma$-spheres. We see that for these values
of $\kappa$ and $\gamma$ it is possible to get a typical
enhancement of 10 times the unlensed intensity at 100 MeV. The
drop to zero at the corners of the diagram for all light curves
corresponding to the largest $\gamma$-sphere is an artifact of the
simulations: the code assigns zero magnification when more than
half of the source is out of the magnification map.
\\

\begin{figure}
\begin{center}
\includegraphics[width=8cm,height=9cm]{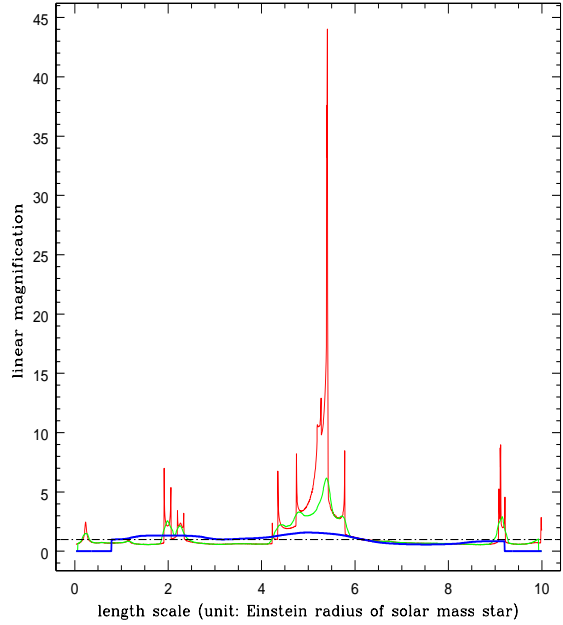}
\end{center}
\caption{Light curve for trajectory number 7, producing one of the
highest levels of magnification out of map given in Figure
\ref{mapk50g0}. The line coding  is as in Figure
\ref{curvask50g0}. } \label{curvas7k50g0}
\end{figure}

The shape of the light curves is also worthy to comment. As an
example, we take trajectory 3. It starts in a region of low
magnification and continues upwards, crossing a region of
relatively high enhancement, where for the innermost
$\gamma$-sphere the magnification is 9 times the intensity of the
unlensed source. There are four caustic crossings there. We can
see that for the innermost $\gamma$-sphere, the caustic crossings
are well separated events, so we have four peaks in the light
curve corresponding to $E=100$ MeV. However, for the larger
$\gamma$-spheres the peaks are smoothed down. We see only two
broadened peaks for $E=1$ GeV, and only one, with almost nil
magnification, in the case of $E=10$ GeV. If the $\gamma$-ray AGN
is within the observing sensitivity, we would see a distinctive
effect during the microlensing event due to the different sizes of
the source at the different energies.\\

The $x$-axis in Figure \ref{curvask50g0} is a linear length scale,
the Einstein radius of a solar mass star, $R_E(M_\odot)=2.23
\;\times 10^{16}$ cm. It can be translated into a time scale as
$t=R_E(M_\odot)/v$, where $v$ is the relative velocity of the
source with respect to the lens, projected onto the source plane.
Typical time scales for microlensing events where discussed in
Section 6.1. To give an example of the similar time scales
predicted with a full caustic pattern plot, we write the time
scale as \be t=\frac {R_E(M_\odot)}{v} = \frac {0.023}{v/c} {\rm
yr}. \ee With a relative velocity $v=5000$ km s$^{-1}$, the length
of each axis in Figure \ref{mapk50g0} is equivalent to 14 years.
The other important time scale involved in a microlensing event is
the rise time to a peak of maximum magnification. This will depend
on the size of the different $\gamma$-spheres, and is given by \be
\tau = \frac R{v}= \frac {0.023}{v/c} \frac R{R_E(M_\odot)} {\rm
yr},\ee where, again, we have scaled it with the Einstein radius
corresponding with one solar mass. Then, the innermost
$\gamma$-spheres (having $R/{R_E} \sim 1/100$) will have a rise
time scale of about 5 days, well within an observing EGRET viewing
period. The largest $\gamma$-spheres, with $R\sim R_E$, can have a
rise time of about 1 yr.  Higher (lower) velocities would imply
lower (higher) time scales. \\

\begin{figure*}
\begin{center}
\end{center}
\caption{Magnification map for lensing with parameters
$\kappa=0.8$ and $\gamma=0.0$. For details, see text. The
corresponding file is $magpat_k80_g00.gif$.} \label{mapk80g0}
\end{figure*}

In order to explore the maximum possible magnification that this
caustic pattern produces, we have selected an extra trajectory
that crosses exactly over a conjunction of several caustics
(trajectory 7 in Figure \ref{mapk50g0}). The light curve is
separately shown in Figure \ref{curvas7k50g0}. We can see that
magnifications up to 45 times the unlensed intensity of the source
are possible in this configuration. The effect of the differential
enhancement is notoriously clear for this case.\\

\begin{figure*}
\begin{center}
\includegraphics[width=16cm,height=15cm]{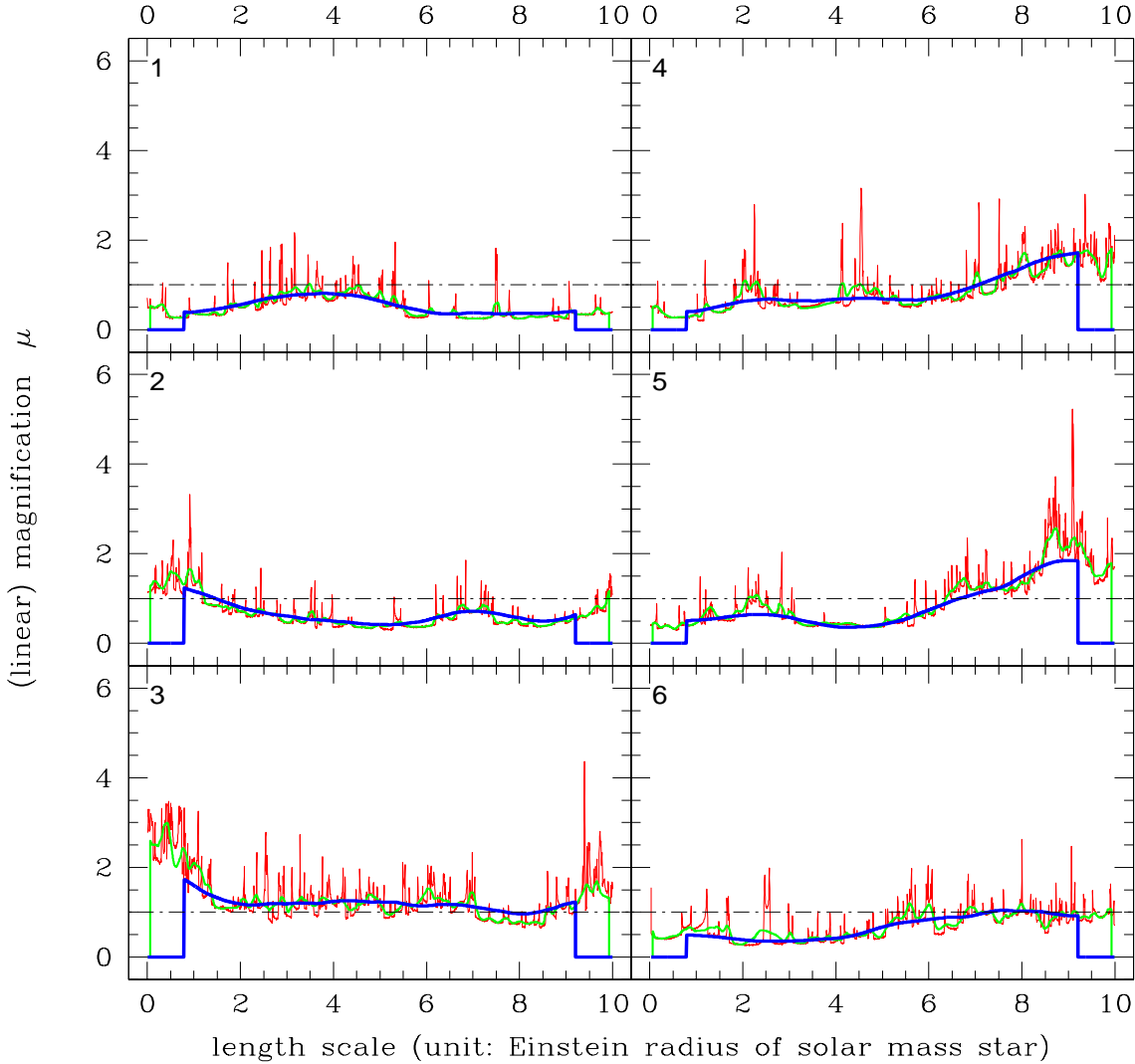}
\end{center}
\caption{Light curves for different source trajectories. Numbers
corresponds to those given in Figure  \ref{mapk80g0}. The line
coding  is as in Figure \ref{curvask50g0}.} \label{curvask80g0}
\end{figure*}

In Figure \ref{mapk80g0} we show the magnification map
corresponding to a higher value of focusing, $\kappa=0.8$. Still
in this case, the shear $\gamma$ is taken equal to zero. As it was
found by Schneider \& Weiss (1987), the critical structure become
more complex with increasing $\kappa$ and is no longer possible to
identify a constellation of compact objects. In addition, we also
see (as in Figure \ref{mapk50g0}) the tendency of the caustic
structure to cluster, generating some crowed critical regions and
some others devoid of high magnification patterns. The explanation
for this was already given by Schneider \& Weiss (1987): The
clustering of caustics is just the non-linear enhancement of
random (Poisson) clustering of the positions of lenses in the
source plane. Over-dense regions tend to attract other over-dense
regions, because gravity is a long range force. \\

\begin{figure}
\begin{center}
\includegraphics[width=8cm,height=9cm]{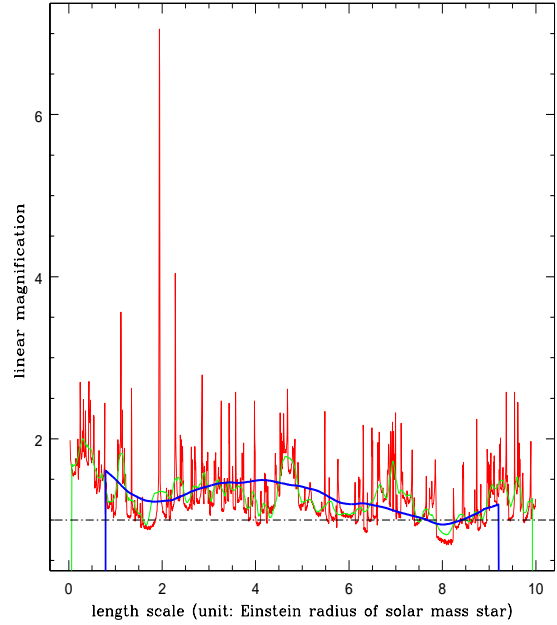}
\end{center}
\caption{Light curve for trajectory number 7, producing one of the
highest levels of magnification out of map given in Figure
\ref{mapk80g0}. The line coding  is as in Figure
\ref{curvask50g0}.} \label{curvas7k80g0}
\end{figure}

The most important feature shown in the map of Figure
\ref{mapk80g0} is that the magnitude of the magnification has been
typically reduced with respect to the $\kappa = 0.5$ case, and
this reduction reach two orders of magnitude with respect to the
single events. The density of caustics is so large that the light
curve is continuously affected by them, producing a less dramatic
combined effect. This effect was first studied by Deguchi \&
Watson (1987): The total magnification is always high, but the
fluctuations decreases beyond $\kappa=0.5$. This effect can be
seen in the light curves presented in Figure \ref{curvask80g0}.
Even for the particularly chosen trajectory, number 7, we find a
maximum magnification of 7. The full light curve is presented in
Figure \ref{curvas7k80g0}. Again, the differential effect is
notorious.\\

\begin{figure*}
\begin{center}
\end{center}
\caption{Magnification map for lensing with parameters
$\kappa=0.2$ and $\gamma=0.16$. For details, see text. The
corresponding file is $magpat_k20_g16.gif$.} \label{mapk20g16}
\end{figure*}

\begin{figure*}
\begin{center}
\includegraphics[width=16cm,height=15cm]{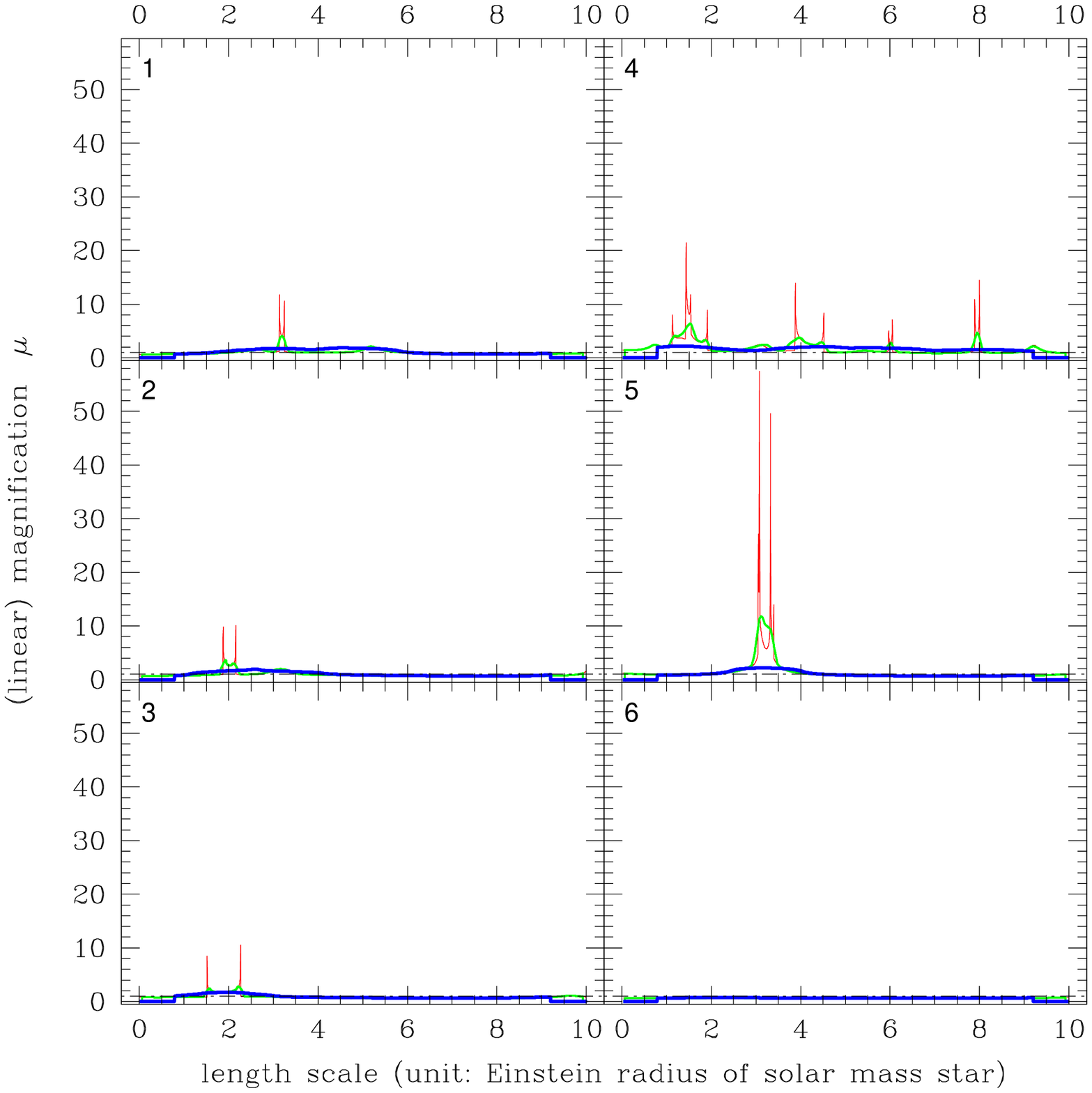}
\end{center}
\caption{Light curves for different source trajectories. Numbers
corresponds to those given in Figure \ref{mapk20g16}. The line
coding  is as in Figure \ref{curvask50g0}.} \label{curvask20g16}
\end{figure*}

In Figure \ref{mapk20g16} we show the magnification pattern for
the case $\kappa=0.2$ and $\gamma=0.16$. The presence of shear
modifies qualitative the magnification map. Here, most of the map
is devoid of magnification, and so, many of the common
trajectories (numbers 1 to 6) cross large regions of very low or
nil magnification (see particularly trajectory 6). However, those
trajectories actually crossing the caustics produce enhancements
in intensity typically between 10 an 20 times the unlensed value.
These effects can be seen in the six panels of Figure
\ref{curvask20g16}, where the differential magnification for the
different $\gamma$-spheres is also noticeable. In addition, the
enhancements of intensity are usually well separated (see for
instance trajectory 4). The enhancements themselves are
reminiscent of those produced by single events considered in
Section 6, in those cases where the presence of shear was
significant. \\

\begin{figure}
\begin{center}
\includegraphics[width=8cm,height=9cm]{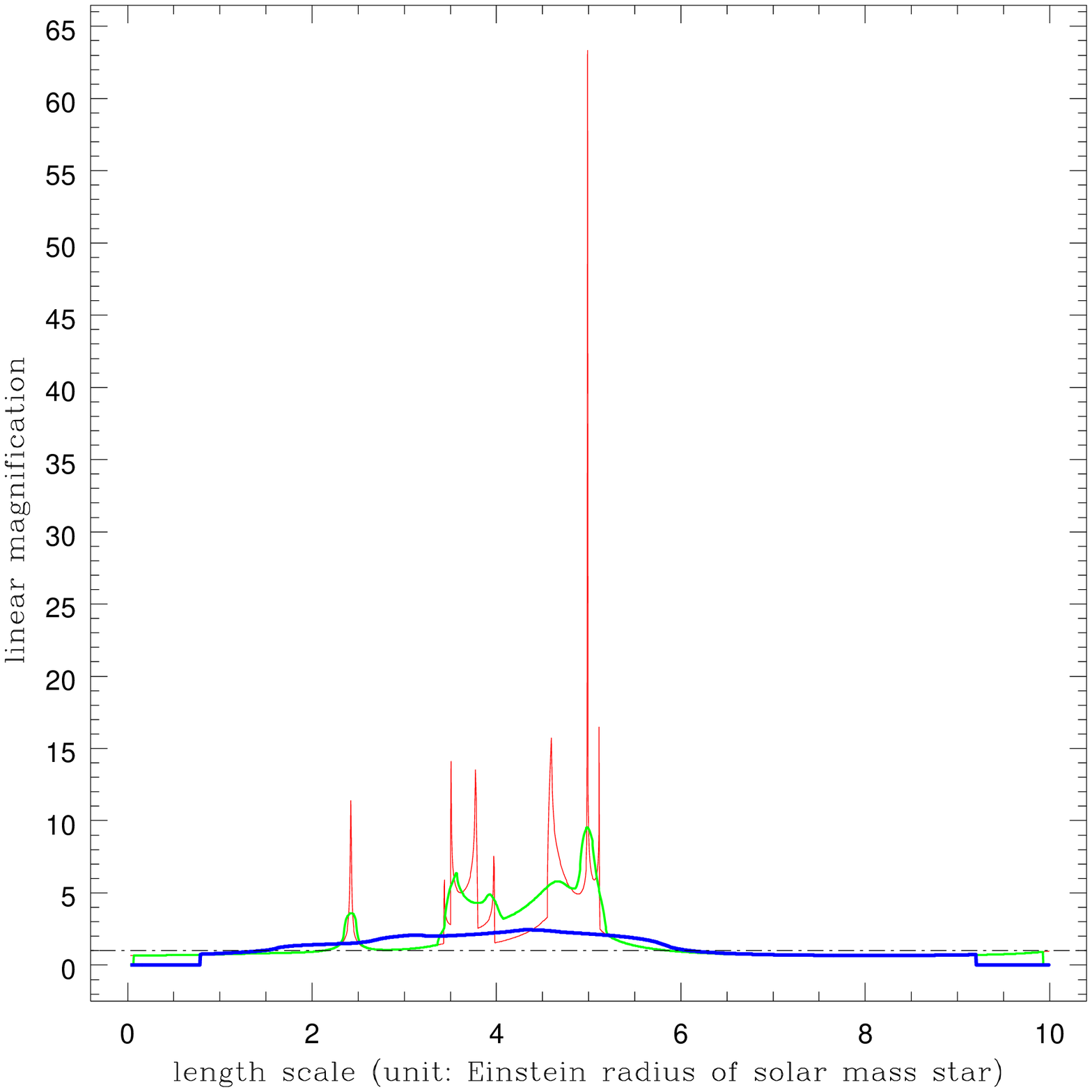}
\end{center}
\caption{Light curve for trajectory number 7, producing one of the
highest levels of magnification out of map given in Figure
\ref{mapk20g16}. The line coding  is as in Figure
\ref{curvask50g0}.} \label{curvas7k20g16}
\end{figure}

From the point of view of unidentified $\gamma$-ray sources,
interposed galaxies with low values of $\kappa$ and $\gamma$ are
probably the most interesting case for the application of the
model. In Figure \ref{curvas7k20g16} we show the case for the
hand-selected light curve. In that case, the magnification reaches
a factor of 65. The probability for this trajectory is less than
those of average enhancement. However, even if the probability is
reduced by, say, a factor $1/A^2$, it is possible to expect many
cases of high magnification, like that presented in Figure
\ref{curvas7k20g16}.\\

\begin{figure*}
\begin{center}
\end{center}
\caption{Magnification map for lensing with parameters
$\kappa=0.9$ and $\gamma=0.4$. For details, see text. The
corresponding file is $magpat_k90_g40.gif$.} \label{mapk90g40}
\end{figure*}

\begin{figure*}
\begin{center}
\includegraphics[width=16cm,height=15cm]{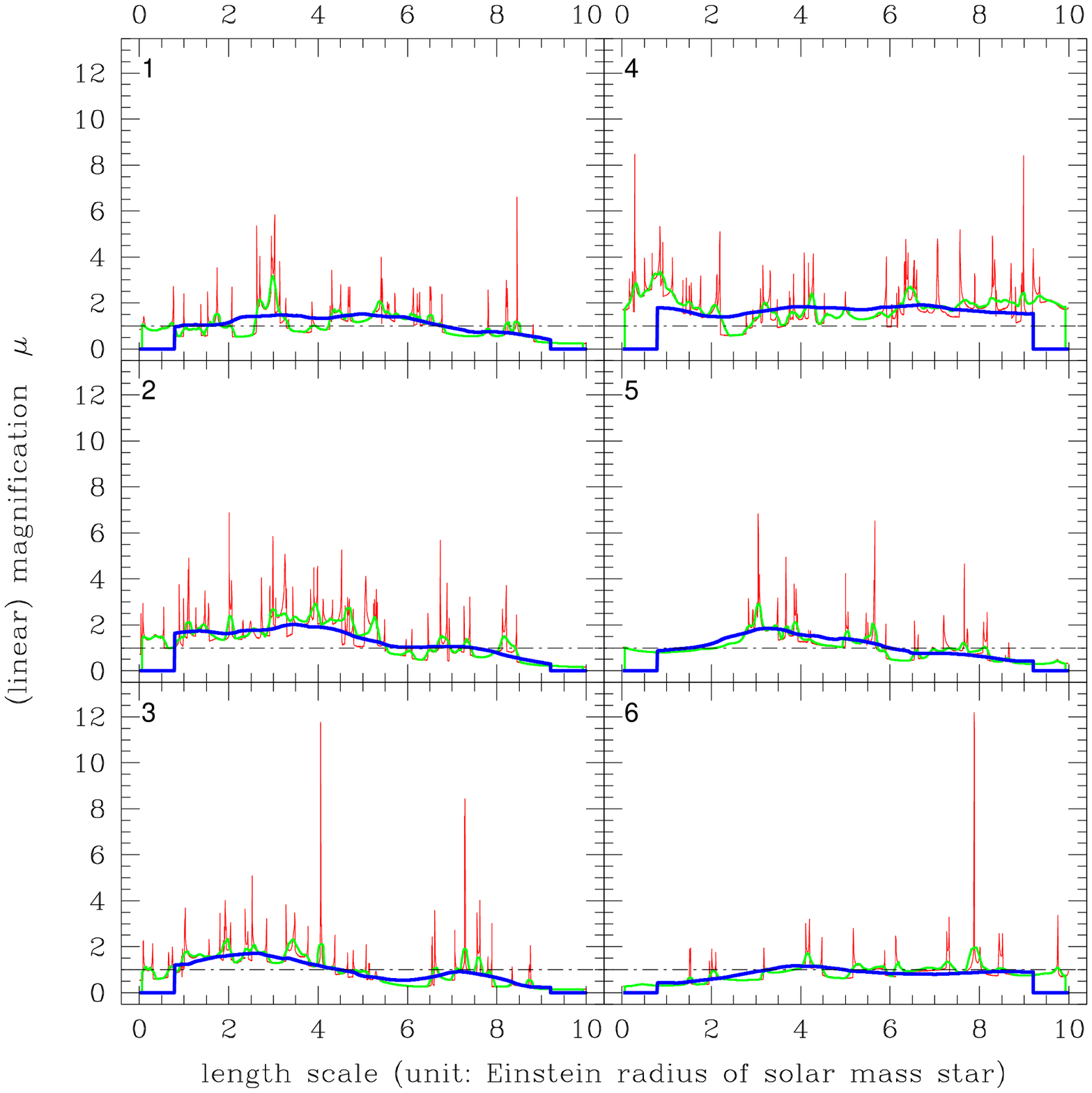}
\end{center}
\caption{Light curves for different source trajectories. Numbers
corresponds to those given in Figure \ref{mapk90g40}. The line
coding  is as in Figure \ref{curvask50g0}.} \label{curvask90g40}
\end{figure*}

Finally, in Figure \ref{mapk90g40} we show the magnification map
corresponding to the case $\kappa=0.9$ and $\gamma=0.4$. Again,
the high value of $\kappa$ makes the critical structure highly
complex. Typically, the magnification values are below a factor of
10 of the unlensed intensity, although some trajectories are found
(see, for example, curves number 3 and 4 of Figure
\ref{curvask90g40}) where a factor of $\sim 10$ is reached in two
well separated regions. Trajectory 7 (whose light curve is given
in Figure \ref{curvas7k50g0}) shows an enhancement of 22 times the
unlensed intensity.

\begin{figure}
\begin{center}
\includegraphics[width=8cm,height=9cm]{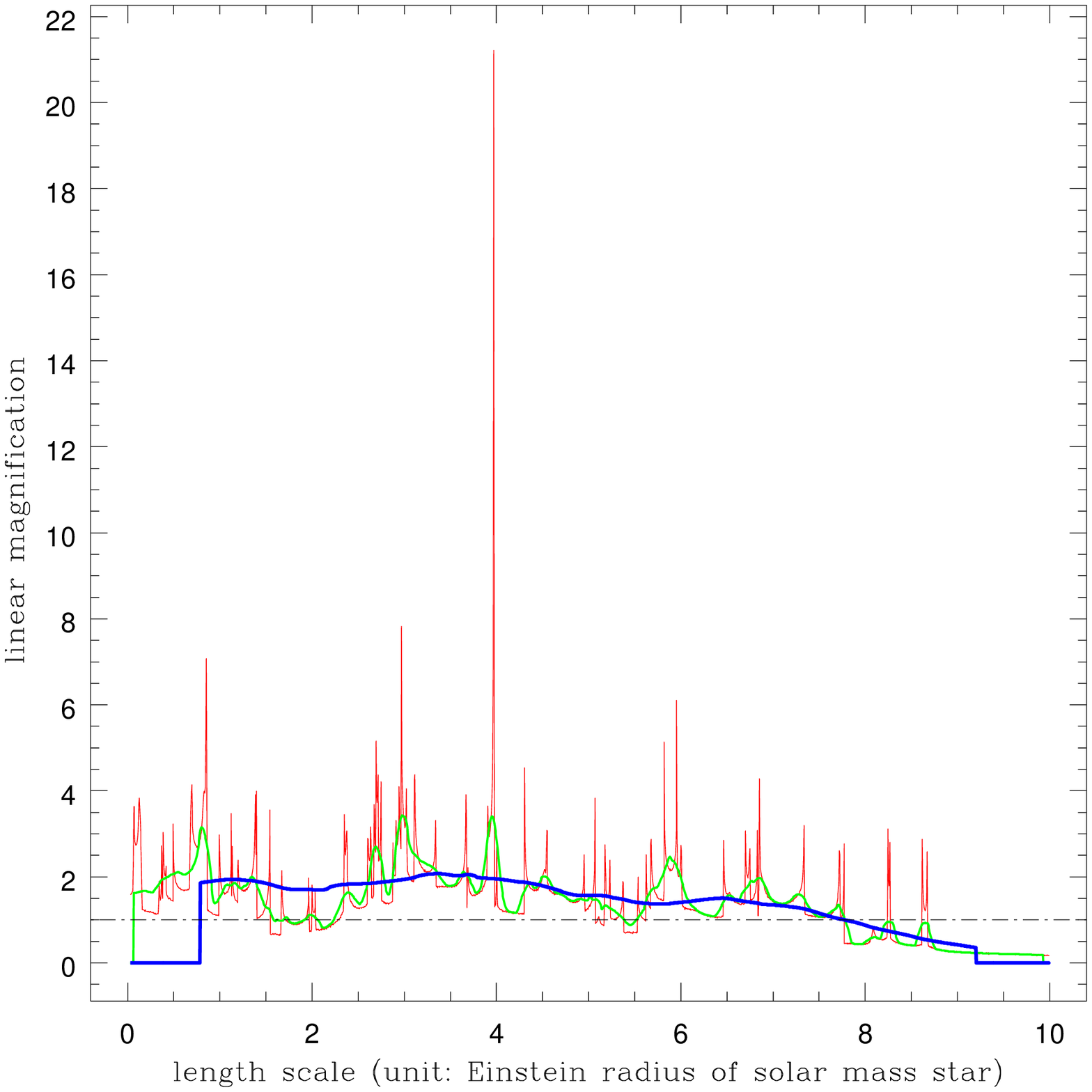}
\end{center}
\caption{Light curve for trajectory number 7, producing one of the
highest levels of magnification out of map given in Figure
\ref{mapk90g40}. The line coding  is as in Figure
\ref{curvask50g0}.} \label{curvas7k90g49}
\end{figure}

\subsection{Speculations on
astrophysical applications}

Let us assume that a correct alignment of source, lens, and
observer has been produced and that microlensing is operative. A
possible astrophysical application of the effects we have
discussed in the previous sections is to constrain the exponent in
the relationship between size and energy for $\gamma$-ray spheres
in the AGNs , $R \propto E^\alpha$. The smaller the source, the
higher is the peak $\gamma$-ray luminosity, and shorter will
result the rise time scale ($\tau=R/v$). Therefore, by observing
throughout the peak at different energies, one could determine the
relative size of the source at such energies, and then test the
radius--energy theoretical relationship. This could be done within
the range of GLAST capabilities. If, somehow, we know the relative
velocity and the redshifts, we could in even determine the sizes
of the different emitting regions. This would directly impact on
the underlying $\gamma$-ray models (e.g. Becker \& Kafatos 1995,
Blandford \& Levinson 1995).

\section{Summary and final comments}

To summarize, we have shown in this paper that
\begin{itemize}
  \item Some of the high-latitude unidentified $\gamma$-ray sources
  (both variable and non-variable)
  could be weak $\gamma$-ray emitting AGNs that are magnified
  through gravitational microlensing by stars in foreground
  galaxies.\\
  \item Although small, the probability of gravitational
  microlensing could be enough to explain a handful of the EGRET
  detections, and maybe many of the forthcoming GLAST
  detections.\\
  \item During a $\gamma$-ray
variability event produced by microlensing of a blazar, there is a
peculiar spectral evolution that could be detected in principle by
the next generation of $\gamma$-ray observatories with fine
spectral capabilities, like GLAST, or even by ground $\gamma$-ray
telescopes located at sufficient altitude, like 5@5 (e.g.
Aharonian et al. 2001).
\end{itemize}
It should be clearly stated that this model can not account for
all unidentified $\gamma$-ray detections at high latitudes.
However, it is interesting to ask whether the proposed
microlensing scenario could be responsible for $\gamma$-ray
variability of some radio loud AGNs already detected by EGRET.
Indeed, there is one possible case, related with the source 3EG
J1832-2110, which has been identified with PKS 1830-211 (Mattox et
al. 1997, Combi \& Romero 1998). The latter is a flat-spectrum
radio source, proposed to be a gravitational lensed QSO by Pramesh
Rao \& Subrahmanyan (1988). The $\gamma$-ray source is likely
variable, presenting  a value of $I=2.5$, and a steep spectral
index, $\Gamma=2.59\pm 0.13$. Both facts, variability and a steep
spectra, argue against a galactic origin for this source. Mattox
et al. (2001) assign to this pair an a priori probability of 0.998
of being correct, and list it within the most likely AGN
identification of the Third EGRET Catalog.\\

High resolution radio images obtained from several interferometric
arrays have revealed that the source has a ring-like structure
with two bright components on sub-arcsecond scales (Jauncey et al.
1991). This suggests a close alignment of the lensed source behind
the lensing object. Two absorption systems have been detected at
$z \sim 0.89$ (Wiklind \& Combes 1996) and $z \sim 0.193$ (Lovell
et al. 1996), so it seems likely that the image of the background
QSO (with a redshift $z > 0.885$, Mattox et al. 2001) is lensed by
two different extragalactic objects, what would undoubtedly
enhance the optical depth and the number of expected
single microlensing events.\\

Indeed, Combi \& Romero (1998) have already proposed that the
$\gamma$-ray emission of 3EG J1832-2110 (then 2EG J1834-2138)
could be produced by gravitational microlensing, using exactly the
same ideas we have deepen in this paper. They have found that
assuming a redshift $z_s = 1$ for the background source and $z_l
\sim 0.89$ for the lens, a MACHO-like object with $M \sim 0.02
M_\odot$ and moving with a low velocity of only $v\sim 1000$ km
s$^{-1}$ would be enough to produce the observed variability. For
this to be possible, the size of the $\gamma$-ray emitting region
should be of about 1.5 10$^{15}$ cm, in good agreement with the
source sizes used in this paper. These results can be slightly
modified by new measurements of the quasar redshift, but will not
change substantially (Oshima et al. 2001).\\

 It is likely then that the first
realization of this proposed mechanism have been already observed
for the pair 3EG J1832-2110/PKS 1830-211. One thing should be
remarked, though: in this case the background source is already a
strong radio emitter --what indeed facilitates the
identification--. This was not the general case we have considered
here, where the sources are weak enough as to yield no significant
lower energy counterparts. A complete microlensing model for the
$\gamma$-ray variability of PKS 1830-211, based on the discussion
presented in this paper, will be presented in a forthcoming
publication.

\section*{Acknowledgments}

This work has been supported by Universidad de Buenos Aires
(UBACYT X-143, EFE), CONICET (DFT, and PIP 0430/98, GER), ANPCT
(PICT 98 No. 03-04881, GER), Princeton University (DFT) and
Fundaci\'{o}n Antorchas (through separates research grants to GER
and DFT). GER was on leave from IAR during part of this research.
We acknowledge insightful comments by Drs. E. Turner, R. Hartman,
and D. Thompson in an early stage of this research.

\section*{Appendix: Solutions to the 1D lens equation}

\bigskip Replacing Eqs. (\ref{p15}) and (\ref{p16}) in Eq. (\ref{p17})
we obtain \ben A_{0}=\frac{1}{\left| 1-\kappa \right| }\left(
\frac{\left( \frac{\varepsilon
Y}{2}+\sqrt{\frac{Y^{2}}{4}+\varepsilon }\right) ^{4}}{\left|
\left( \frac{\varepsilon Y}{2}+\sqrt{\frac{Y^{2}}{4}+\varepsilon
}\right) ^{4}-1\right| }+ \right. \nonumber \\ \left. \frac{\left(
\frac{\varepsilon Y}{2}-\sqrt{\frac{Y^{2}}{4} +\varepsilon
}\right) ^{4}}{\left| \left( \frac{\varepsilon Y}{2}-\sqrt{
\frac{Y^{2}}{4}+\varepsilon }\right) ^{4}-1\right| }\right),
\label{a} \een and \ben A_{0}=\frac{1}{\left| 1-\kappa \right|
}\left( \frac{\left( \varepsilon Y+ \sqrt{Y^{2}+4\varepsilon
}\right) ^{4}}{\left| \left( \varepsilon Y+\sqrt{
Y^{2}+4\varepsilon }\right) ^{4}-16\right| }+\right. \nonumber \\
\left.\frac{\left( \varepsilon Y- \sqrt{Y^{2}+4\varepsilon
}\right) ^{4}}{\left| \left( \varepsilon Y-\sqrt{
Y^{2}+4\varepsilon }\right) ^{4}-16\right| }\right) .\label{b}
\een Let us now analyze different situations, starting with the
case $\varepsilon =1$, $Y>0$. For it we get, \ben
A_{0}=\frac{1}{\left| 1-\kappa \right| }\left( \frac{\left( Y+\sqrt{Y^{2}+4}%
\right) ^{4}}{\left( Y+\sqrt{Y^{2}+4}\right) ^{4}-16}+\right. \nonumber \\ \left.
\frac{\left( Y-\sqrt{%
Y^{2}+4}\right) ^{4}}{16-\left( Y-\sqrt{Y^{2}+4}\right)
^{4}}\right). \label{c1} \een If we have the case $\varepsilon
=1$, $Y<0$, we get \ben
A_{0}=\frac{1}{\left| 1-\kappa \right| }\left( \frac{\left( Y+\sqrt{Y^{2}+4}%
\right) ^{4}}{16-\left( Y+\sqrt{Y^{2}+4}\right) ^{4}}+\right. \nonumber \\ \left.
\frac{\left( Y-\sqrt{%
Y^{2}+4}\right) ^{4}}{\left( Y-\sqrt{Y^{2}+4}\right)
^{4}-16}\right) \label{c2} \een For the case in which $\varepsilon
=-1$, $Y>2$, the magnification results in \ben
A_{0}=\frac{1}{\left| 1-\kappa \right| }\left( \frac{\left( -Y+\sqrt{Y^{2}-4}%
\right) ^{4}}{16-\left( -Y+\sqrt{Y^{2}-4}\right) ^{4}}+\right.
\nonumber \\ \left.
\frac{\left( -Y-\sqrt{%
Y^{2}-4}\right) ^{4}}{\left( -Y-\sqrt{Y^{2}-4}\right)
^{4}-16}\right). \label{c3} \een And lastly, for the case in which
$\varepsilon =-1$, $Y<-2$, \ben
A_{0}=\frac{1}{\left| 1-\kappa \right| }\left( \frac{\left( -Y+\sqrt{Y^{2}-4}%
\right) ^{4}}{\left( -Y+\sqrt{Y^{2}-4}\right) ^{4}-16}+\right. \nonumber \\ \left.
\frac{\left( -Y-\sqrt{%
Y^{2}-4}\right) ^{4}}{16-\left( -Y-\sqrt{Y^{2}-4}\right)
^{4}}\right). \label{c4} \een The four cases can be jointly
written as \ben A_{0}=\frac{1}{\left| 1-\kappa \right|
}sg(Y)\varepsilon \left( \frac{\left(
\varepsilon Y+\sqrt{Y^{2}+4\varepsilon }\right) ^{4}}{\left( \varepsilon Y+%
\sqrt{Y^{2}+4\varepsilon }\right) ^{4}-16}+\right. \nonumber \\
\left.\frac{\left( \varepsilon Y-\sqrt{ Y^{2}+4\varepsilon
}\right) ^{4}}{16-\left( \varepsilon Y-\sqrt{ Y^{2}+4\varepsilon
}\right) ^{4}}\right).  \label{d} \een Simplifying the expression
between the parentheses, we have \ben \frac{\left( \varepsilon
Y+\sqrt{Y^{2}+4\varepsilon }\right) ^{4}}{\left( \varepsilon
Y+\sqrt{Y^{2}+4\varepsilon }\right) ^{4}-16} + \frac{\left(
\varepsilon Y-\sqrt{Y^{2}+4\varepsilon }\right) ^{4}}{16-\left(
\varepsilon Y-\sqrt{Y^{2}+4\varepsilon }\right) ^{4}}= \nonumber
\\ \frac{\varepsilon Y^{2}+2}{Y\sqrt{ Y^{2}+4\varepsilon }}.
\label{e} \een Then (using that $\varepsilon ^{2}=1$), we obtain
\ben A_{0}=\frac{1}{\left| 1-\kappa \right|
}sg(Y)\frac{Y^{2}+2\varepsilon }{Y \sqrt{Y^{2}+4\varepsilon }}=
\nonumber \\ \frac{1}{\left| 1-\kappa \right| }\frac{
Y^{2}+2\varepsilon }{sg(Y)Y\sqrt{Y^{2}+4\varepsilon }}, \label{f}
\een and as $sg(Y)Y=\left| Y\right| $,  Eq. (\ref{p18}) follows.

\end{document}